\documentclass[a4paper,titlepage,11pt]{article}
\usepackage[utf8]{inputenc}
\usepackage{geometry}
\usepackage[english]{babel}
\usepackage{amsmath, amsfonts, graphicx}
\usepackage{mathtools}
\usepackage{graphicx}
\usepackage{float}
\usepackage{pstricks}
 \usepackage{mciteplus}
\usepackage{hyperref}
\usepackage[labelfont=bf,font={small}]{caption}
\usepackage{pgfplots}
\usepackage{tikz}
\usepackage{axodraw4j}
\usetikzlibrary{positioning, arrows}
\usetikzlibrary{external}
\usepackage{footmisc}
\usepackage{multirow}
\usepackage{url}
\usepackage[breakable, theorems, skins]{tcolorbox}
\usepackage{enumerate}
\usepackage{physics}
\usepackage{bm}
\usepackage{epstopdf}

  \newlength{\abstractwidth}
\def\nspc{{\hspace{-2pt}}}

	\newcommand{\sltr}{\text{SL}(2,\mathbb{R})}

		\newcommand{\NS}{\text{NS}}
		\newcommand{\R}{\text{R}}
		\newcommand{\sixj}[6]{\Bigl\{\mbox{\small$\!\begin{array}{ccc} #1 \! & \!\! #3 \! & \!\! #5 \nspc \\[-1mm]  #2 \!  & \!\! #4 \!  & \!\! #6 \nspc \end{array}\!$}\!\Bigr\}}
			\newcommand{\lb}{\langle}
\newcommand{\rb}{\rangle}
			\newcommand{\U}{\scriptscriptstyle \text{U}}
	\newcommand{\M}{\scriptscriptstyle \text{M}}
	  \newcommand{\be}{\begin{equation}}
  \newcommand{\bea}{\begin{eqnarray}}
  \newcommand{\eea}{\end{eqnarray}}
  \newcommand{\beq}{\begin{equation}}
  \newcommand{\ee}{\end{equation}}
  \newcommand{\eeq}{\end{equation}}

		\numberwithin{equation}{section}

\begin{document}

\begin{titlepage}

\setcounter{page}{1} \baselineskip=15.5pt \thispagestyle{empty}

\vfil

${}$
\vspace{1cm}

\begin{center}

\def\thefootnote{\fnsymbol{footnote}}
\begin{changemargin}{0.05cm}{0.05cm} 
\begin{center}
{\Large \bf Degenerate operators in JT and Liouville (super)gravity}
\end{center} 
\end{changemargin}

~\\[1cm]
{Thomas G. Mertens\footnote{\href{mailto:thomas.mertens@ugent.be}{\protect\path{thomas.mertens@ugent.be}}}}
\\[0.3cm]

{\normalsize { \sl Department of Physics and Astronomy
\\[1.0mm]
Ghent University, Krijgslaan, 281-S9, 9000 Gent, Belgium}}\\[3mm]

\end{center}


 \vspace{0.2cm}
\begin{changemargin}{01cm}{1cm} 
{\small  \noindent 
\begin{center} 
\textbf{Abstract}
\end{center} 
We derive explicit expressions for a specific subclass of Jackiw-Teitelboim (JT) gravity bilocal correlators, corresponding to degenerate Virasoro representations. On the disk, these degenerate correlators are structurally simple, and they allow us to shed light on the 1/C Schwarzian bilocal perturbation series. In particular, we prove that the series is asymptotic for generic weight $h\notin - \mathbb{N}/2$. Inspired by its minimal string ancestor, we propose an expression for higher genus corrections to the degenerate correlators. We discuss the extension to the $\mathcal{N}=1$ super JT model. On the disk, we similarly derive properties of the 1/C super-Schwarzian perturbation series, which we independently develop as well. As a byproduct, it is shown that JT supergravity saturates the chaos bound $\lambda_L = 2\pi/\beta$ at first order in 1/C. We develop the fixed-length amplitudes of Liouville supergravity at the level of the disk partition function, the bulk one-point function and the boundary two-point functions. In particular we compute the minimal superstring fixed length boundary two-point functions, which limit to the super JT degenerate correlators. We give some comments on higher topology at the end.
}
\end{changemargin}
 \vspace{0.3cm}
\vfil
\begin{flushleft}
\today
\end{flushleft}

\end{titlepage}
\newpage
\tableofcontents

\setcounter{footnote}{0}

\section{Introduction}

Jackiw-Teitelboim (JT) gravity is a remarkable solvable theory of 2d quantum gravity \cite{Jackiw:1984je,Teitelboim:1983ux,Almheiri:2014cka,Jensen:2016pah,Maldacena:2016upp,Engelsoy:2016xyb}. The recent understanding of the significance of higher genus \cite{Saad:2019lba} and the relation to the black hole information paradox \cite{Saad:2019pqd, Almheiri:2019qdq, Penington:2019kki} have shown that one needs to understand and solve the gravitational theory in quite some detail to fully grasp the fundamental questions in quantum gravity. In this sense, JT gravity is relatively unique and it would be very beneficial if we could extend our knowledge to related models and deformations to learn of the generality of the proposed resolutions. In this sense, we refer to the exciting recent papers \cite{Witten:2020ert,Maxfield:2020ale,Witten:2020wvy}.
\\~\\
In this work, we study a useful subset of boundary correlation functions in JT gravity that is technically simpler to handle and that can be used to understand more deeply some of the structural aspects. This class of correlators plays the same role as degenerate Virasoro representations in 2d Virasoro CFT. We search for and find the same structure in the supersymmetric $\mathcal{N}=1$ version of JT gravity. 

Let us first review the general structure of boundary correlation functions within JT gravity. As well-known, 2d gravity has no bulk degrees of freedom, and by suitable choice of boundary term, gets all of its dynamics from a fluctuating wiggly boundary curve, representing a reparametrization of the boundary circle $F(\tau)$ \cite{Almheiri:2014cka,Jensen:2016pah,Maldacena:2016upp,Engelsoy:2016xyb}. The Lagrangian reduces to the Schwarzian derivative, and one can study correlators (at lowest topology) fully from just this Schwarzian system. Schwarzian quantum mechanics can be described as the 0+1 dimensional theory, described by a higher-derivative action of the form:
\begin{equation}
\label{SSch}
S[f] = -C\int d\tau \, \left\{F,\tau\right\}, \qquad \left\{F,\tau\right\} \equiv \frac{F'''}{F'} - \frac{3}{2}\left(\frac{F''}{F'}\right)^2,
\end{equation}
where $F(\tau)$ describes a time reparametrization subject to specific boundary/periodicity conditions to describe the physics of interest. The coupling constant $C$ has units of length and is inversely proportional to the 2d Newton constant $C \sim 1/ G_N$.\footnote{$C$ gets its units of length from the conformal symmetry breaking parameter in nAdS$_2$/nCFT$_1$. } Most studied is the thermal theory where one writes $F(\tau) = \tan \frac{\pi}{\beta}f(\tau)$ where $f(\tau+\beta) = f(\tau) + \beta$ describes a reparametrization of $S^1$. This theory has been studied and solved by several different techniques, both for the partition sum as for a certain class of correlation functions, composed of bilocal operator insertions of the type:
\begin{equation}
\label{bil}
\mathcal{O}_h(\tau_1,\tau_2) \equiv \left(\frac{F'(\tau_1)F'(\tau_2)}{(F(\tau_1)-F(\tau_2))^2}\right)^{h} .
\end{equation}
This operator can be viewed as a reparametrized matter CFT two-point function, labeled by the real number $h$, the weight of the matter CFT operator. \\
One can study correlation functions of these operators by perturbing $f(\tau) = \tau + \epsilon(\tau)$ for a periodic function $\epsilon(\tau)$ and then study the $1/C$ expansion in the Schwarzian coupling constant. This corresponds physically to an expansion in boundary graviton fluctuations $\epsilon$. One-loop results and the first subleading corrections to the four-point function and its chaotic Lyapunov behavior were studied in \cite{Maldacena:2016upp}. Schwarzian perturbation theory has applications also for higher-point functions \cite{Haehl:2017pak} and for matter correlators in 2d de Sitter space \cite{Maldacena:2019cbz,Cotler:2019nbi}. Higher loop corrections were recently analyzed in \cite{Qi:2019gny}. 

By relating this system as a particular limit of known dynamical systems, one also has access to the exact answers for the correlation functions. In particular, we have the well-known results for the one-loop exact partition function \cite{Maldacena:2016upp,Cotler:2016fpe,Stanford:2017thb}:
\begin{equation}
Z = \left\langle \mathbf{1}\right\rangle_{\beta} = \left(\frac{2\pi C}{\beta}\right)^{3/2}e^{\frac{2\pi^2 C}{\beta}},
\end{equation}
the one-loop exact Schwarzian derivative (or stress tensor) expectation value \cite{Stanford:2017thb,Mertens:2017mtv}:\footnote{Subtracting the zero-temperature answer, $n$-Schwarzian derivative correlators are $n$-loop exact.}
\begin{equation}
\label{sch1pt}
\left\langle \left\{F,\tau\right\}\right\rangle_\beta \equiv \left\langle \left\{\tan \frac{\pi}{\beta}f , \tau\right\}\right\rangle_{\beta} = \frac{1}{\beta Z} \frac{\partial Z}{\partial C} = \frac{2\pi^2}{\beta^2} + \frac{3}{2C \beta},
\end{equation}
and the bilocal disk correlators \cite{Bagrets:2016cdf,Bagrets:2017pwq,Mertens:2017mtv,Mertens:2018fds,Blommaert:2018oro,Blommaert:2018iqz,Kitaev:2018wpr,Yang:2018gdb,Iliesiu:2019xuh}:
\begin{align}
\label{sch2pt}
\left\langle \mathcal{O}_h(\tau,0)\right\rangle_\beta &= \left\langle \left(\frac{F'_1F'_2}{(F_1-F_2)^2}\right)^h\right\rangle_{\beta} =  \left\langle \left(\frac{f'_1f'_2}{\frac{\beta^2}{\pi^2}\sin \frac{\pi}{\beta}(f_1-f_2)^2}\right)^h\right\rangle_\beta \nonumber\\
 &= \frac{1}{Z}\frac{1}{(2C)^{2h}} \int d\mu(k_1) \int d\mu(k_2)\, e^{ - \tau \frac{k_1^2}{2C}}\,  e^{-(\beta-\tau) \frac{k_2^2}{2C}}\frac{ \Gamma( h \pm i k_1 \pm i k_2)}{{2\pi^2}\, \Gamma(2h)},
\end{align}
where $d\mu(k)= dk k \sinh(2\pi k)$ and the $\pm$-notation denotes the product of all cases. The semi-classical (large $C$) gravitational content of correlators like this was studied in detail in \cite{Lam:2018pvp}, see also \cite{Goel:2018ubv}. The bilocal correlators get contributions from all orders in $G_N \sim 1/C$, and have non-perturbative content as well of the order $\sim e^{\#/G_N}$. The last statement will be proven in this work.
\\~\\
In \cite{Saad:2019pqd}, see also \cite{Blommaert:2019hjr,Blommaert:2020seb},  the contributions of including higher genus handles to the disk geometry was argued to lead to the replacement:
\begin{align}
\label{2genexpi}
\left\langle \mathcal{O}_h(\tau,0)\right\rangle_\beta &= \frac{1}{(2C)^{2h}Z} \int dE_1 dE_2\, \rho(E_1,E_2)\, e^{ - \frac{\tau}{2C} E_1}\,  e^{-\frac{\beta-\tau}{2C} E_2}\frac{ \Gamma( h \pm i \sqrt{E_1} \pm i \sqrt{E_2})}{{8\pi^2}\, \Gamma(2h)},
\end{align}
in terms of the energy variable $E_i = k_i^2$ and where the only new thing is the pair density correlator $\rho(E_1,E_2)$ coming from the matrix ensemble underlying JT gravity \cite{Saad:2019lba}. This pair density correlator has a genus expansion:
\begin{equation}
\rho(E_1,E_2) = \sum_g \rho_g(E_1,E_2) + (\text{non-pert}),
\end{equation}
where $\rho_g(E_1,E_2) \sim e^{\chi S_0}$ is weighted by the Euler character $\chi$ and the (double-scaled) matrix parameter $L \equiv e^{S_0}$. Since $S_0$ can be understood in gravity language as the extremal black hole entropy with $S_0 \sim 1/G_N$, these perturbative higher genus effects are actually non-perturbative in $G_N$, of the same order as the non-perturbative corrections to the disk correlator. Next to all of this, there are further non-perturbative corrections in $e^{-S_0}$ that are very important to understand the very late-time physics, but will not play a major role in this work.
\\~\\
In order to gain a better understanding of all of these features, we focus in this work on the special bilocal operators where $h\in -\mathbb{N}/2$.\footnote{In this case, both numerator and denominator of the ratio of gamma-functions in \eqref{sch2pt} or \eqref{2genexpi} diverge. The numerator does so only along the codimension 1 subspace of the integrals where $k_1 = k_2 \pm i m$ for some (half)integer $m$. This simplifies the expressions.} They originate from non-unitary matter insertions at the holographic boundary, and correspond to the degenerate Virasoro representations, or to the finite dimensional (non-unitary) representations of $\mathfrak{sl}(2)$. Their correlators exhibit a simplified structure which will allow us to investigate structural aspects that are more hard to access for generic $h$, both at the disk level and for the role and nature of higher genus corrections. \\
We will call these operators \emph{degenerate operators}, in analogy with their origins in Virasoro representation theory.
\\~\\
In this work, we will establish the following results.
\begin{itemize}
\item 
\textbf{Schwarzian small $\tau$ series expansion.}
Exploiting the knowledge of the degenerate $h\in -\mathbb{N}/2$ bilocal correlators, we will demonstrate the general series expansion in the time separation $\tau$ between both endpoints of the bilocal operator of the finite-temperature bilocal correlator for \emph{generic} real $h$ has the structure:
\begin{align}
\label{smaltau}
\left\langle \mathcal{O}_{h}(\tau,0)\right\rangle &= \frac{1}{\tau^{2h}}\left[ 1 + \frac{h(h-1)}{6C} \tau + \left(a_0\left\langle T\right\rangle + b_0\right) \tau^2 + \left(\# \left\langle T\right\rangle + b_1\right)\tau^3 \right. \nonumber \\
&\left.+ \left( a_1 \left\langle T^2\right\rangle + \# \left\langle T\right\rangle + b_2\right) \tau^4  + \left( \# \left\langle T^2\right\rangle + \# \left\langle T\right\rangle + b_3\right) \tau^5 \right. \nonumber \\
&\left.+ \left( a_2 \left\langle T^3\right\rangle + \# \left\langle T^2\right\rangle + \# \left\langle T\right\rangle+ b_4\right) \tau^6 + \hdots \right],
\end{align}
with $\left\langle T^n\right\rangle$ the thermal piece of the renormalized (or point-split) multi-Schwarzian derivative correlation function. We give expressions below. The structure of this expansion is readily generalized to higher orders in the $\tau$-expansion.\footnote{The coefficients $a_n$ are determined by the small $\tau$ expansion of the semi-classical ($C\to +\infty$) answer, and the $b_n$ can be determined through the zero-temperature result. The other coefficients can in principle also be deduced by exploiting the fact that the coefficients are polynomials in $h$, and hence knowing a small set of datapoints is sufficient to fix the polynomial. The usual correlators \eqref{sch2pt} cannot be used as datapoints since we are not able to analytically write down this expansion, but the degenerate bilocals ($h\in -\mathbb{N}/2$) can.}

\item 
\textbf{Asymptotic vs convergent Schwarzian perturbation series.}
We will prove that the $1/C$ Schwarzian perturbative series on the disk is asymptotic for any real $h \notin -\frac{\mathbb{N}}{2}$, with non-perturbative effects in $G_N$ of order $e^{\#/G_N}$ that go beyond the boundary graviton expansion. The degenerate values $h \in -\frac{\mathbb{N}}{2}$ on the other hand yield convergent perturbation series. The actual proof is contained in appendix \ref{s:proof}. The situation is summarized in Figure \ref{range}.
\begin{figure}[H]
\centering
\includegraphics[width=0.3\textwidth]{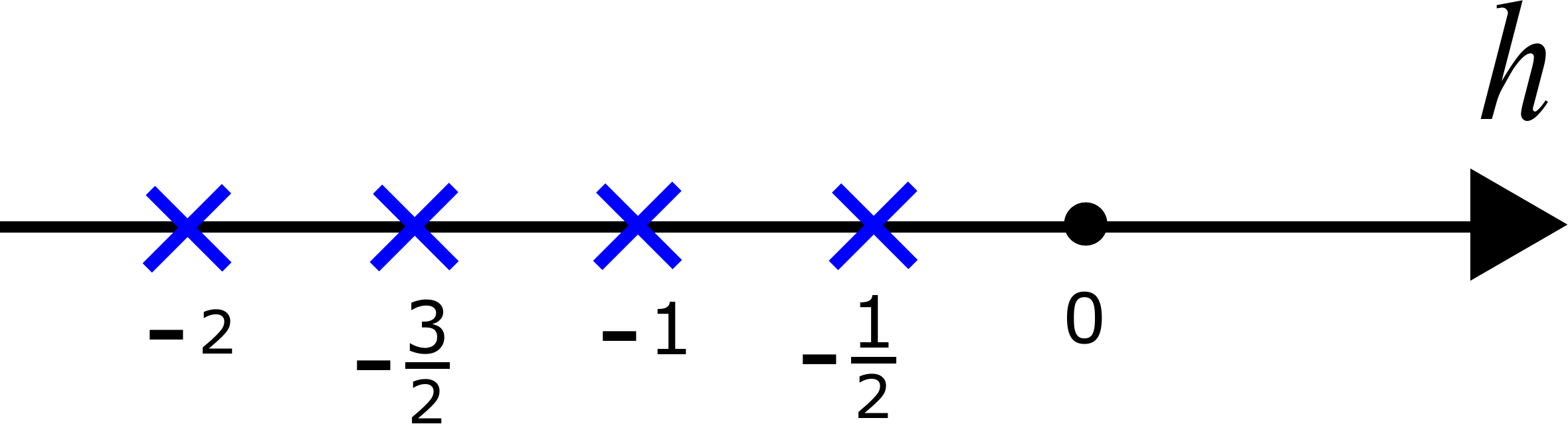}
\caption{The perturbative $1/C$ series is asymptotic, except when $h \in -\mathbb{N}/2$.}
\label{range}
\end{figure}

\item
\textbf{Degenerate operators as limit of minimal string primaries.}
JT gravity can be viewed as a double-scaling limit of the minimal string, consisting of the 2d Liouville CFT, a matter minimal model, and the $bc$ ghosts to cancel the conformal anomaly. This was first noticed in \cite{Saad:2019lba,StanfordSeiberg}, and preliminary remarks concerning boundary correlators were made in the conclusion of \cite{Mertens:2019tcm}. We presented a detailed treatment of this in \cite{Mertens:2020hbs}. 
Related recent results can be found in e.g. \cite{Betzios:2020nry,Johnson:2019eik,Johnson:2020heh,Johnson:2020exp}. Since the minimal string has a matrix model description, the same has to be true for its JT limit, and indeed this was the main result of the impressive work \cite{Saad:2019lba}.\footnote{See also \cite{Okuyama:2019xbv,Okuyama:2020ncd} for more JT computations using matrix model techniques.}

Within this framework, the natural bulk and boundary operators are minimal model primaries dressed by the gravitational Liouville piece. Within the JT limit, the minimal string (genus zero) boundary two-point correlator precisely limit to the degenerate JT correlators we determine in this work. That is, \emph{minimal string primaries are the ancestors of the JT degenerate operator insertions}. Because of this, they correspond to an integrable subsector of the JT gravity operator insertions.

\item
\textbf{Proposal for higher topology including degenerate operator insertions.}
In JT gravity at higher genus, we will hence draw inspiration from the overarching minimal string framework to define the degenerate bilocal correlator. We will show that it only has the same kind of higher topology than the partition function itself, signaling a departure from the generic $h$ bilocal correlator studied in \cite{Saad:2019pqd}. Diagrammatically, we draw:
\begin{equation}
\includegraphics[width=0.5\textwidth]{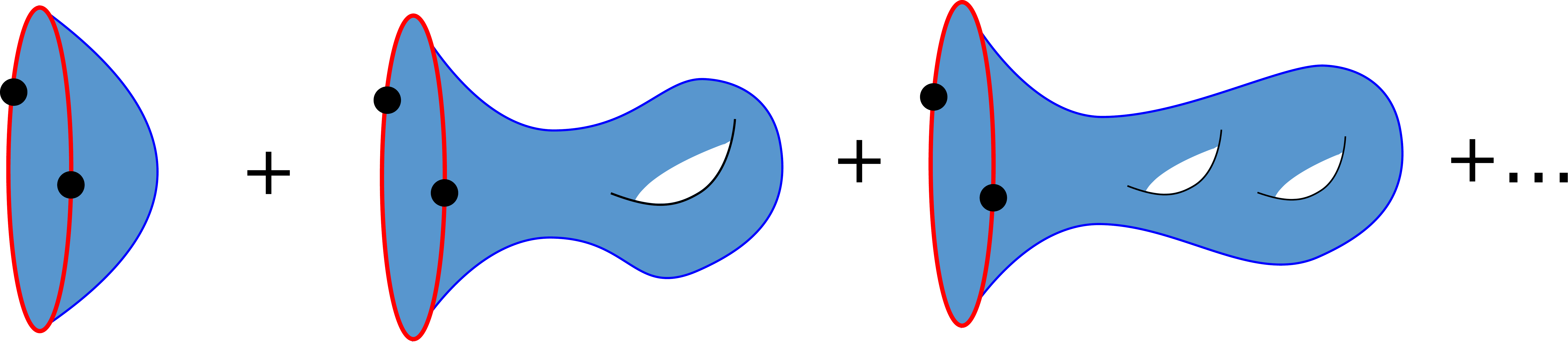}
\end{equation}
for the genus expansion of the degenerate bilocal correlator.
\end{itemize}

It is interesting to try to see to which extent this structure remains intact when going to related solvable models. To that effect, we generalize most of these results to the $\mathcal{N}=1$ super-JT theory in section \ref{s:susy}. This model arises in the universal low-energy description of supersymmetric versions of the SYK model \cite{Fu:2016vas}. As such it is important for a restricted class of condensed matter systems that might be realizable in the lab, and from this perspective should be considerd of similar importance as the bosonic JT model. We take the opportunity to develop the perturbative treatment of the boundary super-Schwarzian model as well in section \ref{s:susypert}. This allows a discussion on the self-energy of a matter field, getting contributions both from the graviton and the gravitino. A byproduct of this development is a quick proof that the leading Lyapunov behavior of the out-of-time-ordered 4-point function saturates the chaos bound, just as in the bosonic JT model.

In order to address higher genus contributions to the degenerate bilocals in super JT gravity, we will first develop our understanding in its overarching Liouville supergravity model. Roughly paralleling the developments in \cite{Mertens:2020hbs}, we construct the fixed length partition function, bulk one-point function and boundary two-point functions, focussing on the presence of a (super) JT limit. Armed with these results, we develop the matrix model perspective on the higher genus degenerate boundary two-point function of the minimal superstring, and find very analogous results to that of the bosonic case.
\\~\\
This work is structured as follows. In \textbf{section \ref{s:deg}} we provide the results on the degenerate JT bilocal correlators. \textbf{Section \ref{s:app1}} provides an application in terms of the Schwarzian small $\tau$ and $1/C$ perturbation series for generic weight $h$, realizing the first two goals mentioned above. We investigate the addition of higher topology to these degenerate correlators from the matrix model and Liouville gravity perspective in \textbf{section \ref{s:hightop}}, which realizes the last two above goals. This concludes the bosonic story.\\
The remainder of the work concerns the generalization of most of these statements to $\mathcal{N}=1$ supergravity. \textbf{Sections \ref{s:susy}} and \textbf{\ref{s:app2}} provide details on the $\mathcal{N}=1$ supersymmetric extensions and in particular the perturbative expansion. We make a detour into Liouville supergravity and the minimal superstring in the next few sections. We start in \textbf{section \ref{s:Lsugra}} with a summary of Liouville supergravities to set the stage, and to initiate our study of fixed length amplitudes. \textbf{Section \ref{s:pfone}} then applies this to the partition function, bulk one-point function, and boundary two-point functions. In \textbf{section \ref{s:matrix}}, we specify to the minimal superstring and give a matrix integral perspective on some of the previous results. The fixed length minimal superstring boundary tachyon correlators are determined, and finally a similar proposal on degenerate $\mathcal{N}=1$ JT supergravity correlators is formulated for arbitrary higher genus. \\
We mention some aspects of the story that are left for the future in the concluding \textbf{section \ref{s:conc}}. Some complementary and technical material is presented in the appendices. In particular, \textbf{appendix \ref{s:degfus}} contains the degenerate 6j symbol when degenerate lines cross in the bulk, given essentially by a Wilson polynomial in the external labels. \textbf{Appendix \ref{s:proof}} provides a thorough investigation on the perturbative content of generic bilocal correlators, in particular it contains a proof that the $1/C$ series is asymptotic generically. \textbf{Appendix \ref{s:ramond}} investigates boundary Ramond operator insertions leading to boundaries that change parity.

\section{Degenerate bilocal correlators in JT gravity}
\label{s:deg}

In this section, we determine closed formulas for the degenerate values of $h \in - \mathbb{N}/2$ of the bilocal operators. 

\subsection{Disk level: Schwarzian bilocals}
As mentioned in the introduction, Schwarzian correlation functions can be computed in several ways. One approach is to use a minisuperspace limit of Liouville CFT between identity branes with Liouville vertex operators \cite{Mertens:2017mtv}. Another is by using SL$(2,\mathbb{R})$ group theoretical methods to describe JT gravity in its first-order BF formulation, described independently in \cite{Blommaert:2018oro,Blommaert:2018iqz} and \cite{Iliesiu:2019xuh}.\footnote{These two versions of the BF model are not entirely the same, but lead to the same final results at least on the disk topology. We focus on the first of these.} At a technical level, the computation that is required is the same: the Liouville minisuperspace wavefunctions can be viewed as solutions to the $\mathfrak{sl}(2,\mathbb{R})$ Casimir eigenvalue equation upon imposing the gravitational constraints (the Hamiltonian reduction). This makes it a so-called parabolic matrix element or Whittaker function.
From either approach, one finds the wavefunction:\footnote{The normalization is a bit technical: from the Liouville perspective, there is an additional normalization factor $1/\Gamma(2ik)$ that makes this set of eigenfunctions an orthonormal basis with flat measure $d\phi$. This normalization factor conspires to give the Schwarzian density of states $k\sinh 2\pi k$. It is not interesting for our purposes and we don't write it. From the group theoretical perspective, the normalization of \eqref{wfbos} is the correct normalization for the so-called parabolic matrix element for SL$^+(2,\mathbb{R})$ subsemigroup to be used to find gravity from BF theory \cite{Blommaert:2018iqz}.}
\begin{equation}
\label{wfbos}
\psi_k(\phi) = K_{2ik}(e^{\phi}),
\end{equation}
in terms of the momentum label $k$, associated to the continuous principal series representation of $\mathfrak{sl}(2,\mathbb{R})$ with $j=-1/2+ik$. \\
The operator insertion is easily deduced from the Liouville perspective as the Liouville primary vertex operator $e^{2 h \phi}$. From the group-theoretical perspective, one needs the matrix elements of the discrete highest weight irreps of SL$(2,\mathbb{R})$, which turn out to be also given by $e^{2 h \phi}$ \cite{Blommaert:2018oro}. \\
In order to compute the vertex functions (the Gamma's) within a bilocal correlation function \eqref{sch2pt}, we include two such wavefunctions \eqref{wfbos}, and one operator insertion, and integrate over the auxiliary variable $\phi$:
\begin{equation}
\label{verdeg}
\int_{-\infty}^{+\infty} d\phi K_{2i k_1}(e^{\phi}) K_{2i k_2}(e^{\phi}) e^{2 h\phi} = 2^{2h-3}\frac{\Gamma(h \pm ik_1 \pm ik_2)}{\Gamma(2h)}.
\end{equation}
So far, this has been the standard story. For $h \in - \mathbb{N}/2$ however, the operator insertion is special in the following sense. \\
Scaling $\phi \to \phi/b$, degenerate Liouville vertex operators are of the form $V_{m,n} \equiv e^{2\alpha \frac{\phi}{b}}$ with $2\alpha = Q - m/b - nb$. To find a well-defined classical limit $b\to 0$, we set $m=1$ and $n \geq 1$, which with $n = 2j+1$ becomes $V_{1,2j+1} = e^{-2 j \phi}$, where $j = \frac{1}{2}, 1 , \frac{3}{2} \hdots$. Comparing to the non-degenerate primaries, this corresponds to effectively setting the weight $h = -j$. \\
From group theory, setting the irrep label $h =-j \in - \frac{\mathbb{N}}{2}$ is selecting the finite-dimensional (but non-unitary) irreps of SL$(2,\mathbb{R})$ of dimension $2j+1$. Some more details on this were presented in appendix D of \cite{Mertens:2019tcm}. From both perspectives it is clear that this choice will have a special structural significance.

Instead of \eqref{verdeg}, the vertex functions reduce to a linear combination of Dirac delta-functions:
\begin{equation}
\label{vertfunc}
\sum_{m=-j}^{+j} c_m^{j}(k_1,k_2) \delta (k_1 - k_2 + i m),
\end{equation}
with a set of momentum-dependent combinatorial prefactors $c_m^{j}(k_1,k_2)$. The explicit form can be determined using the 1d fusion property
\begin{equation}
\label{besselprop}
\frac{K_{\alpha}(x)}{x} = \frac{1}{2\alpha}\left( K_{\alpha+1}(x) - K_{\alpha-1}(x)\right),
\end{equation}
and the orthonormality property:
\begin{equation}
\label{orthoprop}
\int d\phi K_{2i k_1}(e^{\phi}) K_{2i k_2}(e^{\phi}) e^{-\phi} = \frac{1}{32 i k_1 k_2 \sinh(2\pi k_2)}\left(\delta(k_1-k_2 + \frac{i}{2}) - \delta(k_1-k_2 - \frac{i}{2})\right).
\end{equation}
The evaluation of \eqref{verdeg} for $h \in - \mathbb{N}/2$ requires repeated use of \eqref{besselprop}, leading to the schematic \eqref{vertfunc}.
\\~\\
After what is essentially a tedious combinatorial exercise, one finds the coefficients:
\begin{align}
c_m^j (k_1,k_2) &= \frac{1}{k_1 \sinh 2\pi k_1} (-)^{m+j}\left(\begin{array}{c}
2j \\
m+j \\
\end{array}\right)
\frac{(2ik_2 + 2m)}{(2ik_2 - j  + m)_{2j+1}},
\end{align}
where the last factor contains the Pochhammer symbol $(x)_n \equiv \Gamma(x+n)/\Gamma(x)$. This last factor represents the inverse of a polynomial in $k$ of order $2j$.
Despite appearances, the vertex function \eqref{vertfunc} is symmetric under $k_1 \leftrightarrow k_2$.
Inserting these into the finite-temperature two-point function, we get:
\begin{equation}
\label{degmain}
\boxed{
\left\langle \mathcal{O}_{h=-j}(\tau,0)\right\rangle = \frac{(2C)^{2j}}{Z}\int d\mu(k)e^{-\frac{\beta}{2C}k^2}\sum_{m=-j}^{+j} (-)^{m+j} e^{\frac{\tau m^2}{2C}} e^{\frac{-i m k \tau}{C}}\left(\begin{array}{c}
2j \\
m+j \\
\end{array}\right)
\frac{(2ik + 2m)}{(2ik - j  + m)_{2j+1}}},
\end{equation}
instead of the generic \eqref{sch2pt}. We first write some explicit examples, and afterwards we will discuss some properties of this expression.

The simplest example of $h =-1/2 =-j$ was explicitly written in \cite{Mertens:2019tcm}. In this case the $k$-integral can be done in terms of elementary functions:\footnote{For this special case $h=-1/2$, the formula \eqref{msy} given later simplifies enormously and matches the first subleading term in this expansion. This specific value seems to be the only case when that formula can be that much simplified, suggesting that this is the only case for which a very simple (and closed) expression can be found.}
\begin{align}
\label{halfexact}
\left\langle \mathcal{O}_{h=-1/2}(\tau,0)\right\rangle &= \left(\frac{\beta}{\pi}\sin \frac{\pi}{\beta}\tau \right) e^{\frac{\tau}{8C}(1-\frac{\tau}{\beta})} = \left(\frac{\beta}{\pi}\sin \frac{\pi}{\beta}\tau \right) \left[1 + \frac{\tau(\beta-\tau)}{8\beta C} + \hdots \right].
\end{align}
A second example is that of $h =-1$:
\begin{equation}
\label{oneexact}
\frac{(2C)^{2}}{Z}\int d\mu(k) e^{-\frac{\beta}{2C}k^2}\left[e^{\frac{\tau}{2C}}\left(\frac{e^{-\frac{ik\tau}{C}}}{(2ik+1)2ik} + \frac{e^{\frac{ik\tau}{C}}}{(2ik-1)2ik}\right) - \frac{2}{(2ik+1)(2ik-1)}\right].
\end{equation}

Let us now make some remarks on this result.
\begin{itemize}
\item
The appearance of a binomial expansion in \eqref{degmain} is no surprise since the bilocal operator itself can be expanded as such:
\begin{equation}
\mathcal{O}_{-j}(\tau_1,\tau_2) = \left(\frac{(F(\tau_1)-F(\tau_2))^2}{F'(\tau_1)F'(\tau_2)}\right)^{j} = \sum_{m=-j}^{+j}(-)^{m+j} \left(\begin{array}{c}
2j \\
m+j \\
\end{array}\right) \frac{F(\tau_1)^{j+m}F(\tau_2)^{j-m}}{F'(\tau_1)^{j}F'(\tau_2)^{j}}.
\end{equation}
A further indication of the simplified structure is found by transforming to the free field variable $F'=e^{\phi}$ \cite{Bagrets:2016cdf,Bagrets:2017pwq},\footnote{This transformation is closely related to the B\"acklund transformation in Liouville CFT, which can be seen from the dimensional reduction of Liouville to the Schwarzian theory \cite{Mertens:2018fds}.} transforming the Schwarzian action \eqref{SSch} into
\begin{equation}
S[\phi] = \frac{C}{2} \int_{0}^{\beta}d\tau (\partial_\tau \phi )^2,
\end{equation}
with constraint $\int_0^\beta d\tau e^\phi = +\infty$ and operator insertion:
\begin{equation}
\mathcal{O}_{-j}(\tau_1,\tau_2) = e^{-j\phi(\tau_1)}\left(\int_{\tau_1}^{\tau_2}d\tau e^\phi \right)^{2j}e^{-j\phi(\tau_2)},
\end{equation}
which since $2j$ is integer, is only a product of (complex) plane waves in a free theory. One can readily evaluate the $\phi$ path integral explicitly in this way and make contact with our main expression \eqref{degmain}. Since we have just plane waves in a free theory (up to the constraint), these operators can be viewed as an integrable subclass of the bilocal operators in JT gravity.
\item 
The zero-temperature result, and its small-separation expansion can be obtained by expanding \eqref{degmain} as $k\to 0$, and yields very simple closed expressions:
\begin{alignat}{2}
\label{zero}
&&\frac{1}{(2C)^{2}}\left\langle \mathcal{O}_{h=-1}(\tau,0)\right\rangle &= e^{\frac{\tau}{2C}}\left(\frac{\tau}{C}-2\right) +2 = \frac{\tau^2}{(2C)^2} + \mathcal{O}(\tau^3), \\
&&\frac{1}{(2C)^{3}}\left\langle \mathcal{O}_{h=-3/2}(\tau,0)\right\rangle &= e^{\frac{9\tau}{8C}}\left(\frac{3}{4}\frac{\tau}{C} - \frac{3}{2}\right)  - 3 e^{\frac{\tau}{8C}}\left(-\frac{1}{4}\frac{\tau}{C} - \frac{1}{2}\right)  =\frac{\tau^3}{(2C)^3} + \mathcal{O}(\tau^4), \nonumber \\
&&\frac{1}{(2C)^{4}}\left\langle \mathcal{O}_{h=-2}(\tau,0)\right\rangle&= e^{\frac{2\tau}{C}}\left(\frac{1}{3}\frac{\tau}{C} - \frac{11}{18}\right)  - 4 e^{\frac{\tau}{2C}}\left(-\frac{1}{3}\frac{\tau}{C} + \frac{2}{9}\right)  + \frac{3}{2} = \frac{\tau^4}{(2C)^4} + \mathcal{O}(\tau^5). \nonumber
\end{alignat}
\item 
In the semi-classical regime where $C \gg \tau,\beta$, the integral in \eqref{degmain} is dominated by large $k$ at its saddle $k \approx \frac{2C \pi}{\beta}$. The expression then evaluates to:
\begin{equation}
\sum_{m=-j}^{+j}(-)^{m+j} e^{-2i m \frac{\pi}{\beta} \tau}\left(\begin{array}{c}
2j \\
m+j \\
\end{array}\right)
\frac{1}{(2i \frac{\pi}{\beta})^{2j}} = \left(\frac{e^{i  \frac{\pi}{\beta} \tau} - e^{-i  \frac{\pi}{\beta} \tau}}{2i \frac{\pi}{\beta}}\right)^{2j} = \left(\frac{\beta}{\pi}\sin \frac{\pi}{\beta} \tau\right)^{2j},
\end{equation}
by evaluating the binomial expansion, and taking the largest term in the $k$-polynomial. This expression is indeed the expected result for the thermal two-point function of a non-unitary CFT operator $\mathcal{O}_{h=-j}$ when turning off dynamical gravity:
\begin{equation}
\left\langle \mathcal{O}_{h=-j}(\tau) \mathcal{O}_{h=-j}(0)\right\rangle_{\scriptstyle \text{CFT}} = \left(\frac{\beta}{\pi}\sin \frac{\pi}{\beta} \tau\right)^{2j}.
\end{equation}
\item
To perform the combinatorial manipulation at higher values of $j$, an alternative diagrammatical option is to deconstruct it into the elementary $j = +1/2$ bilocals:
\begin{align}
\begin{tikzpicture}[scale=0.65, baseline={([yshift=0cm]current bounding box.center)}]
\draw[thick] (0,0) circle (1.5);
\draw[thick] (0,-1.5) -- (0,1.5);
\draw[fill,black] (0,1.5) circle (0.1);
\draw[fill,black] (0,-1.5) circle (0.1);
\draw (-0.5,0) node {\small $j$};
\end{tikzpicture}
\quad
= \quad 
\begin{tikzpicture}[scale=0.65, baseline={([yshift=0cm]current bounding box.center)}]
\draw[thick] (0,0) circle (1.5);
\draw[thick] (0,-1.5) -- (0,1.5);
\draw[fill,black] (0,1.5) circle (0.1);
\draw[fill,black] (0,-1.5) circle (0.1);
\draw[thick] (0,-1.5) arc (228:132:2);
\draw[thick] (0,-1.5) arc (-48:48:2);
\draw (-0.3,0) node {\small $j_1$};
\draw (-1.1,0) node {\small $j_2$};
\draw (1.1,0) node {\small $j_3$};
\end{tikzpicture}
\end{align}
where $j=j_1+j_2+j_3$. 
The identity that ensures equivalence of this holds for $h > 0$, where it is an analogue of the Barnes lemmas. For the case $h \in -\mathbb{N}/2$ of interest, this requires an independent explicit check, and one can readily see that it is true in this case as well. Such diagrams are closely related to the loop gas diagrams of Kostov \cite{Kostov:2002uq}.
\end{itemize}

\subsection{Diagrammatics, and out-of-time ordered correlators} 
Given these degenerate vertex functions \eqref{vertfunc}, the diagrammatic language of Schwarzian correlators developed in \cite{Mertens:2017mtv} immediately extends to diagrams including degenerate bilocal lines, where multiple of these insertions are easily accommodated. We will draw degenerate bilocal lines with a dashed line, and non-degenerate bilocal lines with a full line. A particularly interesting diagram is that of crossing bilocal lines, corresponding to an out-of-time ordered correlator: 
\begin{align}
\begin{tikzpicture}[scale=0.65, baseline={([yshift=0cm]current bounding box.center)}]
\draw[thick] (0,0) circle (1.5);
\draw[thick] (0.7,-1.3) -- (-0.7,1.3);
\draw[dashed] (-0.7,-1.3) -- (0.7,1.3);
\draw[fill,black] (0.7,-1.3) circle (0.1);
\draw[fill,black] (-0.7,1.3) circle (0.1);
\draw[fill,black] (-0.7,-1.3) circle (0.1);
\draw[fill,black] (0.7,1.3) circle (0.1);
\draw (0.8,-0.7) node {\small $h$};
\draw (0.8,0.7) node {\small $j$};
\end{tikzpicture}
\qquad\qquad\qquad
\begin{tikzpicture}[scale=0.65, baseline={([yshift=0cm]current bounding box.center)}]
\draw[thick] (0,0) circle (1.5);
\draw[dashed] (0.7,-1.3) -- (-0.7,1.3);
\draw[dashed] (-0.7,-1.3) -- (0.7,1.3);
\draw[fill,black] (0.7,-1.3) circle (0.1);
\draw[fill,black] (-0.7,1.3) circle (0.1);
\draw[fill,black] (-0.7,-1.3) circle (0.1);
\draw[fill,black] (0.7,1.3) circle (0.1);
\draw (0.8,-0.7) node {\small $j_2$};
\draw (0.8,0.7) node {\small $j_1$};
\end{tikzpicture}
\end{align}
where we drew an example of a degenerate line crossing a non-degenerate line, and an example of two degenerate lines crossing.
Each such crossing carries a SL$(2,\mathbb{R})$ 6j-symbol. If we take at least one operator pair to be a degenerate pair, we require the degenerate 6j-symbols. Since this development is a bit orthogonal to our main story, we develop the expressions in Appendix \ref{s:degfus}. On a technical level, the main conclusion is that the 6j symbol is given in terms of a Wilson polynomial \cite{groenevelt} (the non-degenerate 6j-symbol was a Wilson function). On a physical level, the main conclusion is that the degenerate 6j symbol does not encode gravitational shockwaves and chaos, which matches indeed with these operators representing an integrable subsector of the JT model.

\section{Application: Schwarzian perturbation theory}
\label{s:app1}
As one of our main applications, we will show that knowledge of the degenerate bilocal correlators on the disk, combined with the structure of the Schwarzian $1/C$ perturbation expansion, allows us to learn a few lessons on the small $\tau$ series expansion for \emph{any} value of $h$. We will moreover discuss the nature of the $1/C$ perturbation series (asymptotic vs convergent).

\subsection{Review: Schwarzian perturbation theory}
\label{s:rev}
In this subsection, we provide a brief recap of the perturbative treatment of Schwarzian QM. We need only one elementary result from the Schwarzian perturbative expansion, which is that the coefficient of each term in the series expansion is a polynomial in the weight $h$ of the bilocal operator. \\
Setting $f(\tau) = \tau + \epsilon(\tau)$, and expanding in $\epsilon$, one writes for the Lagrangian:
\begin{equation}
\label{expa}
\left\{\tan \frac{\pi}{\beta}f(\tau),\tau\right\} = \frac{2\pi^2}{\beta^2} + \left(\epsilon''' + \frac{4\pi^2}{\beta^2}\epsilon'\right) + \left(\frac{2\pi^2}{\beta^2}\epsilon'^2-\frac{3}{2}\epsilon''^{2} - \epsilon' \epsilon''' \right) + \mathcal{O}(\epsilon^3),
\end{equation}
with propagator
\begin{equation}
\label{propa}
\left\langle \epsilon(\tau)\epsilon(0)\right\rangle = \frac{1}{2\pi C}\left(\frac{\beta}{2\pi}\right)^{3} \left[1 - \frac{1}{2}(u-\pi)^2 + \frac{\pi^2}{6} + \frac{5}{2}\cos u + (\tau-\pi) \sin u\right], \qquad u = 2\pi \tau/\beta,
\end{equation}
and vertices from the cubic and higher powers in the $\epsilon$-expansion. Notice that the vertices are $\beta$-independent. The bilocal operator is also expanded as
\begin{equation}
\label{opexpa}
\left(\frac{F'_1F'_2}{(F_1-F_2)^{2}}\right)^{h} = \frac{(1+\epsilon_1')^{h}(1+\epsilon_2')^{h}}{(\frac{\beta}{\pi} \sin \frac{\pi}{\beta}(\tau_{12} + \epsilon_1-\epsilon_2))^{2h}}.
\end{equation}
From \eqref{expa}, one reads that each propagator carries a factor of $1/C$ and each vertex a factor of $C$. For example, Maldacena, Stanford and Yang computed the first-order correction in $1/C$ in their equation (4.36) for a generic bilocal correlator \cite{Maldacena:2016upp}:\footnote{This correction is always positive for $h \geq 1$, changes sign for $h \in \left[0.7337,1\right]$ and is always negative for $h \in \left[0, 0.7337\right]$ and again positive for $h <0$.}

\vspace{-0.4cm}
{\small
\begin{align}
\label{msy}
&\left\langle \mathcal{O}_{h}(\tau,0)\right\rangle = \left\langle \mathcal{O}_{h}(\tau,0)\right\rangle_{C\to+\infty}\times  \\
&\left[ 1 + \frac{\beta}{4\pi^2 C}\left(h \frac{(u^2-2\pi u +2 -2\cos u +2(\pi-u) \sin u)}{4\sin^2\frac{u}{2}} + \frac{h^2}{2}\left(-2 + \frac{u}{\tan\frac{u}{2}}\right)\left(-2 + \frac{u-2\pi}{\tan\frac{u}{2}}\right)\right)\right] \hspace{-0.1cm}+ \hspace{-0.1cm} \mathcal{O}\left(\hspace{-0.1cm}\frac{1}{C^2}\hspace{-0.1cm}\right), \nonumber
\end{align}
}
\normalsize
where $u=2\pi \tau/\beta$. The zero-temperature limit of this formula is readily taken:
\begin{equation}
\left\langle \mathcal{O}_{h}(\tau,0)\right\rangle_{\beta\to+\infty} = \frac{1}{\tau^{2h}} \left[ 1 + \frac{h(h-1)}{6C}\tau \right] + \mathcal{O}\left(1/C^2\right).
\end{equation}
The leading correction is $\sim h(h-1) \sim m^2$, the mass$^2$ of the bulk field dual to the inserted boundary operators. We can interpret it as the one-loop self-energy of the bulk particle due to graviton interactions. We will give an analogous interpretation for the one-loop self-energy contribution from the gravitino in the supersymmetric case in section \ref{s:susy}.
\\~\\
Contemplating the expansion \eqref{opexpa} one also finds that a diagram with $n$ connections to the external endpoints is contributing a polynomial in $h$ of order $n$ without constant term. E.g. the diagrams contributing to the zero'th, first and second order term for the bilocal correlator $G(\tau_1,\tau_2)$ are schematically drawn as:
\begin{equation*}
\includegraphics[width=\textwidth]{fgraphsv3.pdf}
\end{equation*}
The first line contains the Schwarzian diagrams contributing at leading order $1$ (free result) and the first subleading correction $\sim 1/C$ corresponding to the one-loop gravitational self-energy. The second line represents the six $1/C^2$ diagrams. Each contribution is a polynomial in $h$ and we have indicated to which monomials in $h$ they contribute. Dashed lines represent virtual fermions $\psi(\tau)$ coming from the non-trivial Schwarzian path integral measure. The solid lines are CFT matter lines that are external to the actual Schwarzian theory; we choose to draw them nonetheless to emphasize the physical process.

Achieving a more systematic understanding of higher loop corrections is complicated by the following facts.
The Schwarzian model has a non-trivial path integral measure \cite{Stanford:2017thb}.\footnote{This can be found from several perspectives: from a Virasoro coadjoint orbit perspective see \cite{Stanford:2017thb}, for a derivation from the flat Liouville path integral measure, see \cite{Mertens:2018fds}. Finally, it also follows from the natural measure on the SL$(2,\mathbb{R})$ group.} One can flatten the measure by exponentiating it and integrating in an additional fermionic variable $\psi(\tau)$ that contributes to loop diagrams as illustrated above. Secondly, the number of vertices increases with each order. Both of these cause the perturbative series to be extremely unwieldy beyond leading order.\footnote{Some results have been obtained at second order \cite{Qi:2019gny}.}

\subsection{Small $\tau$-expansion}
In order to shed light on the perturbative expansion at higher orders, we will exploit the degenerate result \eqref{degmain}. The strategy is simple: we can Taylor expand the integrand of \eqref{degmain} as a power series in $\tau/C$ and then perform the momentum $k$-integral exactly. This directly gives the small $\tau$-series expansion. 
In terms of the (renormalized) multi-stress tensor Schwarzian expectation values:\footnote{These can be found by setting to zero all contact terms in the multi-Schwarzian derivative correlator, corresponding to a point-splitting procedure, see e.g. \cite{Iliesiu:2020zld} for a recent application.}
\begin{align}
\left\langle T\right\rangle = \frac{2\pi^2}{\beta^2} + \frac{3}{2\beta C}, \quad \left\langle T^2\right\rangle &= \frac{4\pi^4}{\beta^4} + \frac{10\pi^2}{\beta^3 C} + \frac{15}{4\beta^2 C^2}, \quad \left\langle T^3\right\rangle &= \frac{8\pi^6}{\beta^6} + \frac{42 \pi^4}{\beta^5 C} + \frac{105}{2\beta^4 C^2} + \frac{105}{8 \beta^3 C^3} \nonumber
\end{align}
one obtains for the specific examples of $h=-1/2$ and $h=-1$:
{\small
\begin{align}
\label{precisionhalf}
\left\langle \mathcal{O}_{h =-1/2}(\tau,0)\right\rangle &= \tau + \frac{1}{8C}\tau^2 - \left(\frac{1}{12}\left\langle T\right\rangle - \frac{3}{384 C^2}\right) \tau^3 - \left(\frac{1}{96C}\left\langle T\right\rangle - \frac{1}{3072 C^3}\right) \tau^4 \nonumber \\
&\hspace{-1cm}+ \left(\frac{1}{480}\left\langle T^2\right\rangle - \frac{1}{1536C^2}\left\langle T\right\rangle + \frac{1}{98304 C^4}\right)\tau^5 + \left(\frac{1}{3840 C}\left\langle T^2\right\rangle - \frac{1}{36864 C^3}\left\langle T\right\rangle + \frac{1}{3932160 C^5}\right) \tau^6 \nonumber\\
&\hspace{-1cm}+\left(-\frac{1}{1179648 C^4}\left\langle T\right\rangle + \frac{1}{61440 C^2}\left\langle T^2\right\rangle - \frac{1}{40320}\left\langle T^3\right\rangle + \frac{1}{188743680 C^6}\right) \tau^7 + \mathcal{O}(\tau^8).
\end{align}

}%
{\small
\begin{align}
\label{precisionone}
\left\langle \mathcal{O}_{h =-1}(\tau,0)\right\rangle &= \tau^2 + \frac{1}{3C}\tau^3 - \left(\frac{1}{6}\left\langle T\right\rangle-\frac{1}{16C^2}\right) \tau^4 - \left( \frac{1}{15C}\left\langle T\right\rangle - \frac{1}{120C^3}\right) \tau^5 \hspace{4cm}\nonumber \\
&\hspace{-1cm}+ \left( \frac{1}{90} \left\langle T^2\right\rangle - \frac{1}{72}\left\langle T\right\rangle + \frac{1}{1152}\right) \tau^6 +\left( \frac{1}{210} \left\langle T^2\right\rangle - \frac{1}{504}\left\langle T\right\rangle + \frac{1}{13440} \right) \tau^7 \nonumber \\
&\hspace{-1cm}- \left(  \frac{1}{2520} \left\langle T^3\right\rangle -   \frac{1}{960} \left\langle T^2\right\rangle +\frac{1}{4608}\left\langle T\right\rangle - \frac{1}{184320}\right) \tau^8 + \mathcal{O}(\tau^9).
\end{align}
}%

The multi-Schwarzian derivative correlators $\left\langle T^n\right\rangle$ always appear in this specific structural fasion. Since this applies to any $h \in - \mathbb{N}/2$, and since at every fixed order in the $\tau$-expansion one has a polynomial in $h$ as coefficient, this is sufficient to prove that one has the same expansion structure for any $h \in \mathbb{R}$, where only the numerical coefficients change. This leads to the structural expansion of \eqref{smaltau}.
\\~\\
As a classical function, the bilocal operator
\begin{equation}
\mathcal{O}_h(\tau_1,\tau_2) = \left(\frac{F'(\tau_1)F'(\tau_2)}{(F(\tau_1)-F(\tau_2))^2}\right)^{h}
\end{equation}
 can be series-expanded in $\tau_1-\tau_2$ \emph{before} quantizing. By SL$(2,\mathbb{R})$ invariance, the resulting expansion coefficients need to be local SL$(2,\mathbb{R})$ invariants. This is indeed true in the general expression \eqref{smaltau}, since it is only constructed from expectation values of powers of the Schwarzian derivative. Armed with our explicit expressions above, we can now investigate this equality in more detail, and compare term by term whether the exact expressions agree with the computation of powers of Schwarzian derivatives. The answer is quite involved and depends crucially on renormalization subtleties of the multi-stress tensor composite operators. We present the analysis in appendix \ref{s:paradox}. The upshot is that with a suitable choice of renormalization, one can make sense of the relation between both calculational strategies. 
\\~\\
Let us end with a short remark on a particular higher-point function: the time-ordered four-point function. A priori one would expect a four-point function to depend on three independent time parameters, since the Schwarzian theory is time-invariant, removing an overall time shift. However, we know from various perspectives that it only depends on two time parameters, conveniently chosen to be the time differences within each bilocal operator \cite{Mertens:2017mtv}. The small $\tau$-expansion analyzed here provides yet another way of appreciating this statement, and it is given strength by our understanding of the expansion of the classical bilocal in appendix \ref{s:paradox}. Performing a double series expansion for the classical bilocals, they are expandable into Schwarzian derivatives, their powers, and their derivatives.\footnote{An example is written in \eqref{pt} for $h=-1/2$.} We write schematically:
\begin{equation}
\small
\left\langle \left(\frac{F'(\tau_1)F'(\tau_2)}{(F(\tau_1)-F(\tau_2))^2}\right)^{h}\left(\frac{F'(\tau_3)F'(\tau_4)}{(F(\tau_3)-F(\tau_4))^2}\right)^{h}\right\rangle = \sum_{n,m,n_i,m_i}C^{n}_{n_1n_2}C^{m}_{m_1m_2} \left\langle T^{(n_1)}(\tau_2)^{n_2} T^{(m_1)}(\tau_4)^{m_2}\right\rangle \tau_{12}^{2n} \tau_{34}^{2m},
\end{equation}
where $\tau_{ij} \equiv \tau_i - \tau_j$ and the $C^n_{n_1n_2}$ are the expansion coefficients. This depends explicitly on the time differences $\tau_{12}$ and $\tau_{34}$ within each bilocal. Quantum-mechanically, the correlators of Schwarzian derivatives would now provide the link between both bilocals, such that also $\tau_{24}$ appears. However, due to the fact that correlation functions of Schwarzian derivatives are always time-independent, this does not happen, and the four-point correlator only depends on two instead of three time parameters.\footnote{Some examples are written in appendix \ref{s:paradox}. The contact terms in the multi-Schwarzian derivative correlators are truly zero in this case, since we will assume both bilocals have no coincident points.} \footnote{For out-of-time ordered correlators this simplification does not happen, and we cannot just perform the small $\tau$-expansion on any bilocal bridging other operators.}

\subsection{Asymptotic vs convergent perturbation series}
It is apparent from the above formulas (especially \eqref{halfexact} and \eqref{zero}) that the degenerate two-point function has a \emph{convergent} $1/C$ perturbative expansion. In fact, these degenerate values of $h \in - \mathbb{N}/2$ are the only values of $h$ for which a convergent perturbative series is achieved. All other values of $h \notin - \mathbb{N}/2$ correspond to asymptotic series in $1/C$. Since this is a somewhat technical discussion, we present the proof in Appendix \ref{s:proof}. \\
Physically, the $1/C$ expansion is interpretable in terms of multi-boundary graviton exchanges. We then learn that the generic correlators contain non-perturbative gravitational physics not captured by these graviton exchanges. In particular, we estimate the size of non-perturbative corrections to be $\sim e^{- \#/G_N}$. This is of the same order as the higher genus corrections to amplitudes, to which we turn next.

\section{Minimal string and matrix interpretation}
\label{s:hightop}
Our next goal is to analyze this degenerate two-point function on surfaces of higher topology. A guide towards achieving this will be the embedding of JT gravity within minimal string theory, where the boundary tachyon vertex operators limit to our degenerate operator insertions in JT gravity, a statement we will make explicit below. For the minimal string, the matrix interpretation and how it links back to higher genus corrections has been studied extensively in the literature, and we will use it to guide us towards a proposal for higher topological corrections to the degenerate boundary correlators.

\subsection{Motivation from JT gravity: heuristic argument}
We start by giving a heuristic argument why higher topological corrections for amplitudes with degenerate operator insertions behave differently than in the non-degenerate case. \\
One can consider bilocal correlators in JT gravity that include higher topological corrections and the doubly non-perturbative random matrix completion. Let us first review the story for a generic $h >0$ correlator. The rule is to slice up the initial disk region along each of the bilocal lines, and then to add higher topology to the resulting cut surface \cite{Saad:2019pqd,Blommaert:2019hjr}. As an example, the diagrams for the corrections to the boundary two-point function can be found by first cutting the disk along the bilocal line:
\begin{align}
\raisebox{-10mm}{\includegraphics[width=0.55\textwidth]{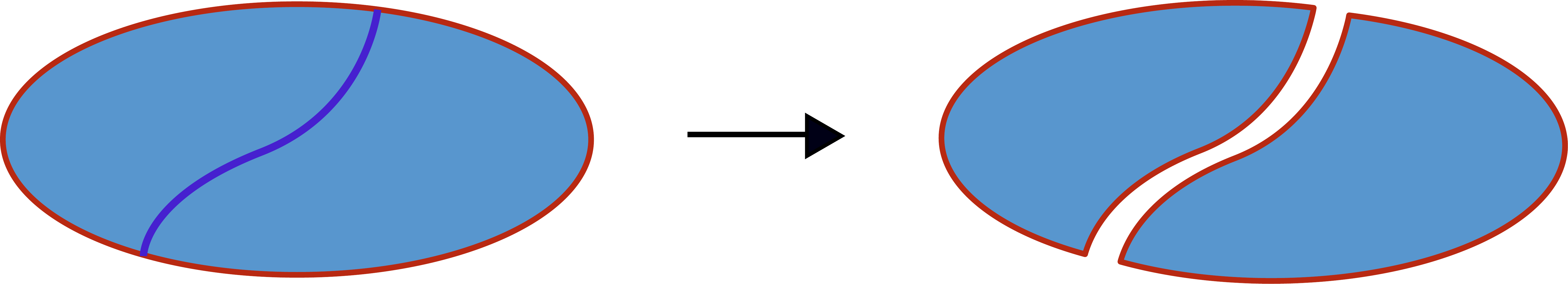}}
\end{align}
and then we include higher genus contributions to these two disk pieces. These fall into two classes, schematically of the types:
\begin{align}
\label{highcut}
\raisebox{-10mm}{\includegraphics[width=0.55\textwidth]{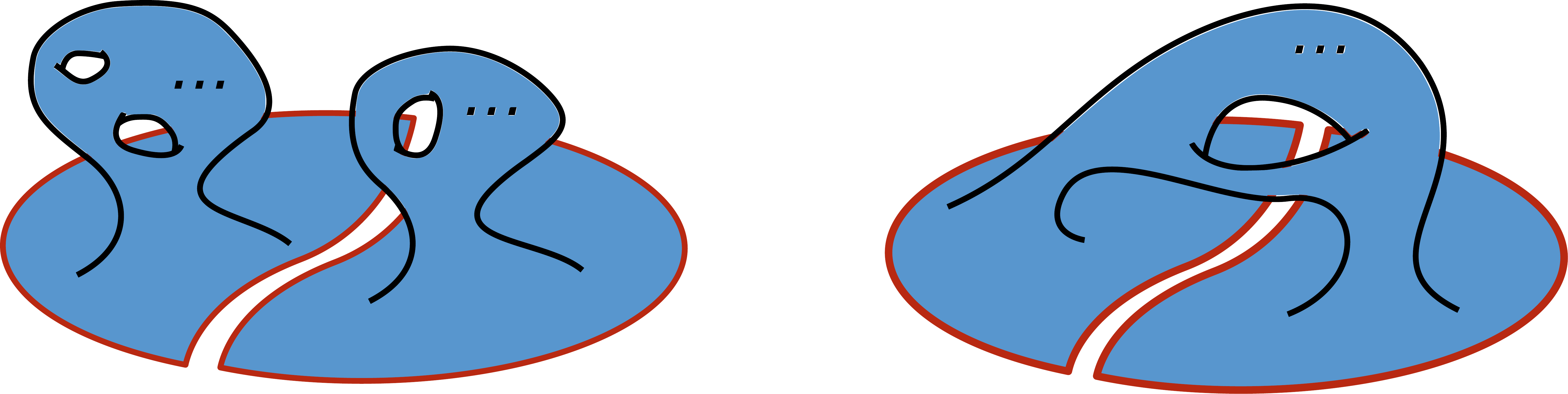}}
\end{align}
In formulas, and working in units where $C=1/2$, the higher genus expansion of the bilocal correlator was argued to be of the form \eqref{2genexpi}, which we retake here:
\begin{align}
\label{2genexp}
\left\langle \mathcal{O}_h(\tau_1,\tau_2)\right\rangle_\beta &= \frac{1}{Z} \int dE_1 dE_2 \rho(E_1,E_2)\, e^{ - \tau E_1}\,  e^{-(\beta-\tau) E_2}\frac{ \Gamma( h \pm i \sqrt{E_1} \pm i \sqrt{E_2})}{{8\pi^2}\, \Gamma(2h)},
\end{align}
where the two-level spectral density is replaced by the random matrix answer:
\begin{equation}
\label{rmt}
\rho(E_1,E_2) =  \rho(E_1) \rho(E_2) - \frac{\sin^2 \pi \rho(\bar{E})(E_1-E_2)}{\pi^2(E_1-E_2)^2} + \rho(E_2) \delta(E_1-E_2), \qquad E_1 \approx E_2, 
\end{equation}
representing respectively the disconnected pieces (the first class of diagrams in \eqref{highcut}), the annulus connecting both sides of the bilocal line (the second class of diagrams in \eqref{highcut}), and a contact term that has no a priori geometric origin. This geometric connection was made for the spectral form factor in \cite{Saad:2019lba}. The disconnected piece $\rho(E_1) \rho(E_2)$ at genus zero gives back the sinh measure and we get back to \eqref{sch2pt}, where $\rho_{g=0}(E) \sim e^{S_0}\sinh 2 \pi \sqrt{E}$. In the regime where $e^{S_0} \gg 1$, higher genus corrections are suppressed, except when $E_1 \approx E_2$ which can compensate for the $e^{-S_0}$ suppression.
\\~\\
We notice that when taking the limit $h \to 0$ in the above formula \eqref{2genexp}, we do not agree with setting $h=0$ from the outset in which case no bilocal line is present at all and the surface is not cut open to begin with. For degenerate bilocal lines where $h \in - \mathbb{N}/2$, we similarly propose to \emph{not} slice up the surface along the line. This means e.g. that a disk with a single degenerate bilocal line has only the same class of higher genus corrections as the disk itself:
\begin{align}
\raisebox{-10mm}{\includegraphics[width=0.25\textwidth]{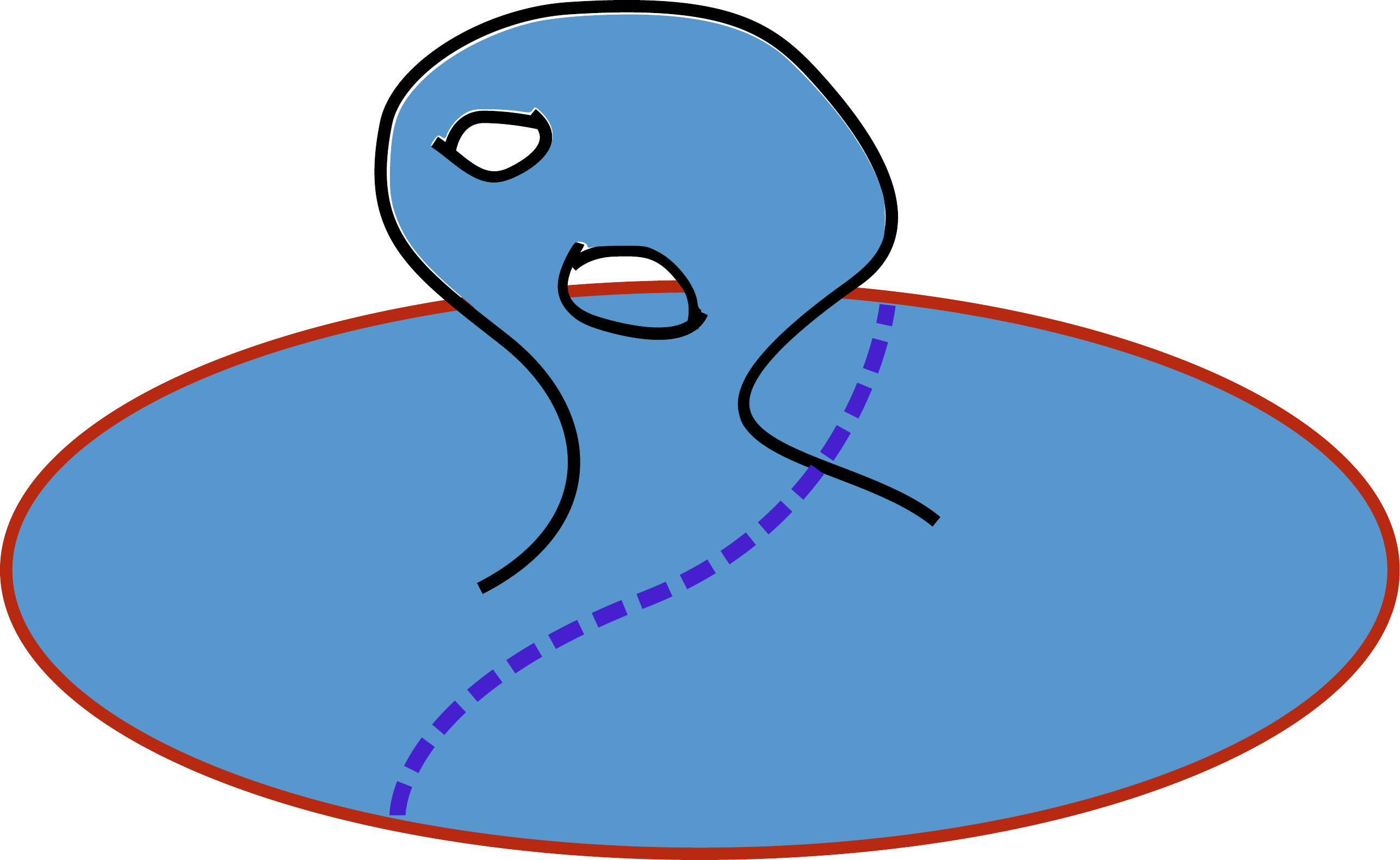}}
\end{align}
where we will depict a degenerate bilocal line by a dashed line in the figures. We will make this graphical presentation more precise in formulas below. \\ 
As a first naive argument, we notice that the pair density correlator $\rho(E_1,E_2)$ \eqref{rmt} does not make sense when it connects regions separated by a degenerate bilocal line. Supposing one starts with \eqref{rmt} for a degenerate line (where hence $E_1$ and $E_2$ are related as in \eqref{vertfunc}), then the effects are not important if $S_0 \gg 1$. If $j$ is half-integer, then since $m\neq 0$, one never has $E_1 \approx E_2$, and higher topology is always suppressed (provided $S_0 \gg 1$): only the disconnected disks contribute. If $j$ is integer, then the $m=0$ term in the sum is the dominant one: the first two terms of \eqref{rmt} cancel by level repulsion in random matrix theory, and the last (plateau) term diverges as $\sim \delta(0)$. This illustrates that the pair correlator $\rho(E_1,E_2)$ does not seem to be the natural quantity when considering degenerate operator insertions.

\subsection{Minimal string: boundary tachyon correlators}
In order to properly understand how to treat higher genus corrections to these amplitudes, we will take inspiration from the minimal string theory of which JT gravity is a parametric double-scaling limit \cite{Saad:2019lba,Mertens:2020hbs}. Minimal string theory consists of a $(p,q)$ minimal model combined with the Liouville CFT and $bc$ ghosts to form a non-critical string theory with $c_L + c_M + c_{gh} = 0$. The Liouville central charge is parametrized as $c_L =1+ 6Q^2$, where $Q= b+b^{-1}$ and $b^2 = p/q$. This and related 2d models have a long history as toy models for quantum gravity, see e.g. \cite{Ishiki:2010wb,Bourgine:2010ja,Hosomichi:2008th,Kostov:2002uq,Zamolodchikov:2005fy,Belavin:2008kv}. The minimal string only has degenerate Virasoro primary matter operators that are dressed by the Liouville sector into physical vertex operators. E.g. the tachyon boundary vertex operators can be written as
\begin{equation}
\mathcal{B}_{r,s} \equiv c\, e^{\beta \phi} \mathcal{O}_{r,s}, \qquad 1 \leq r \leq p-1, \quad 1 \leq s \leq q-1,
\end{equation}
in terms of the degenerate matter operator $\mathcal{O}_{r,s}$, the Liouville vertex operator $e^{\beta \phi}$ and the $bc$ ghost $c$.\footnote{Operators are identified as $\mathcal{O}_{r,s} \equiv \mathcal{O}_{p-r,q-s}$, effectively halving the number of independent matter primaries.}
For the particular case of $p=2$ and $q = 2\mathfrak{m}-1$, a single-matrix description is possible, and we only have $\mathcal{B}_{1,s}$ for $1 \leq s \leq \mathfrak{m} -1$. The JT limit corresponds to taking $\mathfrak{m} \to \infty$ ($b\to 0$), for which one obtains an infinite discrete set of physical operator insertions.
The minimal string two-point function is essentially captured by the Liouville boundary two-point function:
\beq
\lb B_{\beta_1}^{s_1s_2}(x)B_{\beta_2}^{s_2s_1}(0)\rb = \frac{\delta(\beta_2 + \beta_1-Q)+ d(\beta|s_1,s_2)\delta(\beta_2-\beta_1)}{|x|^{2\Delta_{\beta_1}}},
\eeq
with $B_\beta^{s_1s_2} = e^{\beta \phi}$ a boundary Liouville primary operator between boundary segments with labels $s_1$ and $s_2$. We define the quantity\footnote{There is an implicit product over all four sign combinations of the $S_b$ in this and in subsequent similar equations.}
\beq \label{liouvillebdy2pt}
d(\beta|s_1,s_2) = (\pi \mu \gamma(b^2)b^{2-2b^2})^{\frac{Q-2\beta}{2b}} \frac{\Gamma_b(2\beta-Q)\Gamma_b^{-1}(Q-2\beta)}{S_b(\beta \pm i s_1 \pm i s_2)},
\eeq
with the boundary FZZT cosmological constant $\mu_B$ related to the label $s$ as $\mu_B \equiv x = \kappa \cosh 2\pi b s$ where $\kappa = \sqrt{\mu}/\sqrt{\sin \pi b^2}$. The double sine function $S_b$ and b-deformed Gamma-function $\Gamma_b$ are given in appendix \ref{app:spec}. In our case the Liouville label $\beta = b - \beta_M$, in terms of the matter label $\beta_M$:
\beq 
\beta_{M(r,s)} =  \frac{1-s}{2}b + \frac{r-1}{2b}.
\eeq
For the $(2,2\mathfrak{m}-1)$ minimal string with $\mathfrak{m} \in \mathbb{N}$, we hence have $\beta_M = - jb$ for $j= 0,\frac{1}{2}, \hdots \frac{\mathfrak{m}-2}{2} $ which can be identified with a discrete finite irrep $\mathfrak{sl}(2)$ spin label $j$. We now focus on this particular minimal string model, with these operator insertions. 
\\~\\
For such boundary operators, we have the corresponding Liouville parameter $\beta = b +bj$. The Liouville boundary two-point function \eqref{liouvillebdy2pt} can then be rewritten in terms of elementary functions:
\begin{align}
\label{Lioures}
\frac{1}{S_b(b+bj \pm is_1 \pm is_2)} = \frac{\cosh \frac{2\pi}{b}s_1 + (-)^{2j+1}\cosh \frac{2\pi}{b}s_2}{4^j\prod_{n=-j}^{j} \left( \cosh 2\pi b s_1 - \cosh 2\pi b (s_2+in b) \right)}.
\end{align}
It is elementary to check that the r.h.s. is symmetric under $s_1 \leftrightarrow s_2$. We identify the numerator with genus zero resolvents as $R_{g=0,1}(x) = \cosh \frac{2\pi}{b}s$ where $x\equiv \mu_B = \kappa \cosh 2\pi b s$. Following common conventions, we will sometimes denote $\mu_B$ by $x$ to streamline the notation.
\\~\\
Next, we transform this expression to the fixed length basis by applying the integral transform
\begin{equation}
\label{beterL}
-\int_{c_0+i\mathbb{R}} d\mu_{B1} d\mu_{B2}e^{\mu_{B1} \ell_1}e^{\mu_{B2} \ell_2} \times \hdots ,
\end{equation}
for both $s_1$ and $s_2$. Inserting \eqref{Lioures}, we have two terms for which in both cases the first integral is contour deformed to pick up the poles in the denominator. The second integral is then deformed across the branch cut of the $\cosh \frac{2\pi}{b}s_2$ resp. $\cosh \frac{2\pi}{b}s_1$ resolvent. We obtain in the end:
\begin{align}
\label{gendeg}
&\left\langle  \mathcal{B}_{1,2j+1}\mathcal{B}_{1,2j+1} \right\rangle_{\ell_1,\ell_2}  = \frac{1}{4\pi b \kappa Z} \\
&\times\int_\kappa^{+\infty} dx \rho(x) \, e^{-\ell_1 \kappa \cosh 2\pi b s}\sum_{n=-j}^{+j} \frac{(2j)!e^{-\ell_2 \kappa \cosh 2\pi b (s+i nb)}}{\prod_{\stackrel{m=-j}{m\neq n}}^{j} (\cosh 2\pi b (s+i nb) - \cosh 2\pi b (s+imb))}  + (\ell_1 \leftrightarrow \ell_2), \nonumber
\end{align}
which is the same expression as that found in \cite{Mertens:2020hbs}. The prefactors are chosen to match the UV $\ell_2 \approx 0$ behavior $\to \ell_2^{2j}$ with unit coefficient. The advantage of organizing the calculation like this is that now the resolvent and spectral density of the underlying matrix integral make explicit appearances, where the discontinuity of the resolvent is the spectral density $\rho(x(s))$
\begin{equation}
\text{Disc}[R(x)] = R(x+i\epsilon) - R(x-i\epsilon) = -2\pi i\rho(x), \qquad \rho_0(x(s)) = \sinh \frac{2\pi}{b}s,
\end{equation}
and where $x=\kappa \cosh 2 \pi b s$. 

\subsection{A proposal for higher topology}

We will use this suggestive way of writing the amplitude as the \emph{definition} of an insertion in the random matrix integral. In particular, this expression gets its higher genus contributions only from the single-boundary resolvent contributions $R_{g,n=1}(x)$, computable e.g. from Eynard's topological recursion relations \cite{Eynard:2004mh,Eynard:2007kz,Eynard:2015aea}. This is a strong restriction on the allowed topologies contributing to the amplitude. Aside from this restriction, the resulting amplitude has a similar structure as the proposed JT bilocal correlator (for $h\notin -\mathbb{N}/2$) at higher genus in \cite{Saad:2019pqd}, where indeed also only the spectral densities $\rho(E_1, \hdots E_n)$ are adjusted to accommodate for different topology. It would be very valuable to verify this proposal by an explicit computation of the genus one result in the minimal string continuum language, which seems out of reach at the moment.
In matrix language, where the resolvent is $R(x) = \left\langle \text{Tr}\frac{1}{x-H}\right\rangle$ in terms of the random matrix $H$, we can then identify \eqref{Lioures} as the genus zero contribution of the matrix insertion:\footnote{There are obviously other options that agree on the lowest genus zero result but differ beyond that. We believe this is the most natural one as it will have all the expected properties. The matrix operator used in \cite{Ishiki:2010wb} is slightly different, but would also only give single-boundary resolvent contributions $R_{g,n=1}(x)$, which is our main point.}
\begin{equation}
\frac{1}{\prod_{\stackrel{n=-j}{n\neq 0}}^{j}(x_1-x_{2+inb})}\text{Tr}\left( \frac{1}{(x_1-H)(x_2-H)} \right), \, \text{or} \,\, \frac{1}{\prod_{n=-j}^{j}(x_1-x_{2+inb})}\text{Tr}\left( \frac{(x_1+x_2) - 2H}{(x_1-H)(x_2-H)} \right),
\end{equation}
where the left expression is valid for $j \in \mathbb{N}$ and the right expression for $j \in \mathbb{N}+1/2$. We dropped overall factors here, and used the notation $x_{2+inb} \equiv \kappa \cosh 2 \pi b (s_2+inb)$.
\\~\\
Notice the presence of the $(\ell_1 \leftrightarrow \ell_2)$ second term in \eqref{gendeg}. This is in unison with the fact that the amplitude \eqref{beterL} is manifestly invariant under swapping $\ell_1$ and $\ell_2$.
To understand its meaning, we can write this second term suggestively as:
\begin{align}
\frac{1}{4\pi b \kappa Z}\int_\kappa^{+\infty} dx  \, e^{-\ell_1 \kappa \cosh 2\pi b s}\sum_{n=-j}^{+j} \rho(x(s+inb)) (-)^{2n}\frac{(2j)!e^{-\ell_2 \kappa \cosh 2\pi b (s+i nb)}}{\prod_{\stackrel{m=-j}{m\neq n}}^{j} (\cosh 2\pi b (s+i nb) - \cosh 2\pi b (s+imb))},
\end{align}
where $x_{+n} = \cosh 2\pi b (s + in b)$. At genus zero we have $\rho_0(x(s+inb)) = \sinh \frac{2\pi}{b}(s+inb) = (-)^{2n}\rho_0(x(s))$ and both terms in \eqref{gendeg} are equal. At higher genus, this is no longer the case. Within the $s$-coordinates,  this means one can interpret the full answer as
\begin{align}
&\left\langle  \mathcal{B}_{1,2j+1}\mathcal{B}_{1,2j+1} \right\rangle_{\ell_1,\ell_2}  \\
&= \frac{(2j)!}{Z}\int_0^{+\infty} ds \, \sum_{n=-j}^{+j} \rho_{\text{eff}}(s,n) e^{-\ell_1 \kappa \cosh 2\pi b s} \frac{e^{-\ell_2 \kappa \cosh 2\pi b (s+i nb)}}{\prod_{\stackrel{m=-j}{m\neq n}}^{j} (\cosh 2\pi b (s+i nb) - \cosh 2\pi b (s+imb))},\nonumber
\end{align}
with an effective density
\begin{equation}
\label{effdos}
\rho_{\text{eff}}(s,n) = \frac{1}{2}\sinh 2\pi b s \left[\rho(x(s)) + (-)^{2n}\rho(x(s+inb))\right], \qquad \rho_{\text{eff},g=0}(s,n) = \sinh 2 \pi b s \sinh \frac{2 \pi}{b}s.
\end{equation}
When viewed gravitationally, this can be interpreted as adding higher topology to each of the possible sectors of the diagram, leading to a symmetrized result in the end. As an example, the genus one correction is geometrically computed by considering
\begin{align}
\lb \mathcal{B}_{1,2j+1}\mathcal{B}_{1,2j+1} \rb_{g=1} =  \int_0^{+\infty} ds \, \sum_{n=-j}^{+j} \left(\quad \raisebox{-10mm}{\includegraphics[width=0.15\textwidth]{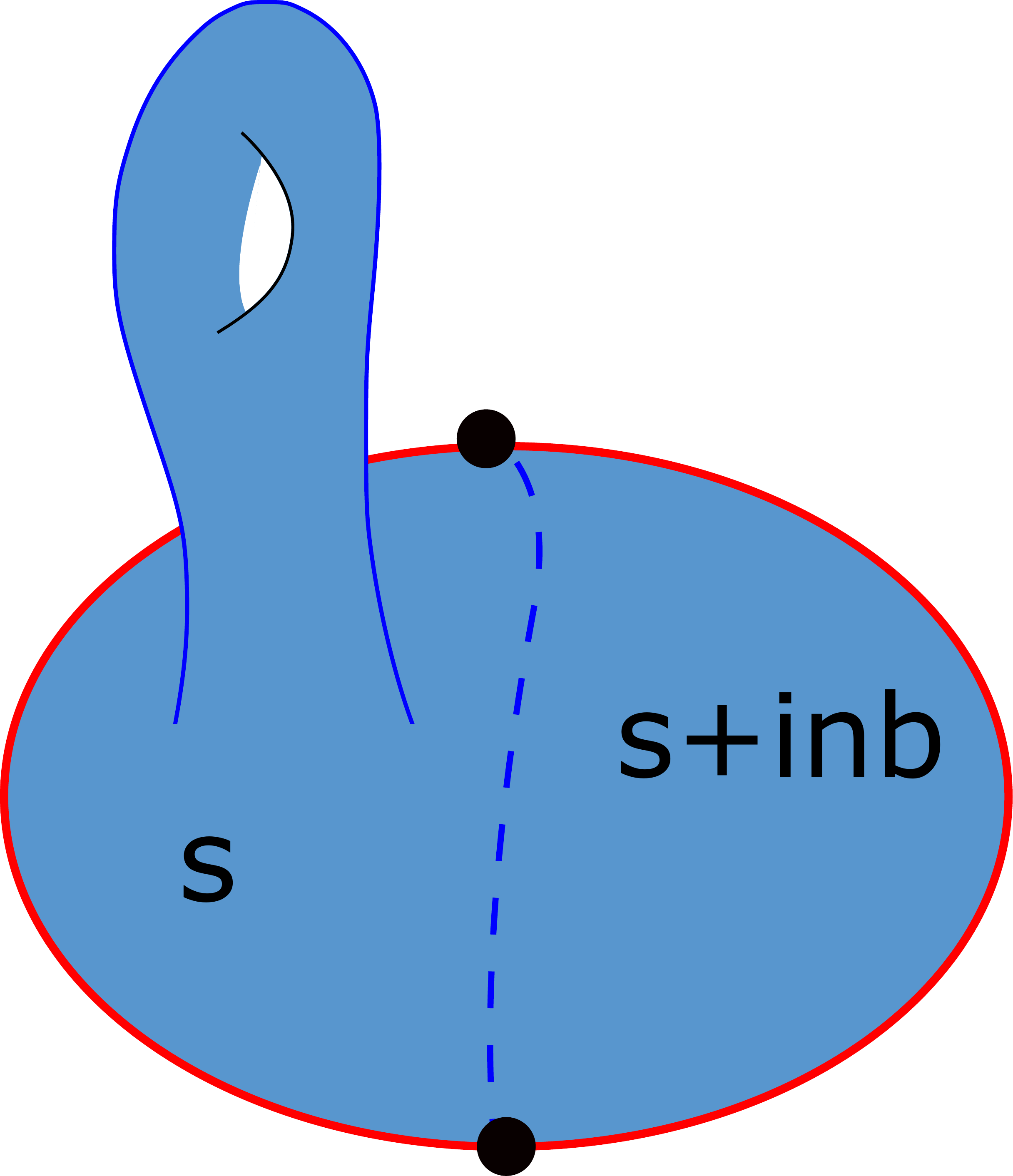}}
+ (-)^{2n} \quad  \raisebox{-10mm}{\includegraphics[width=0.15\textwidth]{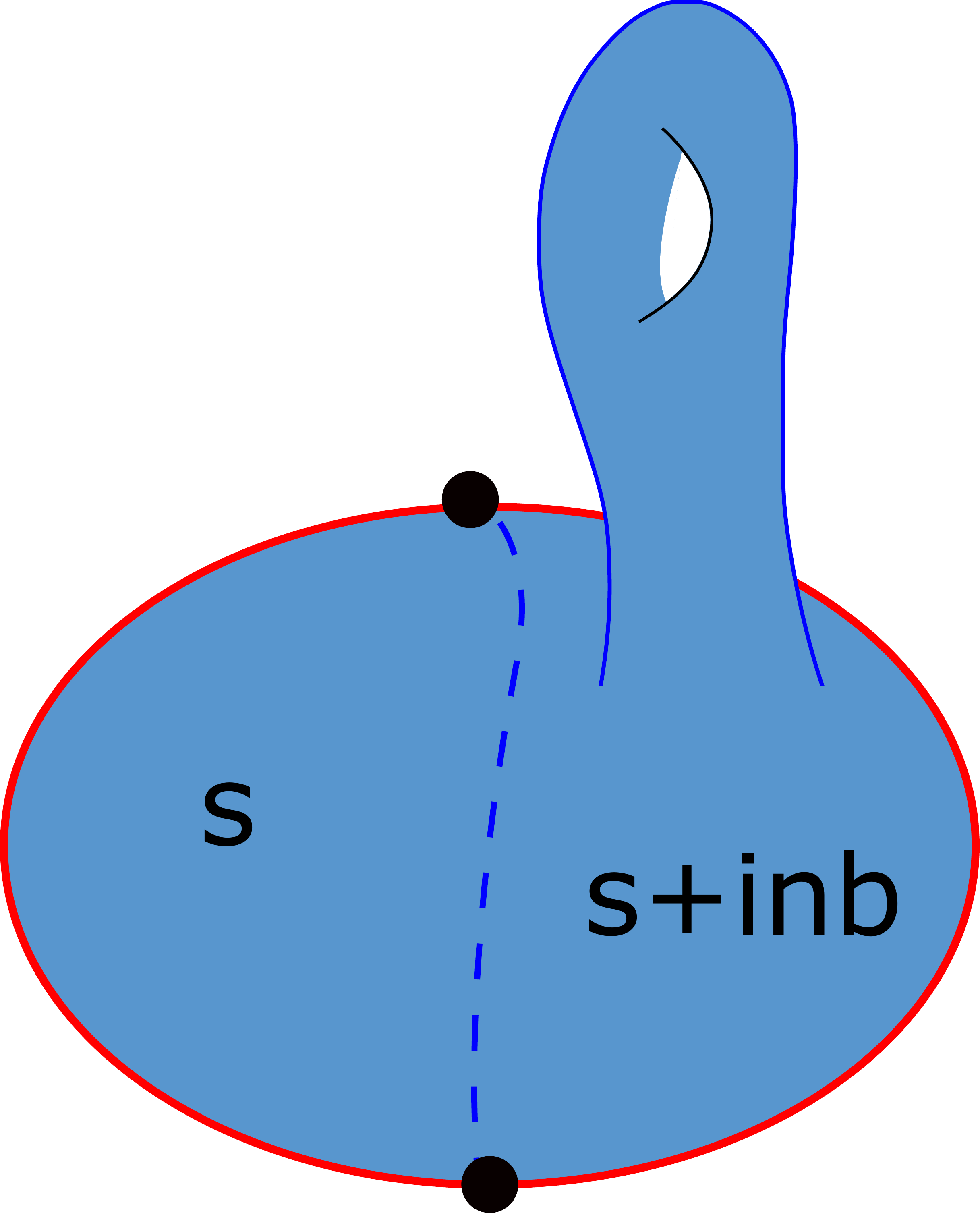}} \right)
\end{align}
Since we sum over handles in each sector of the diagram (instead of multiplying), it is conceptually convenient to lump the contributions together when drawing the diagram. The gravitational interpretation is then geometrically:
\begin{align}
\lb \mathcal{B}_{1,2j+1}\mathcal{B}_{1,2j+1} \rb = \quad \raisebox{-10mm}{\includegraphics[width=0.7\textwidth]{highergenusDegv2.pdf}}
\end{align}
where the meaning of the handles is encoded in \eqref{effdos}. For practical computations, it is more convenient to compute \eqref{gendeg} directly and add the $\ell_1 \leftrightarrow \ell_2$ contribution by hand.

\subsection{JT limit}
In the JT limit, we set $s_i = bk_i$ for finite $k_i$. The denominator of \eqref{Lioures} becomes
\begin{align}
\cosh 2\pi b s \quad \to  \quad  1+ 2\pi^2b^4 k^2.
\end{align}
For the resolvents in the numerator of \eqref{Lioures}, we focus on the spectral region close to the edge at $x = -\kappa$, by parametrizing $x = -\kappa+ 2\pi^2 \kappa b^4z_{\rm JT}^2$: 
\begin{align*}
\begin{tikzpicture}[scale=1.2]
\draw[->,thick, -latex] (0,-0.8) -- (0,0.8); 
\draw[->,thick, -latex] (-3,0) -- (1,0); 
\draw[->,thick, -latex] (-0.9,0.6) -- (-0.9,0.05); 
\draw[red!80!black, thick,decoration = {zigzag,segment length = 2mm, amplitude = 0.5mm},decorate] (-3,0) -- (-1,0);
\filldraw[red!80!black] (-1,0) circle (0.07);
\node at (-1,-0.4) {\scriptsize $-\kappa$};
\node at (1,1) {\large $x$};
\draw[thick] (0.8,0.75+0.4) -- (0.8,0.75) -- (0.8+0.3,0.75);
\end{tikzpicture}
\end{align*}

Plugging this in the resolvent, and focusing on the $(2,2\mathfrak{m}-1)$ minimal string for which $b^2 = 2/(2\mathfrak{m}-1)$, we get:
\begin{align}
R_{g=0,1}(z_{\rm JT}) =  \cosh \frac{1}{b^2} \text{arccosh}(-1+ 2\pi^2 b^4 z_{\rm JT}^2) \quad \to  \quad  \sin 2\pi z_{\rm JT},
\end{align}
where $z_{\rm JT}^2 = -k^2$. In terms of the uniformizing coordinate $k$, we can hence write the JT resolvent at lowest order in the genus expansion as:
\begin{equation}
\label{tree}
R_{g=0,1}(k) = \sinh 2\pi k.
\end{equation}
The full one-boundary JT resolvent is denoted as $R(k)$, and has a genus expansion:
\begin{equation}
R(k) = \left\langle \text{Tr} \frac{1}{k^2-H}\right\rangle = \sum_{g} R_{g,1}(k).
\end{equation}
Transforming the JT limit of \eqref{Lioures} to the length basis gives the expression:
\begin{align}
\label{beter}
\int_{c_0+i\mathbb{R}} dk_1^2dk_2^2 \, \frac{R(k_1) + (-)^{2j+1} R(k_2)}{\prod_{n=-j}^{j}(k_1^2 - (k_{2}+in)^{2})} e^{k_1^2 \ell_1}e^{k_2^2 \ell_2}.
\end{align}
Deforming the first contour picks up all of the poles. Deforming the second contour picks up the discontinuity across the branch cut, where the discontinuity of $R$ between $-k^2+i\epsilon$ and $-k^2 - i\epsilon$ is the spectral density, at genus zero given by the expression:
\begin{equation}
\sinh 2 \pi \sqrt{-E+i\epsilon} - \sinh 2 \pi \sqrt{-E-i\epsilon} = 2 \sin 2 \pi \sqrt{E}.
\end{equation}
Finally setting $k^2 \to - k^2$ to have the integral over $\mathbb{R}^+$, we obtain:
\begin{align}
\label{gendegJT}
&\lb \mathcal{B}_{1,2j+1}\mathcal{B}_{1,2j+1} \rb_{g=0}  \nonumber \\
&= \frac{1}{Z}(2j)!\int_0^{+\infty} dk \rho_{g=0}(k) \, e^{-\ell_1 k^2} \sum_{n=-j}^{+j} \frac{e^{-\ell_2 (k+in)^2}}{\prod_{\stackrel{m=-j}{m\neq n}}^{j} ((k+in)^2-(k+im)^2)} +(\ell_1 \leftrightarrow \ell_2).
\end{align}
with $\rho_{g=0}(k) = k\sinh 2 \pi k$. One can check that this formula matches with \eqref{degmain} upon identifying $\ell_1 = \beta-\tau$ and $\ell_2 = \tau$. We have noticed this before in \cite{Mertens:2020hbs}. This shows that the \emph{minimal string tachyon boundary vertex operators limit to the JT degenerate operators}, which is one of our main conclusions. \\

Written as \eqref{beter}, the inclusion of higher genus and non-perturbative random matrix effects is straightforward since we again only adjust the resolvents $R(k)$ in \eqref{beter} into the exact answer, in the end only replacing the spectral measure from the seed value $\rho_{g=0}(k)$ to the random matrix (all-genus) result $ \rho(k)$ in \eqref{gendegJT}.
The vertex functions and structure of the amplitude remain the same. The topological corrections to the degenerate bilocal correlators are hence in one-to-one to those of the partition function (i.e. the $h=0$ case), in particular there is no pair density correlator $\rho(k_1,k_2)$ contribution, and hence no handles connecting opposite sides of the bilocal line in this case. \\

\begin{figure}[H]
\centering
\includegraphics[width=0.65\textwidth]{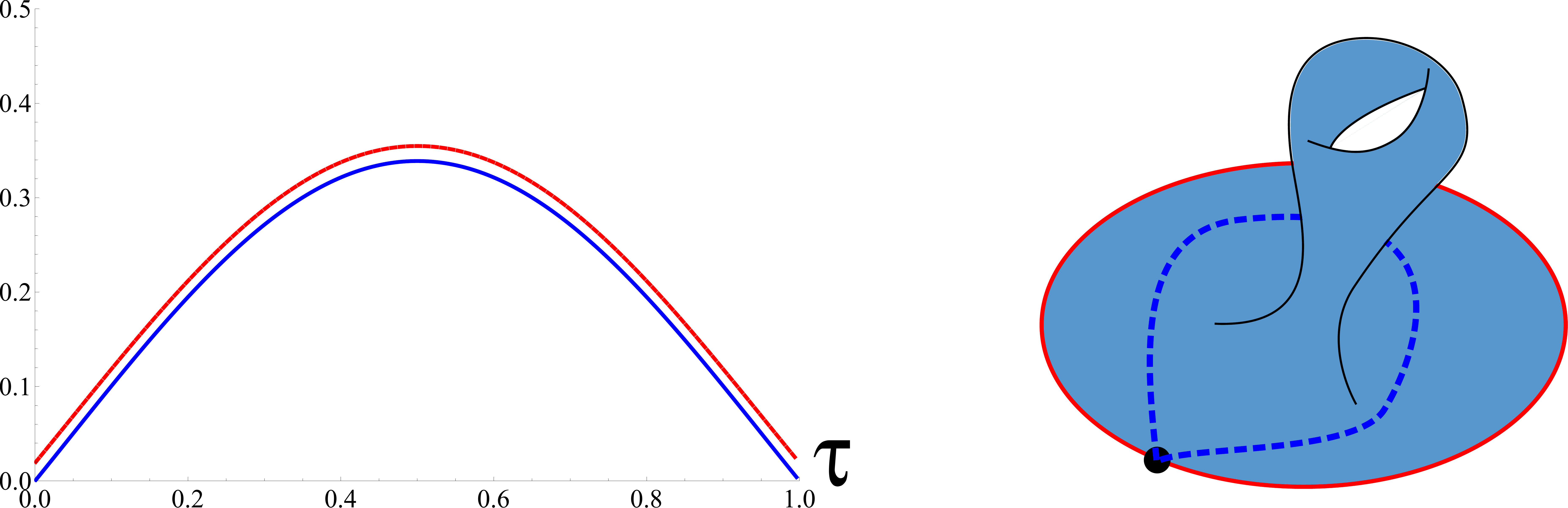}
\caption{Left: degenerate two-point correlator for $j=1/2$ and $\beta=1$ and $C=1/2$. Blue (bottom): genus zero result \eqref{halfexact}. Red (top): genus zero \eqref{halfexact} plus genus one \eqref{halfg1} result, for a suitable choice of genus counting parameter $e^{-S_0}$. Right: $\tau \to 0$ limit can be non-trivial when wrapping around higher topology.}
\label{HighgenDeg}
\end{figure}

To be slightly more explicit, it is convenient to relate the single-density expectation value $\rho$ to the resolvent $W$ in the notation of \cite{Saad:2019lba}. We can write:
\begin{equation}
\rho_{g}(k) = \frac{1}{2}\left(W_{g,1}( 2\pi i k + \epsilon) + W_{g,1}(- 2\pi i k + \epsilon)\right),
\end{equation}
the $\epsilon$'s playing the important role of regulators in the Laplace integrals.\footnote{For instance, at genus 1 we have:
\begin{equation}
\rho_{g=1}(k) = \frac{1}{2} \left[\frac{3-2\pi^2(2\pi (k+ i \epsilon))^2}{24(2\pi (k+ i \epsilon))^4}\right] + \frac{1}{2}\left[\frac{3-2\pi^2(2\pi (k- i \epsilon))^2}{24(2\pi (k- i \epsilon))^4}\right] = \text{Pv}\left[\frac{3-2\pi^2(2\pi k )^2}{24(2\pi k )^4}\right],
\end{equation}
where the principal value Pv is taking the real part of this expression, and is effectively removing divergences as $k \to 0$: $\text{Pv}f(k) \equiv 1/2(f(k+i\epsilon) + f(k-i\epsilon))$.}
 E.g. for $j=1/2$ and using the explicit expressions for $W_{g=1,1}$ from \cite{Saad:2019lba}, we have the genus one correction ($C = 1/2$):
\begin{align}
\label{halfg1}
\frac{1}{2Z} \int_0^{+\infty} dk \rho_{g=1}(k) \frac{\sin k \tau}{k} e^{-\beta k^2}e^{\frac{\tau}{4}} + (\tau \leftrightarrow \beta-\tau).
\end{align}
The sum of the genus zero and one result for this particular $j=1/2$ is plotted in figure \ref{HighgenDeg}.
Since the genus zero correlator vanishes as $\tau=0$ or $\tau=\beta$, this is where the higher genus corrections are quite visible and non-trivial. In terms of bilocal lines connecting the two boundary operators, the reason is the possibility of the line encircling higher topology and becoming noncontractible, effectively becoming sensitive to non-UV physics due to the minimal distance to wrap around the higher topology defect.
\\~\\
Beyond some fixed genus amplitudes, the full non-perturbative spectral density for the JT matrix model was recently computed numerically in \cite{Johnson:2020exp}. The above then shows that this information is sufficient to determine the class of degenerate boundary correlators.
\\~\\
At late real times $t=-i\tau$, the degenerate correlators \eqref{degmain} increase exponentially without bound. In particular, the role of strongly suppressed ($e^{-S_0} \ll 1$) higher topological corrections is less interesting here as it can never stabilize the late time behavior.

\section{Degenerate bilocal correlators in JT supergravity}
\label{s:susy}
In the remainder of this work, we generalize the discussion to $\mathcal{N}=1$ JT supergravity and its boundary super-Schwarzian quantum mechanics. Since several of the analogous prerequisites (i.e. super-Schwarzian perturbation theory, and the fixed length amplitudes of Liouville supergravity) have not been developed in the literature before, we will do this here as well. This is automatically a rather technical endeavor. \\ 
The reader only interested in the bosonic story, can safely move to the concluding section at this point.
\\~\\
In this section, we first review JT supergravity in terms of the super-Schwarzian boundary action and the bilocal boundary operator. We then study the correlator for degenerate bilocal operators and exploit these to reach a similar conclusion on the structure of the perturbative small $\tau$-series \eqref{smaltau} on the disk.

\subsection{Set-up: $\mathcal{N}=1$ super-JT disk correlators}

JT supergravity on the disk can be analogously written in terms of a boundary $\mathcal{N}=1$ super-Schwarzian theory \cite{Forste:2017kwy,Cardenas:2018krd}, with action \cite{Fu:2016vas}
\begin{equation}
\label{SSSch}
S[f,\eta] = C \int d\tau d\theta \, \text{Sch}(\tau,\theta), \qquad C \sim 1/G_N.
\end{equation}
The super-Schwarzian is defined in $\mathcal{N}=1$ ($\tau,\theta$) superspace by:
\begin{equation}
\text{Sch}(\tau,\theta) \equiv \text{Sch}_f(\tau) + \theta \text{Sch}_b(\tau) = \frac{D^4\theta'}{D\theta'} - 2 \frac{D^3\theta'D^2\theta'}{(D\theta')^2},
\end{equation}
with $D= \partial_\theta + \theta \partial_\tau$ the superderivative and $\theta'= \sqrt{\partial_\tau F}\left(\theta + \eta + \frac{1}{2}\theta\eta\partial_\tau \eta\right)$, with $F(\tau) = \tan \frac{\pi}{\beta} f(\tau)$. This action describes the dynamics of the superframe $(f,\eta)$ of a boundary super-clock. \\
Written in component fields (the reparametrization $f(\tau)$ and its superpartner $\eta(\tau)$), one writes
\begin{align}
\label{susyact}
\text{Sch}_b(\tau) &= \frac{1}{2}\left[\left\{F,\tau\right\} + \eta \eta''' + 3 \eta'\eta'' - \left\{F,\tau\right\}\eta\eta'\right], \\
\text{Sch}_f(\tau) &= \eta'' + \frac{1}{2}\eta\eta'\eta'' + \frac{1}{2}\eta \left\{F,\tau\right\}.
\end{align}
The action \eqref{SSSch} is then written in bosonic space as:
\begin{equation}
S[f,\eta] = C\int d\tau \text{Sch}_b = \frac{C}{2}\int d\tau \left[\left\{F,\tau\right\} + \eta \eta''' + 3 \eta'\eta'' - \left\{F,\tau\right\}\eta\eta'\right],
\end{equation}
Considering the thermal disk theory, one is interested in the set of superreparametrizations of the supercircle. This is defined by the Euclidean path integral:
\begin{equation}
Z = \int_{\mathcal{M}}{{\mathcal D}f}{{\mathcal D}\eta}\, e^{-S[f,\eta] },
\end{equation}
over the space
\begin{equation}
{\mathcal M} = {\text{SDiff}}_{\mathcal{N}=1}(S^1)/\text{OSp}(1|2),
\end{equation}
describing a circle time reparametrization $f(t)$, satisfying $f(t+\beta)=f(t)+\beta$ and its fermionic superpartner $\eta(t)$ satisfying antiperiodic boundary conditions $\eta(t+\beta) = -\eta(t)$, as required for fermions around the thermal circle.
\\~\\
Under super-reparametrizations, the inverse superdistance, which is at the same time the elementary $h =1/2$ superspace two-point function, is transformed as \cite{Fu:2016vas}
\begin{equation}
\frac{1}{\tau_1-\tau_2-\theta_1\theta_2} \to \frac{D_1 \theta'_1 D_2 \theta'_2}{\tau_1'-\tau_2'-\theta_1'\theta_2'},
\end{equation}
where $\tau' = F(\tau + \theta \eta)$. For higher $h$, the classical bilocal operator is of the form:
\begin{equation}
\label{susyhigher}
\mathcal{O}_{h}(\tau_1,\tau_2, \theta_1,\theta_2) \equiv \left( \frac{D_1 \theta'_1 D_2 \theta'_2}{\tau_1'-\tau_2'-\theta_1'\theta_2'}\right)^{2h},
\end{equation}
expandable in its four components by expanding in the $\theta_i$'s. The two fermionic components give zero in a single bilocal correlator by fermion conservation. The bottom and top components are non-zero.
The (one-loop) exact partition function $Z$ and bilocal correlation functions for this model are known.
\\~\\
The disk JT supergravity partition function $Z$ is \cite{Fu:2016vas,Stanford:2017thb,Mertens:2017mtv}:
\begin{equation}
Z = \int_{0}^{+\infty}dk 2 \cosh 2\pi k e^{-\beta\frac{k^2}{2C}} = \left( \frac{2\pi C}{\beta} \right)^{1/2} e^{\frac{2\pi^2 C}{\beta}},
\end{equation}
the super-Schwarzian derivative one-point function is \cite{Mertens:2017mtv}
\begin{equation}
\label{schS1pt}
\left\langle \text{Sch}_b(\tau) \right\rangle_{\beta} \equiv \left\langle T\right\rangle = \frac{1}{\beta Z} \frac{\partial Z}{\partial C} = \frac{2\pi^2}{\beta^2} + \frac{1}{2C \beta}.
\end{equation}
Again the $n$ multi-Schwarzian derivative correlator is $n$-loop exact. The $\mathcal{N}=1$ super-Schwarzian correlation functions of bilocal operators were determined in \cite{Mertens:2017mtv} using super-Liouville techniques. The answer for the (bottom component) two-point function is
\begin{align}
\label{super2pt}
G^{\mathcal{B}}_h(\tau) \equiv \left\langle \mathcal{O}_{h}(\tau,0)\right\rangle &= \frac{1}{Z}\frac{1}{\pi^{2}(2C)^{2\ell}}\int dk_1 dk_2 e^{-\tau \frac{k_1^2}{2C} - (\beta-\tau) \frac{k_2^2}{2C}} \cosh \left(2\pi k_1\right) \cosh \left(2\pi k_2\right)  \\
&\times \frac{\Gamma\bigl(\textstyle \frac{1}{2}+h \pm i(k_1-k_2)\bigr)\, \Gamma\bigl(h \pm i(k_1+k_2)\bigr) + (k_2 \to -k_2)}{\Gamma(2h)}, \nonumber 
\end{align}
and its superpartner (i.e. top component) is
\begin{align}
\label{ssuper2pt}
G^{\mathcal{T}}_h(\tau) &= \frac{1}{Z}\frac{1}{\pi^{2}(2C)^{2h1}}\int dk_1 dk_2 e^{-\tau \frac{k_1^2}{2C} - (\beta-\tau) \frac{k_2^2}{2C}} \cosh \left(2\pi k_1\right) \cosh \left(2\pi k_2\right) \\[2mm]
&\times \frac{\left(k_1+k_2\right)^2\Gamma\bigl(\textstyle\frac{1}{2}+h \pm i(k_1-k_2)\bigr) \Gamma\bigl(h \pm i(k_1+k_2)\bigr) + (k_2 \to -k_2)}{\Gamma(2h)}. \nonumber
\end{align}
In both equations, the factor on the second line will be called the vertex function in what follows, and this will get replaced by a different expression for the degenerate values of $h$. Some more details and properties are described in appendix \ref{a:corr}.

\subsection{Disk level: super-Schwarzian bilocals}
The computation of the degenerate bilocals in this case proceeds along similar lines as for the bosonic case. We refer to appendix \ref{a:super1} for the details and some examples. The resulting vertex function for the bottom component of the bilocal operator, replacing the factor on the second line of \eqref{super2pt}, is given by:
\begin{equation}
\label{vertfuncS}
\sum_{m=-j}^{j} c_{mB}^{j}(k_1,k_2) \delta (k_1 - k_2 + i m) + \sum_{m=-j+1/2}^{j-1/2} c_{mT}^{j}(k_1,k_2) \delta (k_1 - k_2 + i m),
\end{equation}
where we defined both bottom (B) and top (T) contributions:\footnote{This choice of words is motivated by the semi-classical large $C$ regime, where for the bottom component $G^\mathcal{B}$ the bottom $c_{mB}$ piece dominates and conversely for the superpartner (= top) component $G^\mathcal{T}$.}
\begin{align}
\label{botco}
c_{mB}^j(k_1,k_2) &= \frac{1}{\cosh 2\pi k_1} (-)^{m+j}\left(\begin{array}{c}
2j\\
m+j \\
\end{array}\right) \frac{1}{(2ik_2 - j +\frac{1}{2} + m)_{2j}} \\
\label{topco}
c_{mT}^j (k_1,k_2) &=  \frac{2j}{\cosh 2\pi k_1} (-)^{m+j-1/2}\left(\begin{array}{c}
2j-1 \\
m+j-1/2 \\
\end{array}\right) \frac{1}{(2ik_2 - j + m)_{2j+1}}.
\end{align}
Inserting these in \eqref{super2pt}, we get the degenerate boundary two-point functions. 
\\~\\
Let us make some comments on these formulas.
\begin{itemize}
\item
The super-Schwarzian bilocal operator \eqref{susyhigher} for $j =-h \in \mathbb{N}/2$ has for its bottom component:
\begin{equation}
\left(\frac{F_1-F_2}{F_1'^{1/2}F_2'^{1/2}}\right)^{2j}\left[1 - j \eta_1\eta_1'\right]\left[1 - j \eta_2 \eta_2'\right] - 2 j \eta_1 \eta_2 \left(\frac{F_1-F_2}{F_1'^{1/2}F_2'^{1/2}}\right)^{2j-1},
\end{equation}
and is expandable as two terms that each form a binomial expansion of order $2j$ and $2j-1$ respectively, matching with the expansions in \eqref{botco} and \eqref{topco} respectively.
\item
If one takes the semi-classical $C\to +\infty$ limit, we again take $k$ large in the inverse polynomials in \eqref{botco} and \eqref{topco}. Since $c^j_{mT}$ has one extra power of $1/k$, this piece is subdominant, and the bottom part $c^j_{mB}$ gives the expected answer $\left(\frac{\beta}{\pi}\sin \frac{\pi}{\beta}\tau \right)^{2j}$.
\item
One can also determine the degenerate superpartner correlator, this is the top component of the bilocal operator \eqref{susyhigher}. We give the results in Appendix \ref{a:super}.
\item
As for the bosonic case, the $1/C$ expansions of these equations yield convergent series. The argument in appendix \ref{s:proof} should be easily generalizable to provide a proof that for all other values of $h$ the $1/C$ series is asymptotic only.
\end{itemize}

\section{Application: super-Schwarzian perturbation theory}
\label{s:app2}
In this section we apply the degenerate boundary correlators determined above to shed light on the super-Schwarzian perturbative expansion. In order to truly compare, we need to independently develop this perturbative approach to the super-Schwarzian as well, which we do below in subsection \ref{s:susypert}. As a byproduct of developing the perturbative approach, we give a quick argument that $\mathcal{N}=1$ JT supergravity saturates the chaos bound, in the sense that the out-of-time ordered four-point function has maximal Lyapunov growth.

\subsection{Small $\tau$-expansion}
As explicit applications, the supersymmetric degenerate bilocal correlators \eqref{halfSexact} and \eqref{oneSexact} have the following small $\tau$ expansions:

{\small \begin{align}
\label{Sprecisionhalf}
\left\langle \mathcal{O}_{h=-1/2}(\tau,0)\right\rangle &= \tau + \frac{1}{16C}\tau^2 - \frac{1}{12}\left(\left\langle T\right\rangle-\frac{1}{32C^2}\right) \tau^3 - \left( \frac{1}{128C}\left\langle T\right\rangle - \frac{1}{12288C^3}\right) \tau^4 \nonumber \\
&\hspace{-1.25cm}+ \left( \frac{1}{480} \left\langle T^2\right\rangle - \frac{1}{2560  C^2}\left\langle T\right\rangle + \frac{1}{491520 C^4}\right) \tau^5 +\left( \frac{1}{4608 C} \left\langle T^2\right\rangle - \frac{1}{73728 C^3}\left\langle T\right\rangle + \frac{1}{23592960 C^5} \right) \tau^6 \nonumber \\
&\hspace{-1.25cm}- \left(  \frac{1}{40320} \left\langle T^3\right\rangle -   \frac{1}{86016 C^2} \left\langle T^2\right\rangle +\frac{1}{2752512 C^4}\left\langle T\right\rangle + \frac{1}{1321205760 C^6}\right) \tau^7 + \mathcal{O}(\tau^8),
\end{align}
}%
{\small \begin{align}
\label{Sprecisionone}
\left\langle \mathcal{O}_{h=-1}(\tau,0)\right\rangle &= \tau^2 + \frac{5}{24C}\tau^3 - \frac{1}{6}\left(\left\langle T\right\rangle-\frac{7}{256C^2}\right) \tau^4 - \left( \frac{49}{960C}\left\langle T\right\rangle - \frac{17}{6144C^3}\right) \tau^5 \hspace{2cm}\nonumber \\
&\hspace{-1.25cm}+ \left( \frac{1}{90} \left\langle T^2\right\rangle - \frac{403}{46080 C^2}\left\langle T\right\rangle + \frac{341}{1474560 C^4}\right) \tau^6 +\left( \frac{107}{26880 C} \left\langle T^2\right\rangle - \frac{1381}{1290240 C^3}\left\langle T\right\rangle + \frac{13}{786432 C^5} \right) \tau^7 \nonumber \\
&\hspace{-1.25cm}- \left(  \frac{1}{2520} \left\langle T^3\right\rangle -   \frac{1303}{1720320 C^2} \left\langle T^2\right\rangle +\frac{1213}{11796480 C^4}\left\langle T\right\rangle + \frac{5461}{5284823040 C^6}\right) \tau^8+\mathcal{O}(\tau^9),
\end{align}
}%
where
\begin{align}
\left\langle T\right\rangle = \frac{2\pi^2}{\beta^2} + \frac{1}{2\beta C}, \qquad \left\langle T^2\right\rangle = \frac{4\pi^4}{\beta^4} + \frac{6\pi^2}{\beta^3 C} + \frac{3}{4\beta^2 C^2}, \qquad \left\langle T^3\right\rangle = \frac{8\pi^6}{\beta^6} + \frac{30 \pi^4}{\beta^5 C} + \frac{45}{2\beta^4 C^2} + \frac{15}{8 \beta^3 C^3} \nonumber
\end{align}
are the (renormalized) thermal super-Schwarzian multi-stress tensor correlators. This leads to the same structure of the small $\tau$-expansion  \eqref{smaltau} as in the bosonic case. Since the expansion coefficients are polynomials in $h$ by the perturbative expansion, and we can determine this structure for all $h \in - \mathbb{N}/2$,  this is sufficient to uniquely determine these polynomials. So this structure of the perturbative expansion holds for generic real values of $h$.
\\~\\
Notice that no simplification in this perturbative series occurs due to the presence of supersymmetry: $\mathcal{N}=1$ supersymmetry in 1d is not sufficient to argue for non-renormalization theorems. \\
We will very explicitly see this at one-loop in the next subsection.

\subsection{Super-Schwarzian perturbation theory}
\label{s:susypert}
\begin{figure}[H]
\centering
\includegraphics[width=0.6\textwidth]{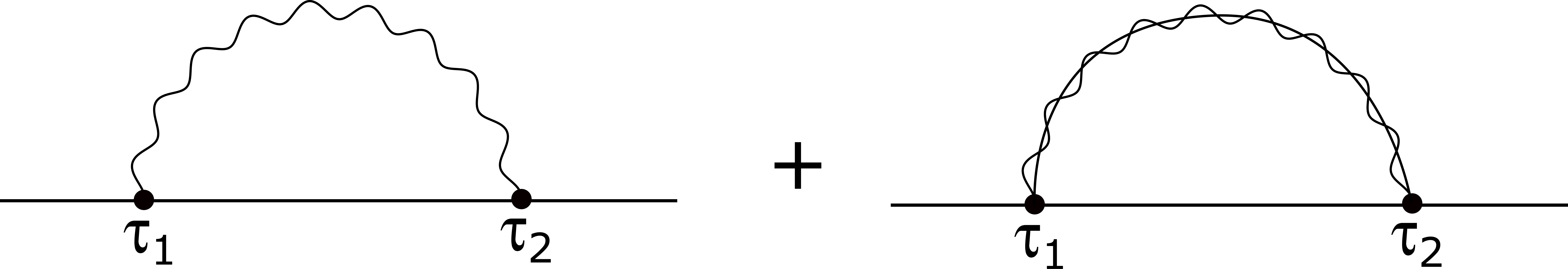}
\caption{Two Feynman graphs contributing at order $1/C$ in the supersymmetric theory.}
\label{fgraphS}
\end{figure}

From the two explicit expressions \eqref{Sprecisionhalf} and \eqref{Sprecisionone}, and the fact that the coefficient of the second term in that expansion is a quadratic homogeneous polynomial in $h$, we can write the answer for the one-loop self-energy $\Sigma$ for arbitrary real $h$ as:
\begin{equation}
\label{selfenS}
\Sigma = \frac{h(h-1/4)}{6C} =  \frac{h(h-1)}{6C} + \frac{h}{8C}.
\end{equation}
In the second equality, we have split the self-energy into a contribution from gravity, and a contribution from the gravitino (Figure \ref{fgraphS}).
To substantiate this result, we now reproduce this same term from perturbation theory within the super-Schwarzian system directly.
\\~\\
The quadratic piece of the boundary gravitino action contained in \eqref{susyact} is 
\begin{equation}
L = (2C) \left( \eta' \eta'' - \frac{\pi^2}{\beta^2} \eta \eta' \right).
\end{equation}
For ease of notation, we set $\beta=2\pi$ from here and restore it in the end. The zero-modes of this action satisfy $\eta''' + \frac{1}{4} \eta' =0$:
\begin{equation}
\eta(\tau) = A e^{i \tau/2} + B e^{-i \tau /2} + D,
\end{equation}
with $D=0$ due to antiperiodicity $\eta(\tau+2\pi) = - \eta(\tau)$. The orthonormal eigenmodes are given by $\psi_n(\tau) = \frac{1}{\sqrt{2\pi}} e^{in \tau}$, with eigenvalue $\lambda_n = 4i C n(n^2-1/4)$ and hence the zero-modes $n=\pm 1/2$. The propagator is then given by:
\begin{equation}
\left\langle \eta(\tau)\eta(0)\right\rangle = \frac{1}{8\pi i C}\sum_{n\in \mathbb{Z}+1/2, \neq \pm 1/2}\frac{1}{n(n^2-1/4)} e^{i n \tau}.
\end{equation}
Choosing a contour $\mathcal{C}$ that encircles all half-integers $\neq \pm 1/2$, one can write this as:
\begin{equation}
\left\langle \eta(\tau)\eta(0)\right\rangle = \frac{1}{8\pi i C} \oint_{\mathcal{C}}\frac{ds}{e^{2\pi i s}+1}\frac{1}{s(s^2-1/4)} e^{i s \tau}.
\end{equation}
Deforming the contour to encircle the poles $0,\pm 1/2$ and the (vanishing) piece at infinity, one evaluates the residue immediately to find:\footnote{The application of this method to the bosonic Schwarzian is described e.g. in \cite{Sarosi:2017ykf}. We have restored the units of $\beta$ here.}
\begin{equation}
\label{fermiprop}
\left\langle \eta(\tau)\eta(0)\right\rangle = -\frac{1}{4C}\frac{\beta^2}{4\pi^2} \left[-2 + 2 \left( 1- 2\frac{\tau}{\beta} \right) \cos \frac{ \pi\tau}{\beta} + \frac{6}{\pi}\sin \frac{\pi \tau}{\beta}\right],
\end{equation}
with special cases:
\begin{align}
\left\langle \eta'(0)\eta(0)\right\rangle = -\frac{1}{4C} \frac{\beta}{2\pi^2}, \qquad \left\langle \eta(\tau)\eta(0)\right\rangle_{\beta \to \infty} = - \frac{1}{4C}  \left[ \frac{\beta}{2\pi^2} \tau - \frac{\tau^2}{4} + \hdots \right].
\end{align}
The bottom component of the bilocal operator for generic $h$ \eqref{susyhigher} is explicitly
\begin{equation}
\label{twobott}
\left(\frac{F_1'F_2'}{(F_1-F_2)^2}\right)^{h}\left[1 + h \eta_1\eta_1'\right]\left[1 + h \eta_2 \eta_2'\right] + 2 h \eta_1 \eta_2 \left(\frac{F_1'F_2'}{(F_1-F_2)^2}\right)^{h+1/2}.
\end{equation}
The $1/C$ correction to a single bilocal operator is then readily found. At this order, the bosonic and fermionic contributions just add up; the bosonic contribution found by using the propagator \eqref{propa} in \eqref{opexpa}, while the new fermionic contribution is found by using \eqref{fermiprop} in \eqref{twobott}. We get:
\begin{align}
\label{msyS}
&\left\langle \mathcal{O}_{h}(\tau,0)\right\rangle = \left\langle \mathcal{O}_{h}(\tau,0)\right\rangle_{C \to + \infty} \nonumber \\
&\times \left[ 1 + \frac{\beta}{4\pi^2 C}\left(h \frac{(u^2-2\pi u +2 -2\cos u +2(\pi-u) \sin u)}{4\sin^2\frac{u}{2}} + \frac{h^2}{2}\left(-2 + \frac{u}{\tan\frac{u}{2}}\right)\left(-2 + \frac{u-2\pi}{\tan\frac{u}{2}}\right)\right)\right. \nonumber \\
&\left.+\frac{\beta}{4\pi^2 C}  h \left(\frac{\pi +  \left( u - \pi  \right) \cos \frac{u}{2}}{\sin \frac{u}{2}}-2\right)\right],
\end{align}
where $u=2\pi \tau/\beta$. The first line is the bosonic answer \eqref{msy} from the $\mathcal{N}=0$ Schwarzian, and the second line contains the gravitino contribution.
Taylor-expanding this in $\tau$ to find the lowest correction in the $\tau/C$ series, we find the answer:
\begin{equation}
\frac{h(h-1/4)}{6 C} = \underbrace{\frac{h(h-1)}{6 C}}_{\text{graviton}} + \underbrace{\frac{3/4 h}{6 C}}_{\text{gravitino}},
\end{equation}
where the gravitino gives a positive contribution to the self-energy, indeed matching with the result we got from analysing the general structure above in \eqref{selfenS}.

\subsection{Lyapunov behavior in $\mathcal{N}=1$ JT supergravity}
\begin{figure}[H]
\centering
\includegraphics[width=0.3\textwidth]{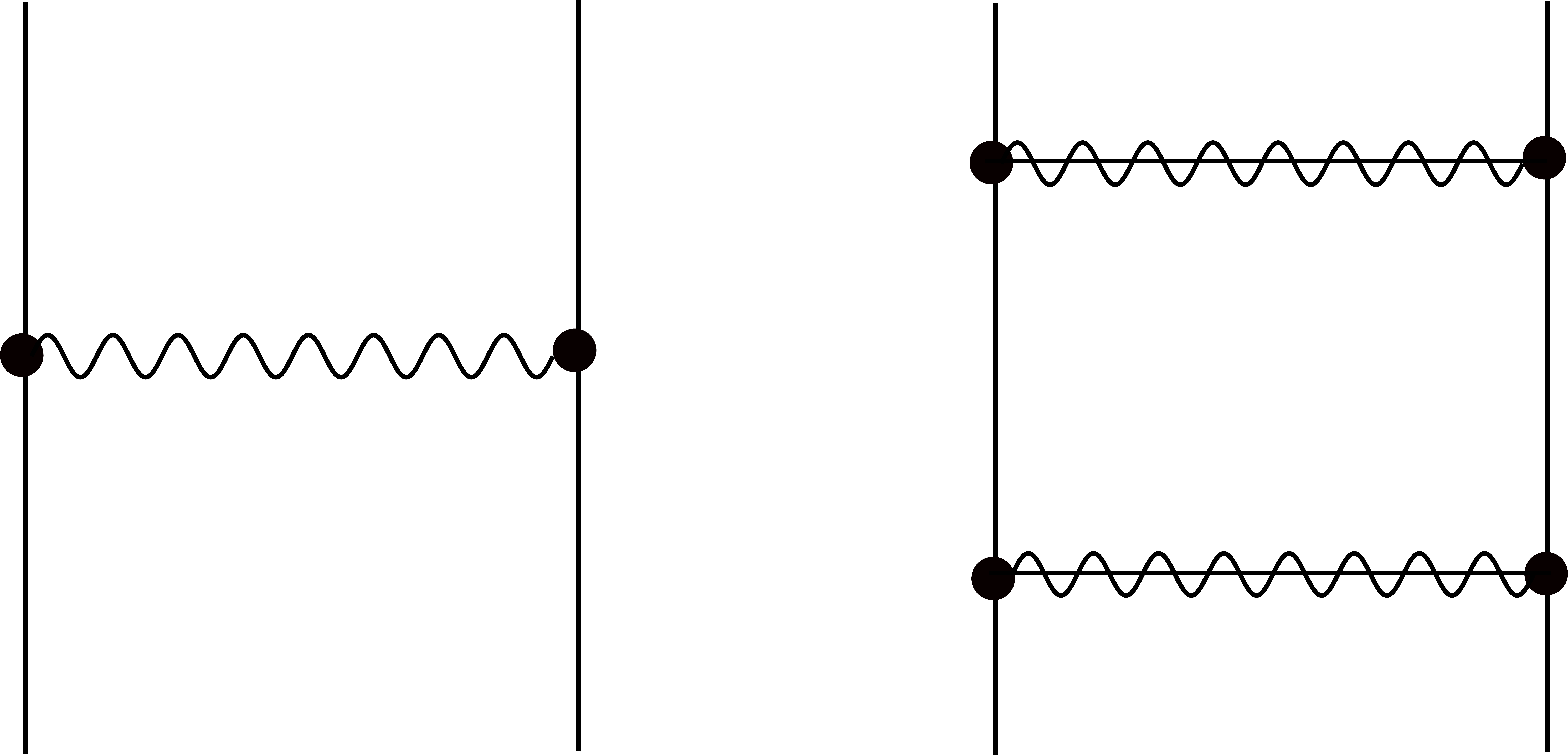}
\caption{Lowest graviton (left) and gravitino (right) contribution to the four-point function. The gravitino contribution only starts at $1/C^2$.}
\label{fgraphLya}
\end{figure}

As an aside, starting with \eqref{twobott}, one can also analyze the lowest fermionic correction to the four-point function. One readily sees that this contains \emph{two} $\eta$-propagators, and is hence suppressed as $1/C^2$, unlike the graviton which has a single propagator and only $1/C$ suppression (Figure \ref{fgraphLya}).
Since this is true for any ordering of the four points, this is also immediately so for the boundary out-of-time ordered four-point function, and hence the leading Lyapunov behavior $\sim \frac{\beta}{C} e^{\frac{2\pi}{\beta}t}$, and the Lyapunov exponent in particular is maximal $\lambda = 2\pi/\beta$ \cite{Maldacena:2015waa} in $\mathcal{N}=1$ JT supergravity. \\
When evaluating the $1/C^2$ correction in the $\mathcal{N}=1$ case, the bosonic contribution was evaluated in \cite{Qi:2019gny}. The fermionic contribution can be explicitly evaluated using the propagator \eqref{fermiprop} twice for each of the possible contractions of the right graph of figure \ref{fgraphLya}. Note that the bosonic partner of the measure fermion $\psi$ only appears at higher order. We will not evaluate it explicitly.

\section{Liouville supergravities}
\label{s:Lsugra}
Next we aim to understand higher genus corrections to these degenerate correlators. In order to do so, and paralleling the bosonic treatment, it is useful to first understand the ancestral minimal superstring for which JT supergravity is found in a parametric limit. 
The next few sections will have the goal to investigate fixed length amplitudes of $\mathcal{N}=1$ Liouville supergravities in their own right and to find out precisely how one obtains a (super) JT parametric limit. We come back to the degenerate bilocal correlators in section \ref{s:matrix} and in particular in subsection \ref{s:mampl}.
\\~\\
This section provides a short review on $\mathcal{N}=1$ Liouville supergravity and the minimal superstring, and a summary on the transformation to the fixed length basis. This section is complemented by some technical review material contained in appendix \ref{s:detliou}, where in \ref{s:lagr} we provide a Lagrangian treatment to prove the transformation to fixed length amplitudes in both the Ramond and the Neveu-Schwarz sector. In \ref{s:liou} we collect the known super-Liouville amplitudes in the FZZT brane basis, to be used later on in the main text.

\subsection{Liouville supergravity and minimal superstring}
We consider the $\mathcal{N}=1$ supersymmetric Liouville model \cite{DiVecchia:1982bz,DHoker:1983xdu,Babelon:1985gw,Arvis:1982tq} with central charge $c_L = \frac{3}{2}+3Q^2$ where $Q=1/b+b$ on a manifold with a circular boundary. We couple this to a matter sector with $c_M < 3/2$ parametrized by $c_M = \frac{3}{2}- 3q^2$. Demanding a cancellation of the superconformal anomaly requires:
\begin{equation}
c_L + c_M = 15 \quad \Rightarrow \quad q=1/b-b.
\end{equation}
Neveu-Schwarz (NS) boundary vertex operators are of the form
\begin{equation}
\mathcal{B} = \bigl( c e^{-\varphi}\bigr) \, e^{\frac{\beta}{2}\phi}\, \mathcal{O}_{\beta_M},
\end{equation}
with $c e^{-\varphi}$ the superghost contribution and $e^{\frac{\beta}{2}\phi}$ the boundary super-Liouville vertex operator, dressing the matter operator $\mathcal{O}_{\beta_M}$. We have the constraint $\Delta_\beta + \Delta_{\beta_M} = 1/2$. With\footnote{\label{fn:time}The parametrization and definition of $\beta_M$ is in parallel to the bosonic case. It follows from consider timelike super-Liouville to describe the $c_M <3/2$ matter CFT. This essentially boils down to taking an ordinary super-Liouville CFT with $b\to ib $ and $Q \to -iq $. Primary boundary vertex operators are then described by $\mathcal{O}_{\beta_M} = e^{\frac{\beta_M}{2} \chi}$ in terms of the (timelike) Liouville field $\chi$, with weight $\Delta_{\beta_M} = \frac{1}{2}\beta_M(q+\beta_M) $.}
\begin{equation}
\Delta_\beta = \frac{1}{2}\beta(Q-\beta), \qquad \Delta_{\beta_M} = \frac{1}{2}\beta_M(q+\beta_M),
\end{equation}
this leads to the solutions $\beta = b- \beta_M$ or $\beta = 1/b + \beta_M$. These choices are related by applying a boundary reflection transformation $\beta \to Q - \beta$, and represents a freedom in dressing the matter operator with given $\beta_M$. We hence focus on the first case only:
\begin{equation}
\label{onsh}
\beta = b - \beta_M.
\end{equation}
Ramond (R) boundary vertex operators are of the form
\begin{equation}
\mathcal{B}^{\epsilon} = \bigl( c e^{-\varphi/2} \bigr) \sigma^{\epsilon} e^{\frac{\beta}{2}\phi}\mathcal{O}_{\beta_M}^{\tilde{\epsilon}},
\end{equation}
with the constraint $\Delta_\beta + \Delta_{\beta_M} = 5/8$ and 
\begin{equation}
\Delta_\beta = \frac{1}{2}\beta(Q-\beta) + \frac{1}{16}, \qquad \Delta_{\beta_M} = \frac{1}{2}\beta_M(q+\beta_M) + \frac{1}{16},
\end{equation}
solvable by the same relation \eqref{onsh}.
\\~\\
If we consider for the matter theory a $(p,q)$ superminimal model,\footnote{$p,q\geq 2$, and $(p,q)$ odd and coprime, or $(p,q)$ even and $(p/2,q/2)$ coprime and $(p-q)/2$ odd. This last restriction follows from modular invariance and is violated for the $(2, 4 \mathbf{k}+2)$ models. We comment on this further on.} then we have only a finite set of tachyon vertex operators, corresponding to taking the degenerate super-Virasoro primaries as matter operators, and dressing these with the appropriate super-Liouville operators \cite{Seiberg:2003nm}. \\
Depending on the parity of $r-s$, open string tachyon vertex operators in the minimal superstring are of the form:
\begin{align}
\label{MM1}
\mathcal{B}_{r,s}^{\NS}= \bigl( c e^{-\varphi} \bigr) e^{\frac{\beta_{r,s}}{2}\phi}\mathcal{O}_{r,s}, \quad r-s \text{ even}, \qquad r=1...p-1, \,\, s=1 ...q-1, \\
\mathcal{B}_{r,s}^{\R} = \bigl( c e^{-\varphi/2}\bigr) \sigma^{\epsilon \bar{\epsilon}}e^{\frac{\beta_{r,s}}{2}\phi}\mathcal{O}_{r,s}, \quad r-s \text{ odd}, \qquad r=1...p-1, \,\, s=1 ...q-1,
\end{align}
where we left the legpole factors implicit. The operator $\mathcal{O}_{r,s}$ is a superminimal model primary operator, and has the identification $\mathcal{O}_{r,s} \equiv \mathcal{O}_{p-r,q-s}$. This is dressed by the super-Liouville primary vertex operator with parameter:
\begin{align}
\beta_{r,s} = (1+s)\frac{b}{2} + (1-r) \frac{1}{2b}.
\end{align}
For $r-s$ even, this operator is in the NS sector of the theory, whereas for $r-s$ odd it is in the R sector.
\\~\\
In the special case where $p=2$, we have an even more restricted class:
\begin{equation}
\mathcal{B}_{1,2j+1}, \qquad \beta_{1,s} = b +jb, \quad j= 0,\frac{1}{2},1 \hdots \begin{cases}
\mathbf{k}-1/2 ,\quad q=4\mathbf{k},\\
\mathbf{k}, \qquad\quad\,\,\,\, q=4\mathbf{k}+2,
\end{cases}
\end{equation}
where all half-integer $j$ give R-sector boundary operators, and all integer $j$ give NS-sector boundary operators. We have already parametrized to the specific superminimal model series of interest to us: $(2,4\mathbf{k})$ or $(2,4\mathbf{k}+2)$. R-sector operators will not play a role in our main story, and their fixed length correlators are studied in appendix \ref{s:ramms}.
\\~\\
We will not be bothered too much by the precise matter sector in the remainder of this and the next section, since we only focus on cases where the amplitude factorizes into a Liouville piece, a matter piece and a superghost piece. The only length dependence comes from the super-Liouville piece and this will be our focus. The main effect of the matter and ghosts sectors is a cancellation of the dependence on the worldsheet coordinates, much like happens in the bosonic minimal string. 

However, in section \ref{s:matrix} we will investigate the matrix model interpretation of certain of the quantities we computed, and for this we will restrict to the $(2,q)$ superminimal models, with a particular emphasis on $q=4\mathbf{k}$, with $\mathbf{k}\in \mathbb{N}$.

\subsection{Transform to fixed length basis}
We next present a short summary of how the transform to the fixed length basis works. The starting point is the Lagrangian description of (conformal) boundaries in $\mathcal{N}=1$ Liouville SCFT, for which we refer the reader to appendix \ref{s:lagr}. \\
It is known that the Cardy boundary states for the non-degenerate super-Virasoro representations with Liouville momentum $s$ can be labeled as:
\begin{equation}
\left|\NS,s,\eta\right\rangle, \qquad \left|\R,s,\eta\right\rangle,
\end{equation}
in terms of a representation label $s$, related to the boundary FZZT cosmological constant $\mu_B(s)$ by \cite{Fateev:2000ik}:
\begin{equation}
\label{mub}
\mu_B = \kappa \begin{cases}
\cosh \pi  b s ,\qquad \eta=+1,\\
\sinh \pi b s, \qquad \eta=-1,
\end{cases}
\qquad \kappa= \sqrt{\frac{2\mu}{\cos \frac{\pi b^2}{2}}},
\end{equation}
a sign $\eta = \pm 1$ representing a local fermion boundary condition, and a global fermionic boundary condition Neveu-Schwarz (NS) or Ramond (R) as one goes around the boundary circle. Associated to each of these classes of FZZT boundary states, will be a fixed-length amplitude that we will determine in the following. 
\\~\\
For the R-sector FZZT-branes $\left|\R,s,\eta\right\rangle$, the integral transform to go to $\left|\R,\ell,\eta\right\rangle$, or to transfer from the $\mu_B$-basis to the $\ell$-length basis is the following:
\begin{equation}
\label{Rint}
-i\int_{\mathcal{C}} d\mu_B e^{\mu_B^2\ell} \hdots, \qquad \mathcal{C}={\mu_B^2-\eta\kappa^2:-i\infty \to+i\infty},
\end{equation}
The contour $\mathcal{C}$ is half of a hyperbola. This rule is motivated from the Lagrangian perspective in appendix \ref{s:lagr} and applied explicitly to the partition function and the bulk one-point function in sections \ref{s:pf} and \ref{s:bulk} respectively. The contour is deformed to wrap the negative $\mu_B^2$ axis, where the role of taking the discontinuity across the branch cut at negative $\mu_B^2$ is played by adding instead of subtracting the two contributions, leading to the effective transformation:
\begin{equation}
\int_{0}^{+\infty} ds e^{-\frac{\ell}{4\pi} (\cosh(2 \pi b s) -\eta)} \left\{\begin{array}{c} \cosh \pi b s, \, \eta=+1 \\ \sinh \pi b s, \, \eta=-1 \end{array}\right. \hdots
\end{equation}
For the NS-sector FZZT-brane $\left|\NS,s,\eta\right\rangle$, to transfer to the fixed-length boundary state $\left|\NS,\ell,\eta\right\rangle$, one uses the integral transform:\footnote{Both integral transforms \eqref{Rint} and \eqref{NSint} have the same structure as proposals made for suitable macroscopic loop operators in the $\hat{c}=1$ matrix model in \cite{Takayanagi:2003sm,Takayanagi:2004jz}, lending further support for these equations.}
\begin{equation}
\label{NSint}
-i\int_{\mathcal{C}} d\mu_B^2 e^{\mu_B^2\ell} \hdots, \qquad \mathcal{C}={\mu_B^2-\eta\kappa^2:-i\infty \to+i\infty},
\end{equation}
leading to
\begin{equation}
\int_{0}^{+\infty} ds e^{-\frac{\ell}{4\pi} (\cosh(2 \pi b s) -\eta)} \sinh 2 \pi b s \hdots
\end{equation}
where the discontinuity is just as in the bosonic case happening by subtraction. The NS-branes play a somewhat minor role in our story as they behave largely as in the bosonic Liouville story, and in particular they have the same semi-classical (bosonic JT) limit.

\section{Fixed length amplitudes of Liouville supergravity}
\label{s:pfone}
In this section, we discuss the above transformation to find the fixed-length disk amplitudes. We first present a derivation of the marked disk partition function and the bulk one-point functions, paralleling the bosonic treatment of \cite{Mertens:2020hbs}. Then we describe local marking operators and how they act indeed as the identity insertion in the fixed length basis, followed by our treatment of the boundary two-point function. This last part is our main result in this section, and in particular equations \eqref{final1} and \eqref{final2}. \\
At a technical level, the starting point is the super-Liouville amplitudes with FZZT brane boundaries, which are summarized in appendix \ref{s:liou}.

\subsection{Marked partition function}
\label{s:pf}

We first apply the above transform to the marked partition function, and derive fixed-length disk amplitudes. 
\\~\\
The bulk one-point function for the insertion of a bulk cosmological constant operator for FZZT boundary condition $s$ is given by:
\begin{equation}
\partial_\mu Z = \left\langle c\bar{c}e^{-\varphi-\tilde{\varphi}} \bar{\psi} \psi e^{b \phi}\right\rangle = \left[\frac{i \eta}{b^2} \Big(\mu \pi \gamma\left( bQ/2 \right) \Big)^{\frac{1}{2b^2}-\frac{1}{2}} \Gamma\Big(\frac{b^2}{2} - \frac{1}{2}\Big) \Gamma\Big(\frac{3}{2}-\frac{1}{2b^2}\Big) \right] \cosh \pi \left( b-\frac{1}{b} \right) s,
\end{equation}
in terms of the superghost contributions $c\bar{c}e^{-\varphi-\tilde{\varphi}}$ that we will suppress, and the super-Liouville fields $\psi$, $\bar{\psi}$ and $\phi$, where the dependence on the fermionic boundary condition $\eta$ is implicit in the relation between $\mu_B$ and the brane parameter $s$ in \eqref{mub}. This equation is the superpartner of \eqref{NS:1}. Integrating w.r.t. $\mu$, and choosing a more convenient normalization of the amplitude, Seiberg and Shih found the following unmarked FZZT disk partition functions \cite{Seiberg:2003nm}:
\begin{align}
Z(\mu_B)^{\U} &\sim b^2 \cosh \pi b s \cosh \frac{\pi}{b}s - \sinh \pi b s\sinh \frac{\pi}{b}s , \qquad \eta=+1, \\
Z(\mu_B)^{\U} &\sim b^2 \sinh \pi b s\sinh \frac{\pi}{b}s - \cosh \pi b s \cosh \frac{\pi}{b}s , \qquad \eta=-1.
\end{align}
We will analyze the two fermionic boundary conditions $\eta=\pm1$ separately.

\begin{figure}[h]
\centering
\includegraphics[width=0.6\textwidth]{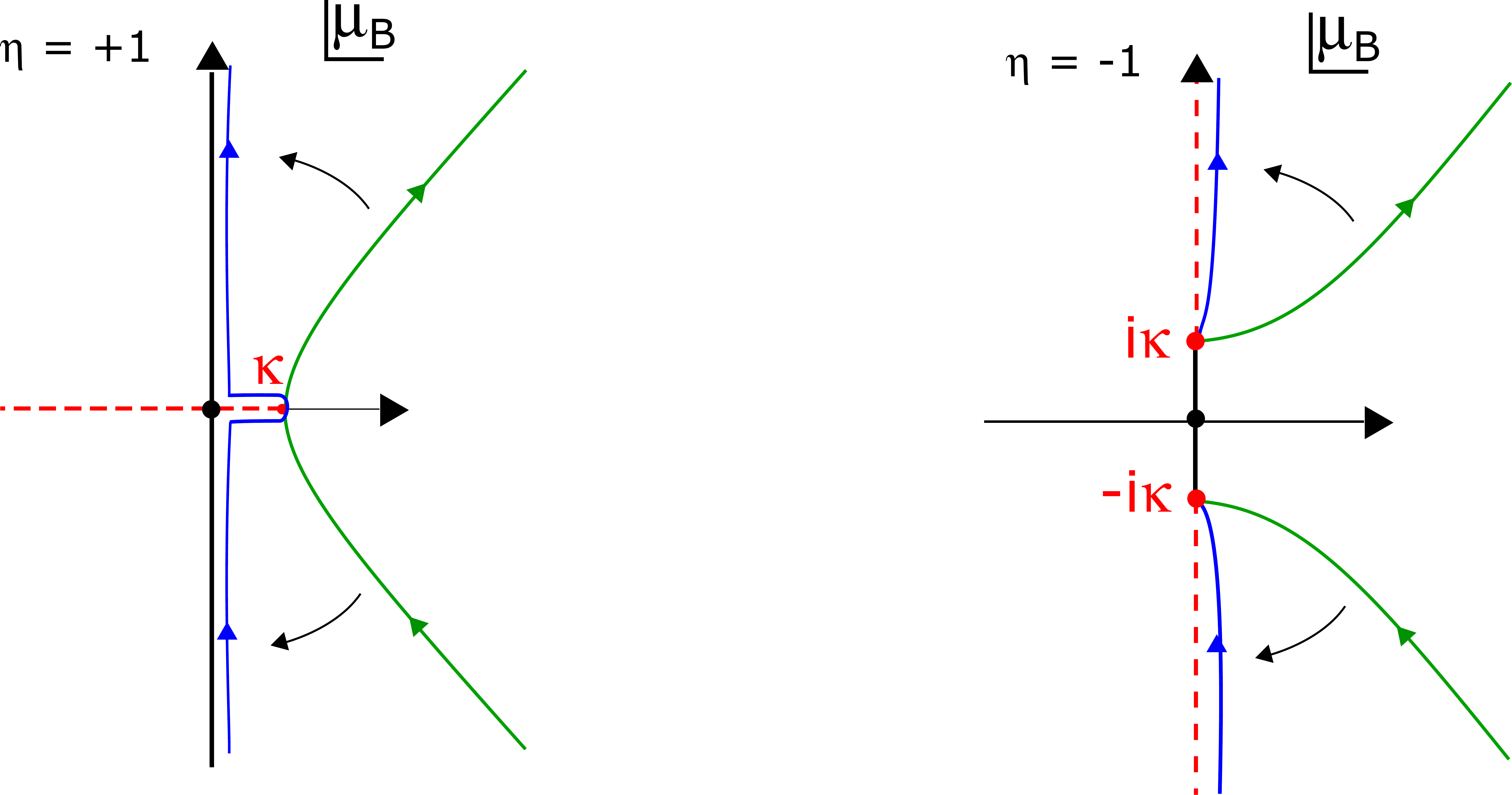}
\caption{Integration contour in the $\mu_B$ plane in the supersymmetric case. The initial contour is in green, and the final one in blue. The red lines denote the branch cuts of the integrand. Left: $\eta=+1$. Right: $\eta=-1$.}
\label{contourDeformSUSY}
\end{figure}

\begin{center}
$\bm{\eta=+1}$
\end{center}
The marked partition function is 
\begin{equation}
\label{mark1}
Z(\mu_B)^{\M} = \partial_{\mu_B}Z(\mu_B)^{\U} = \cosh(\frac{1}{b^2}\text{arccosh}\frac{\mu_B}{\kappa}) .
\end{equation}
The integration contour \eqref{Rint} in the $\mu_B$ plane is a single leaf of a hyperboloid with top at $\mu_B = +\kappa$ (Figure \ref{contourDeformSUSY}).\footnote{The difference here compared to there is in a (re)normalization of $\mu$ and $\mu_B$, effectively mapping $\mu_0 \to \kappa^2$.}
The contour is initially along the vertical line $\mu_B^2 - \eta \kappa^2: -i \infty \to i\infty$. We can contour deform it to hug the imaginary axis.\footnote{The small real segment $(0,\kappa)$ cancels between top and bottom part of the contour, just like it did in the bosonic case. Note that one cannot contour deform to the right since the integral diverges there. Note also that the case with $\eta=-1$ is a disconnected contour in the $\mu_B$-plane.}
Deforming this contour to the imaginary axis, and evaluating, we can write it as:
\begin{equation}
\int_0^{+\infty} d\mu_B e^{-\ell \mu_B^2 } \cosh(\frac{1}{b^2}\text{arcsinh}\frac{\mu_B}{\kappa}),
\end{equation}
where we used:
\begin{equation}
\cosh(\frac{\pi}{b} \left( s+ \frac{i }{2b} \right) ) + \cosh(\frac{\pi}{b} \left(s - \frac{i }{2b} \right) ) =  2 \cos \frac{\pi}{2b^2} \cosh\frac{\pi s}{b}.
\end{equation}
Interpreting the boundary length $\ell=\beta$ as the inverse temperature, we can read this as a disk contribution of the thermal partition function with energy $E= \mu_B^2$ and density of states:
\begin{equation}
\label{dosR}
\rho_{\R,\eta=+1}(E) = \frac{\cosh \frac{1}{b^2}\text{arcsinh} \frac{\sqrt{E}}{\kappa}}{\sqrt{E}},
\end{equation}
starting at $E=0$ with a hard spectral edge $\rho \sim 1/\sqrt{E}$ and asymptotically following a power law as $\rho \sim E^{1/2b^2-1/2}$. For the particular case of the ($2,4\mathbf{k}$) minimal superstring, the series expansion of $\rho_{\R,\eta=+1}(E)$ in powers of $\sqrt{E}$ truncates at $E^{\mathbf{k}-1/2}$. At the thermodynamical saddle, we can write the first law as
\begin{equation}
\sqrt{2E^2+\kappa^2 E} = \frac{1}{b^2\beta},
\end{equation}
at high energies (UV) going as $E \sim 1/\beta$, just like in the bosonic case \cite{Mertens:2020hbs}, and following the (super)JT black hole relation $\sqrt{E} \sim \beta^{-1}$ at lower energies. The UV behavior means we expect to find a holographic bulk that deviates from the asymptotic AdS boundary conditions. It would be interesting to learn of a dilaton supergravity model and (super)potential that is able to generate this law from a black hole solution in the bulk.
\\~\\
Substituting
\begin{equation}
\mu_B = \kappa \sinh \pi b s \quad \Leftrightarrow \quad s = \frac{1}{\pi  b} \text{arcsinh}\frac{\mu_B}{\kappa},
\end{equation}
we get:
\begin{align}
\label{inttodo}
\boxed{
Z_{\R,\eta=+1}  = \int_{0}^{+\infty} ds e^{-\frac{\kappa^2}{2} \ell (\cosh(2\pi b s)-1)} \cosh \pi b s \cosh\frac{\pi s}{b}}.
\end{align}
The quantity $\kappa^2/2$ limits to $\mu$ as $b\to0$ by \eqref{mub}, and can be viewed as the effective bulk cosmological constant $\mu_{\text{eff}}$. Note that deforming the contour in the above way has effectively swapped the local fermionic boundary condition $\eta=\pm 1 \to \eta = \mp 1$. We will see this happen for the correlation functions below as well. The integral \eqref{inttodo} can be evaluated in terms of modified Bessel functions of the second kind as:
\begin{align}
Z_{\R,\eta=+1} = \frac{e^{\frac{\kappa^2}{2} \ell}}{4\pi b} \left( K_{\frac{1}{2}+\frac{1}{2b^2}}\left(\kappa^2 \ell/2 \right) + K_{\frac{1}{2}-\frac{1}{2b^2}}\left(\kappa^2 \ell/2\right)\right).
\end{align}
One can view the transform $\int_{0}^{+\infty} ds e^{- \kappa^2 \ell (\cosh(2 \pi b s)-1)} \cosh \pi b s \hdots $ as the correct supersymmetric version of the transform to the length basis \cite{Fateev:2000ik,Douglas:2003up}. We will apply this kernel also for correlation functions below.

\begin{center}
$\bm{\eta=-1}$
\end{center}
The marked partition function is in this case given by
\begin{equation}
\label{mark2}
Z(\mu_B)^{\M} = \partial_{\mu_B}Z(\mu_B)^{\U}= \sinh(\frac{1}{b^2}\text{arcsinh}\frac{\mu_B}{\kappa}) .
\end{equation}
Looking at the branchcut of arcsinh, we have the discontinuity-like relation:
\begin{equation}
\sinh(\frac{\pi}{b} \left( s+ \frac{i }{2b} \right) ) + \sinh(\frac{\pi}{b} \left( s - \frac{i }{2b} \right) ) =  2 \cos \frac{\pi}{2b^2} \sinh\frac{\pi s}{b},
\end{equation}
leading to
\begin{equation}
\int_{\kappa}^{+\infty} d\mu_B e^{-\ell \mu_B^2 } \sinh(\frac{1}{b^2}\text{arccosh}\frac{\mu_B}{\kappa}),
\end{equation}
with density of states:
\begin{equation}
\label{dosm}
\rho_{\R,\eta=-1}(E) = \frac{\sinh \frac{1}{b^2}\text{arccosh} \frac{\sqrt{E}}{\kappa}}{\sqrt{E}},
\end{equation}
starting at $E=\kappa^2$ with a spectral edge $\rho \sim \sqrt{E-\kappa^2}$ and asymptotically a power law as $\rho \sim E^{1/2b^2-1/2}$. For the ($2,4\mathbf{k}$) minimal superstring, the series expansion of $\rho_{\R,\eta=-1}(E)$ in powers of $\sqrt{E-\kappa^2}$ truncates at $(E-\kappa^2)^{\mathbf{k}-1/2}$. This model generates a semi-classical first law 
\begin{equation}
\sqrt{2E^2 - \kappa^2 E} = \frac{1}{b^2\beta}.
\end{equation}
We now set $\mu_B = \kappa \cosh \pi  b s$ and hence the fixed-length marked disk amplitude is given by the expression:
\begin{align}
\label{inttodo2}
\boxed{
Z_{\R,\eta=-1}(\ell)  = \int_0^{+\infty} ds e^{-\frac{\kappa^2}{2} \ell (\cosh(2\pi b s) +1)} \sinh \pi b s \sinh(\frac{\pi}{b}s)}.
\end{align}
The integral can be done and leads to
\begin{align}
Z_{\R,\eta=-1}(\ell) = e^{-\frac{\kappa^2}{2} \ell}\frac{1}{4\pi b} \left( K_{\frac{1}{2}+\frac{1}{2b^2}}(\kappa^2 \ell/2) - K_{\frac{1}{2}-\frac{1}{2b^2}}(\kappa^2 \ell/2)\right).
\end{align}
So in both cases we have to take the \emph{sum} of both terms of the contour across the cut. To make a distinction with the bosonic case, where we take the genuine discontinuity by subtraction, here we will denote the sum by the \textbf{s}discontinuity, or \textbf{s}Disc for short. This will distinguish the R-sector and the NS-sector branes, the latter requiring taking the genuine discontinuity by subtracting both sides.
\\~\\
In both cases, the final result for the fixed length partition functions matches with the R-sector minisuperspace Liouville QM computation results \cite{Douglas:2003up}. This generalizes the observation made by the Zamolodchikovs \cite{Zamolodchikov:2001ah} to the supersymmetric case that the Liouville disk partition function at fixed length \eqref{inttodo} and \eqref{inttodo2} is exactly computed by the minisuperspace Liouville problem, details of the latter can be found in \cite{Douglas:2003up}.

\subsection{Bulk one-point function}
\label{s:bulk}
Let us insert a (gauge-fixed) bulk operator $\mathcal{T}_\alpha = c \bar{c} \, e^{-\varphi - \tilde{\varphi}}e^{\alpha \phi}$ and transform the bulk one-point function to the fixed length basis. We will parametrize $\alpha = \frac{Q}{2} - \frac{\theta}{2b} =  \frac{Q}{2} + iP$, in terms of what turns out to be a conical defect $\theta$, or a macroscopic primary $P$.
\\~\\
Starting with an NS brane with an NS bulk operator insertion \eqref{NS:1} for $\eta=+1$, we first mark the bulk one-point function, by differentiating w.r.t. $\mu_B^2$ as:
\begin{equation}
\label{deri}
\partial_{\mu_B^2} \cosh  \frac{\theta}{b^2} \text{arccosh} \frac{\mu_B}{\kappa} = \frac{1}{2\mu_B}\theta\frac{\sinh  \frac{\theta}{b^2} \text{arccosh}\frac{\mu_B}{\kappa}}{b^2\sqrt{\mu_B+\kappa}\sqrt{\mu_B-\kappa}}.
\end{equation}
We transform this to the fixed length basis using the NS transformation \eqref{NSint}. Deforming the contour to $\mu_B: -i\infty +\epsilon \to +i \infty +\epsilon$,\footnote{One checks explicitly that there is effectively no branch cut in the region $(0,\kappa)$.} we get again two contributions that have to be added. The $2\mu_B$-factor in the NS transform measure \eqref{NSint} cancels with the explicit factor above in \eqref{deri}. Also, the square root factors in the denominator creates a relative minus sign between both parts of the contour, effectively leading to a subtraction of the pieces on both sides of the branch cut. We hence see that the NS computation is in effect doing the same discontinuity calculation as in the bosonic case of \cite{Mertens:2020hbs}. We end up with:
\begin{equation}
\text{Disc}\left[\partial_{\mu_B} \cosh  \frac{\theta}{b^2} \text{arccosh}\frac{\mu_B}{\kappa} \right] = \theta \frac{\sin \frac{\pi \theta}{2b^2}\, \cosh \frac{\theta}{b^2} \text{arcsinh} \frac{\mu_B}{\kappa} }{b^2\sqrt{\mu_B^2+\kappa^2}}.
\end{equation}
Substituting $\mu_B= \kappa \sinh \pi b s$, we get\footnote{The normalization is somewhat arbitrary. We have chosen it to find the JT bulk defect one-point function in the JT limit.}
\begin{equation}
\label{b11}
\left\langle \mathcal{T}_\alpha\right\rangle_{\ell,\NS,\eta=+1} = \frac{1}{b} \int_{0}^{+\infty} ds e^{-\frac{\kappa^2}{2} \ell(\cosh2\pi b s - 1)} \cosh \frac{\pi \theta}{b}s.
\end{equation}
The case $\eta=-1$ is entirely analogous. We have the discontinuity relation:
\begin{equation}
\text{Disc}[\partial_{\mu_B} \cosh  \frac{\theta}{b^2} \text{arcsinh}\frac{\mu_B}{\kappa}] = \theta \frac{\sin \frac{\pi \theta}{2b^2}\, \cosh \frac{\theta}{b^2} \text{arccosh} \frac{\mu_B}{\kappa} }{b^2\sqrt{\mu_B^2-\kappa^2}}.
\end{equation}
Setting similarly $\mu_B = \kappa \cosh \pi  b s$, we write finally:
\begin{equation}
\label{b12}
\left\langle \mathcal{T}_\alpha\right\rangle_{\ell,\NS,\eta=-1} = \frac{1}{b} \int_{0}^{+\infty} ds e^{- \frac{\kappa^2}{2} \ell(\cosh2\pi b s +1)} \cosh \frac{\pi \theta}{b}s.
\end{equation}
Once again, following the contour deformation argument, we are effectively swapping $\eta = \pm1$.
\\~\\
Both of these integrals \eqref{b11} and \eqref{b12} can be readily evaluated explicitly yielding:
\begin{equation}
\label{bonep}
\left\langle \mathcal{T}_\alpha\right\rangle_{\ell,\NS,\eta} = e^{\eta \frac{\kappa^2}{2} \ell} \frac{1}{2\pi b^2}K_{\frac{\theta}{2b^2}}(\kappa^2 \ell/2).
\end{equation}

The bulk one-point functions of the Ramond sector operators \eqref{R:1} and \eqref{R:2} can be transformed to the length basis in the same way. We do not mark these boundaries further since the bulk Ramond operator creates a branch cut that necessarily already marks the boundary. The calculation is identical to the one for the Ramond partition functions in \ref{s:pf}, and we end up with:
\begin{align}
\label{RR1}
\left\langle \Theta_\alpha\right\rangle_{\ell,\R,\eta=+1} &= \frac{1}{b} \int_{0}^{+\infty} ds e^{-\frac{\kappa^2}{2} \ell(\cosh2\pi b s -1)} \cosh \pi b s \cosh \frac{\pi \theta}{b}s, \\
\label{RR2}
\left\langle \Theta_\alpha\right\rangle_{\ell,\R,\eta=-1} &= \frac{1}{b} \int_{0}^{+\infty} ds e^{-\frac{\kappa^2}{2} \ell(\cosh2\pi b s +1)} \sinh \pi b s \sinh \frac{\pi \theta}{b}s.
\end{align}
The integrals are done as:
\begin{align}
\label{bonep1}
\left\langle \Theta_\alpha\right\rangle_{\ell,\R,\eta=+1} &= \frac{e^{-\frac{\kappa^2}{2} \ell}}{2\pi b^2} \left( K_{\frac{1}{2}+ \frac{\theta}{2b^2}} (\kappa^2 \ell/2) + K_{\frac{1}{2}- \frac{\theta}{2b^2}} (\kappa^2 \ell/2) \right), \\
\label{bonep2}
\left\langle \Theta_\alpha\right\rangle_{\ell,\R,\eta=-1} &=  \frac{e^{+\frac{\kappa^2}{2} \ell}}{2\pi b^2} \left( K_{\frac{1}{2}+ \frac{\theta}{2b^2}} (\kappa^2 \ell/2) - K_{\frac{1}{2}- \frac{\theta}{2b^2}} (\kappa^2 \ell/2) \right).
\end{align}
For the special case $\theta=0$, the bulk insertion corresponds to the gravitationally dressed matter Ramond ground state with $\Delta_{\alpha_M} = \frac{c_M}{24}$, corresponding to the ``middle'' $\alpha_{\frac{p}{2},\frac{q}{2}}$ degenerate label in case of the $(p,q)$ superminimal models with both $p$ and $q$ even.

We can get back to the Ramond disk partition function from here by setting $\theta=1$. This is \emph{not} true in the bosonic model, or in the NS sector, where the bulk one-point function has one additional marking compared to the partition function. We make some comments on this in appendix \ref{app:rampf}.

Setting $\theta = 2iPb$, these bulk insertions are macroscopic holes with label $P$ from the (super)Liouville geometry perspective. These specific bulk insertions are required when gluing disks together. We will write some formulas in the concluding section \ref{s:conc}.

\subsubsection*{Neveu-Schwarz partition function}
In the bosonic Liouville gravity, it was illustrated in \cite{Mertens:2020hbs} that the partition function $Z$ can be found from the bulk one-point function by letting $\theta \to 1$, and simultaneously removing a single marking.
Starting with \eqref{bonep}, we can remove a single marking by dividing by $\ell$. Letting $\theta \to 1$ defines the partition function. We get
\beq
\label{NSpf}
Z_{\NS, \eta}(\ell) = e^{\eta \frac{\kappa^2}{2} \ell} \frac{1}{2\pi b^2 \ell }K_{\frac{1}{2b^2}}(\kappa^2 \ell/2) = b \kappa^2  \int_{0}^{+\infty} ds e^{- \frac{\kappa^2}{2} \ell(\cosh2\pi b s - \eta)} \sinh 2 \pi b s \sinh \frac{\pi}{b}s,
\eeq
with the spectral densities:
\begin{align}
\label{NSspdens}
\rho_{\NS,\eta=+1}(E) &= \sinh \frac{1}{b^2}\text{arcsinh} \frac{\sqrt{E}}{\kappa}, \qquad \eta=+1, \\
\rho_{\NS,\eta=+1}(E) &= \sinh \frac{1}{b^2}\text{arccosh} \frac{\sqrt{E}}{\kappa}, \qquad \eta=-1.
\end{align}
For the particular case of the $(2,4\mathbf{k}+2$) minimal superstring, the series expansions of these quantities in powers of $\sqrt{E}$ and $\sqrt{E-\kappa^2}$ respectively, truncate at $E^{\mathbf{k}+1/2}$ and $(E-\kappa^2)^{\mathbf{k}+1/2}$ respectively. This should be compared to the similar statement for the Ramond partition functions discussed above that show a similar truncation for the $(2,4\mathbf{k}$) minimal superstring.

\subsection{Marking operators}
\label{s:mark}
Now we move on to inserting boundary vertex operators at the boundary of the disk. The simplest such operator is to pick the matter operator to be the identity $\mathbf{1}_M$, and to gravitationally dress this with the super-Liouville (and ghost) pieces. This leads to the following $\beta_M=0 \,(\beta=b)$ boundary operator and its superpartner:
\begin{equation}
B^b(x) = e^{\frac{b}{2}\phi(x)}, \qquad \Lambda^b(x) = \frac{b}{2}(\psi(x)+\eta \bar{\psi}(x)) e^{\frac{b}{2}\phi(x)},
\end{equation}
in terms of the Liouville field $\phi$ and the fermions $\psi$ and $\bar{\psi}$ at the boundary coordinate $x$. We refer to appendix \ref{s:detliou} for the details. \\
Here we show, in analogy with the bosonic case \cite{Mertens:2020hbs}, that in the fixed-length basis these boundary insertions are trivial insertions. We can set $\beta = b$ in the Liouville boundary two-point functions (reflection coefficients) \eqref{fukhoso} and get (up to a prefactor that is chosen with hindsight):\footnote{Using the shift identities \eqref{susyshift}, one can prove the following needed equalities:
\begin{align}
S_{\NS}(b\pm ix) = \frac{\cosh\frac{\pi b x}{2}}{\cosh \frac{\pi x}{2b}}, \qquad S_{\R}(b\pm ix) = \frac{\sinh\frac{\pi b x}{2}}{\sinh \frac{\pi x}{2b}}. 
\end{align}
}
\begin{alignat}{2}
\label{2mark}
d(b|s_+s_+') &&= \frac{1}{\kappa}\frac{\cosh\frac{\pi}{b}s_1 + \cosh \frac{\pi}{b}s_2}{\cosh \pi b s_1 + \cosh \pi  b s_2}, \qquad d'(b|s_+s_+') &= \frac{1}{\kappa}\frac{\cosh\frac{\pi}{b}s_1 - \cosh \frac{\pi}{b}s_2}{\cosh \pi b s_1 - \cosh \pi  b s_2}, \nonumber \\
d(b|s_-s_-') &&= \frac{1}{\kappa}\frac{\sinh\frac{\pi}{b}s_1 + \sinh \frac{\pi}{b}s_2}{\sinh \pi b s_1 + \sinh \pi  b s_2}, \qquad d'(b|s_-s_-') &= \frac{1}{\kappa}\frac{\sinh\frac{\pi}{b}s_1 - \sinh \frac{\pi}{b}s_2}{\sinh \pi b s_1 - \sinh \pi  b s_2},
\end{alignat}
where the notation $d(\beta|s_\eta s_\eta')$ and $d'(\beta|s_\eta s_\eta')$ denote the operator with $\beta =b$ resp. superpartner boundary two-point function on a boundary segmented in two pieces labeled by $s_\eta$ and $s_\eta'$. This notation is further explained in appendix \ref{s:liou}. In appendix \ref{s:markapp} we perform a simple consistency check on these specific relations.
\\~\\
Let us now transform these to the fixed-length basis. We will use this paragraph to argue that to find the fixed length operator insertion of interest, we should take a linear combination of both $d(\beta|s_+s_+')$ and $d'(\beta|s_+s_+')$ amplitudes. Contour deforming both $s$ and $s'$-integrals in the same fashion as in subsection \ref{s:pf}, one has the relation:
\begin{align}
\text{sDisc} &d(b|s_+s_+') = \text{sDisc} d'(b|s_+s_+') = 2\sin \frac{\pi}{2b^2} \nonumber \\
&\times \left[\frac{\sinh \frac{1}{b^2}\text{arcsinh} \frac{\mu_1}{\kappa} + \sinh \frac{1}{b^2}\text{arcsinh} \frac{\mu_2}{\kappa}}{\mu_1+\mu_2}+ \frac{\sinh \frac{1}{b^2}\text{arcsinh}\frac{\mu_1}{\kappa} - \sinh \frac{1}{b^2}\text{arcsinh} \frac{\mu_2}{\kappa}}{\mu_1-\mu_2}\right],
\end{align}
where $\mu_i$ is shorthand notation for $\mu_{Bi}$. Since these are equal, there is no contribution to $d(b|s_+s_+') - d'(b|s_+s_+')$ when $s \neq s'$. However, a more careful treatment is required when $s=s'$. 
By first subtracting both terms in $d(b|s_+s_+') - d'(b|s_+s_+')$ and only in the end evaluating the discontinuity across the cut, we get
\begin{align}
\text{sDisc}\left[d(b|s_+s_+') - d'(b|s_+s_+')\right] 
&= 8\pi \cos \frac{\pi}{2b^2} \cosh \frac{1}{b^2}\text{arcsinh} \frac{\mu_1}{\kappa} \, \delta(\mu_1-\mu_2) ,
\end{align}
giving back the original partition function in the $\eta = +1$ sector \eqref{inttodo}:
\begin{equation}
\label{markedsames}
\int ds e^{-\frac{\kappa^2}{2}\ell(\cosh 2\pi b s -1)} \cosh \pi b s \cosh \frac{\pi}{b}s.
\end{equation}
The linear combination $d(b|s_+s_+') - d'(b|s_+s_+')$ can be identified as the two-point function of the linear combination of operator and superpartner as $\Lambda^b + i B^b$, defined in \eqref{defNs}. Concretely, this means that inserting the operator $(\frac{b}{2}(\psi +\eta \bar{\psi})+ i)e^{\frac{b}{2}\phi}$ has no effect after transforming to the fixed length basis.\footnote{This local boundary operator is to be identified with $M_2(x)$ \eqref{marki2} upon using that $\gamma_0^2=1$ in the two-point function.} \\
The significance of this is that these boundary operators in the fixed length basis act as identity operators and in particular their two-point function is just the partition function. Following the notation used in the bosonic minimal string, we will refer to these operators as \emph{local marking operators}. The triviality of correlators of these operators also holds for more than two marking operators. This is discussed easiest using the matrix description of the minimal superstring, as we do in section \ref{s:matrix}.
\\~\\
Analogously, for the $\eta=-1$ case, one considers
\begin{equation}
d(\beta|s_-s_-') - d'(\beta|s_-s_-') = 2 \frac{\mu_1 \sinh \frac{\pi}{b}s_2 - \mu_2 \sinh \frac{\pi}{b}s_1}{\mu_1^2-\mu_2^2},
\end{equation}
reproducing the $\eta = -1$ sector partition function \eqref{inttodo}, after transforming to the fixed length basis. The interpretation as inserting two marking operators that act as the identity in the fixed-length basis holds true for this case as well.

\subsection{Boundary two-point function}
\label{s:twop}

Here we will consider the boundary two-point function in general Liouville supergravities. This generalizes the preceding discussion on marking operators with $\beta=b$ to generic values of $\beta$. We first consider the super-Liouville sector of the boundary two-point function, since this one will carry almost all of the interesting information. The relevant equations were obtained in \cite{Fukuda:2002bv}. Transitioning to the fixed-length amplitudes proceeds by applying the integral transform \eqref{Rint}. Deforming the contour as for the partition function, we will evaluate the \textbf{s}discontinuity of these expressions. Just as in the previous subsection, it turns out one obtains well-behaved answers by pairing up the formulas for the boundary operator and its superpartner.
\\~\\
We define a Neveu-Schwarz (NS) boundary operator in the Liouville sector, combining the bosonic Liouville vertex operator with its superpartner \eqref{defNs} as:
\begin{equation}
\label{op1}
\mathcal{O}^\beta \equiv \Lambda^\beta + i B^\beta = (\frac{\beta}{2}(\psi + \eta \bar{\psi}) + i ) e^{\frac{\beta}{2}\phi},
\end{equation}
The resulting boundary two-point functions are now straightforward to compute.
\begin{itemize}
\item
For a $\eta=+1$ boundary, the super-Liouville boundary two-point function of two such operators \eqref{op1}, is:
\begin{equation}
\left\langle \mathcal{O}^\beta  \mathcal{O}^\beta  \right\rangle_{++} \equiv D^{\beta}_{s,s'} = d'(\beta|s_+s_+') - d(\beta|s_+s_+').
\end{equation} 
Define the double discontinuity with all plus-signs as:
\begin{align}
\label{Discdef}
\text{sDisc} D^{\beta}_{s,s'} \equiv D^{\beta}_{s+\frac{i}{2b},s'+\frac{i}{2b}} + D^{\beta}_{s-\frac{i}{2b},s'+\frac{i}{2b}} + D^{\beta}_{s+\frac{i}{2b},s'-\frac{i}{2b}} + D^{\beta}_{s-\frac{i}{2b},s'-\frac{i}{2b}},
\end{align}
as we are instructed to do in the $\mu_B$-contour deformation of an amplitude with only NS-sector operator insertions on a Ramond boundary, as in section \ref{s:pf}. Using the supersymmetric shift relations \eqref{susyshift}, one explicitly evaluates this to:
\begin{align}
\label{susdeform}
\text{sDisc } D^{\beta}_{s,s'} = &\left[-16 \cos \frac{\pi}{\beta}\left(\beta-\frac{1}{2b}\right)  \sin \frac{\pi \beta}{b}\right] \nonumber \\
&\hspace{-1cm}\times \cosh \frac{\pi s}{b}\cosh \frac{\pi s'}{b} \left( d'(\beta + \frac{1}{b}|s_-s_-') + d(\beta+ \frac{1}{b}|s_-s_-') \right).
\end{align}
We notice two things. Firstly, the r.h.s. contains a similar linear combination of two-point functions, but with a shift in $\beta \to \beta +1/b$. Secondly, one has in effect changed fermionic boundary conditions from $\eta=+1$ to $\eta=-1$ during this process.

\item
Secondly, we consider a $\eta=-1$ boundary, for which the boundary-two point function $\left\langle \mathcal{O}^\beta  \mathcal{O}^\beta  \right\rangle_{--}$ is given by the expression $D^{\beta}_{s,s'} \equiv d'(\beta|s_-s_-') - d(\beta|s_-s_-')$. Using the same definition \eqref{Discdef},
we obtain:
\begin{align}
\text{sDisc } D^{\beta}_{s,s'} = &\left[-16 \cos \frac{\pi}{\beta}\left(\beta-\frac{1}{2b}\right)  \sin \frac{\pi \beta}{b}\right] \nonumber \\
&\hspace{-1cm}\times \sinh \frac{\pi s}{b} \sinh \frac{\pi s'}{b} \left( d'(\beta + \frac{1}{b}|u_+u_+') + d(\beta+ \frac{1}{b}|u_+u_+') \right),
\end{align}
with a similar qualitative interpretation. Notice the appearance of two sinh measure factors in the second line, to be contrasted with the cosh appearing in \eqref{susdeform}.
\end{itemize}

Next we combine this super-Liouville sector with the matter sector, the superghosts and the overall normalization factor (the legpole factor) of the vertex operator. The full NS boundary tachyon vertex operators are defined as:
\begin{align}
\label{NSope}
\mathcal{B}^{+}_{\beta_M} &=  (\pi \mu \gamma(bQ/2))^{\frac{2\beta-Q}{4b}}\Gamma(\frac{b}{2}(Q-2\beta))\,\, \left(c e^{-\varphi}\right) \, \left[e^{\frac{\beta}{2} \phi} e^{\beta_M \chi} + (\text{superpartner})\right], \\
\mathcal{B}^{-}_{\beta_M} &= \left(\pi \mu \gamma\left(bQ/2\right)\right)^{\frac{2\beta-Q}{4b}}\Gamma(\frac{1}{2b}(Q-2\beta))\,\,\left(c e^{-\varphi}\right) \, \left[ e^{\frac{\beta}{2} \phi} e^{(-q-\beta_M) \chi} + (\text{superpartner})\right].
\end{align}
This expression includes, in order, the normalization and legpole factor, the superghost vertex operator, the Liouville vertex operator, and the matter vertex operator. The latter has been conveniently parametrized through timelike super-Liouville as we pointed out in footnote \ref{fn:time}. As written here, we need to include the superpartner of this expression as well, and take the particular linear combination as in \eqref{op1}. We refrain from writing out this expression in detail.
\\~\\
Including the superghost and matter contributions and imposing the Virasoro constraints in the full theory has two additional effects compared to the pure Liouville SCFT correlators described above.

Firstly, normalizing the matter boundary two-point function as $1$, the main effect of the superghost and matter boundary two-point function is to cancel the dependence on the worldsheet coordinates $x$ in the final result.

Secondly, just as in the bosonic case, the Liouville piece of the boundary two-point function diverges as $\delta(0)$ from the Liouville zero-mode, which is cancelled by the volume of the (super)conformal Killing group of the two-punctured disk. For two NS boundary operators, the ratio is finite and equals $2(Q-2\beta)$ precisely like in the bosonic case.\footnote{For the boundary Ramond two-point function in string theory, the ratio gives a factor independent of the vertex operator labels. It would be interesting to explicitly derive this by integrating the $\mathcal{N}=1$ Liouville boundary three-point function $\left\langle B_b \Theta^{\epsilon\alpha_1} \Theta^{\epsilon\alpha_2}\right\rangle$ w.r.t. $\mu_B$.} \\
The prefactors in the final result are of three kinds: the explicit legpole factors in the vertex operators \eqref{NSope}, the prefactors coming from the contour rotation argument in \eqref{susdeform} and its cousins, and finally the ratio of $\Gamma_{\NS}$ factors present in the super-Liouville two-point functions of \eqref{fukhoso}. It is not difficult to show that these conspire to the simple result of $1/S_{\NS}(2\beta_M)$ for the NS-operator insertion, and $1/S_{\R}(2\beta_M)$ for the R-operator insertion.\footnote{One way to derive this, is to use the shift relations \eqref{gammashift} twice for the numerator $\Gamma_{\NS / \R}(2\beta-Q)$, then write it in terms of $S_{\NS / \R}$ and apply its shift relation twice again \eqref{susyshift}. Use is made throughout of the gamma-function reflection formula in the form
\begin{equation}
\Gamma\left(\frac{1}{2}+z\right)\Gamma\left(\frac{1}{2}-z\right) =  \frac{\pi}{\cos \pi z}.
\end{equation}
}

Combining all of the ingredients, we finally arrive at the fixed-length expression for $\eta=+1$:
\begin{align}
\label{final1}
 &\left\langle \mathcal{B}^{+}_{\beta_M} \hspace{0.1cm}\mathcal{B}^{-}_{\beta_M}\right\rangle_{\ell_1,\ell_2,\eta=+1} = \int ds_1ds_2 \rho_{\R,\eta=+1}(s_1) \rho_{\R,\eta=+1}(s_2) e^{-\kappa^2 \ell_1 \sinh^2 \pi b s_1} e^{-\kappa^2 \ell_2 \sinh^2 \pi b s_2}  \\
&\times\left[ \frac{S_{\R}(\beta_M \pm i( s_1 +  s_2))S_{\NS}(\beta_M \pm i( s_1 -  s_2))}{S_{\NS}(2\beta_M)} + \frac{S_{\NS}(\beta_M \pm i( s_1 +  s_2))S_{\R}(\beta_M \pm i(s_1 -  s_2))}{S_{\NS}(2\beta_M)}\right], \nonumber
\end{align}
with $\rho_{\R,\eta=+1}(s) = \cosh \frac{\pi}{b}s \cosh \pi b s$. For $\eta=-1$, we get the slightly simpler:
\begin{align}
\label{final2}
 \left\langle \mathcal{B}^{+}_{\beta_M} \hspace{0.1cm}\mathcal{B}^{-}_{\beta_M}\right\rangle_{\ell_1,\ell_2,\eta=-1} &= \int ds_1ds_2 \rho_{\R,\eta=-1}(s_1) \rho_{\R,\eta=-1}(s_2) e^{- \kappa^2 \ell_1 \cosh^2 \pi b s_1} e^{-\kappa^2 \ell_2 \cosh^2 \pi b s_2} \hspace{4cm} \\
&\times\left[ \frac{S_{\NS}(\beta_M \pm i s_1 \pm i s_2)}{S_{\NS}(2\beta_M)} + \frac{S_{\R}(\beta_M \pm i s_1 \pm i s_2)}{S_{\NS}(2\beta_M)}\right], \nonumber
\end{align}
with $\rho_{\R,\eta=-1}(s) = \sinh \frac{\pi}{b}s \sinh \pi b s$. These are our final results for the fixed-length boundary two-point function in $\mathcal{N}=1$ Liouville supergravity.

\subsection{Ramond boundary operators}
In an analogous treatment, one can consider Ramond (R) boundary operators as well. Natural operators in the fixed-length basis sum over both chiralities of \eqref{defR}:
\begin{equation}
\label{op2}
\Sigma^\beta \equiv \Theta^{+\beta} + \Theta^{-\beta} =  (\sigma^{\epsilon=+1}+\sigma^{\epsilon=-1}) e^{\frac{\beta}{2} \phi},
\end{equation}
with $\sigma^{\epsilon}$ the boundary spin field, change the fermionic boundary condition $\eta$ of the brane segment in the fixed-length basis.
For our main story, we will not need the Ramond sector boundary operators, so we present the analogous results for their boundary two-point functions in appendix \ref{s:ramond}. We also compare these operator insertions to those in super-Liouville theory between a pair of ZZ-branes where branch cuts from the spin fields signal a change in boundary condition $\eta$.

\subsection{JT supergravity limit}
Here we discuss how a JT (super)gravity limit is achieved, starting with the above determined fixed-length amplitudes in Liouville supergravity. We will make contact with known amplitudes in these limiting JT models. In all cases, the JT limit is a double-scaling limit where we take the Liouville parameter $b\to0$ and let the boundary length segments $\ell \to +\infty$ in a suitable way to be specified below. These results generalize the statements made in \cite{Saad:2019lba,Mertens:2020hbs} about the JT limit of the minimal string, to the supersymmetric case. \\
We retake all of our fixed-length amplitudes one by one.
\begin{center}
\textbf{Partition functions}
\end{center}
The JT limit of the partition functions \eqref{inttodo} and \eqref{inttodo2} is readily evaluated, by letting $s\to 0$ and $\ell \to +\infty $ as
\begin{equation}
s=2bk, \qquad \ell = \frac{\ell_{\rm JT}}{4\pi^2 \kappa^2 b^4},
\end{equation} 
in terms of a finite length $\ell_{\rm JT}$ and finite momentum $k$. We obtain
\begin{align}
\label{SJTpf}
Z_{\R,\eta=+1}(\ell)  &\to \int_{0}^{+\infty} ds e^{-\ell_{\rm JT} k^2} \cosh 2\pi k, \\
Z_{\R,\eta=-1}(\ell)   &\to \int_{0}^{+\infty} ds e^{-\ell_{\rm JT} k^2} k\sinh 2\pi k,
\end{align}
which is for $\eta=+1$ indeed the $\mathcal{N}=1$ supersymmetric Schwarzian partition function \cite{Stanford:2017thb,Mertens:2017mtv}. In the JT limit, the NS-boundary partition functions \eqref{NSpf} for both $\eta = \pm1$ produce the bosonic JT density of states $\rho(k) = k \sinh 2 \pi k$.
\begin{center}
\textbf{Bulk one-point function}
\end{center}
\begin{align}
\left\langle \mathcal{T}_\alpha\right\rangle_{\ell,\NS,\eta}  &\to e^{\kappa^2 \ell \eta}\int_{0}^{+\infty} dk e^{-\ell_{\rm JT} k^2} \cosh 2\pi \theta k.
\end{align}
This corresponds to a conical defect of deficit $2\pi(1-\theta)$. Analogously,
\begin{align}
\left\langle \Theta_\alpha\right\rangle_{\ell,\R,\eta=+1} &\to  e^{-\kappa^2 \ell} \int_{0}^{+\infty} dk e^{-\ell_{\rm JT} k^2} \cosh 2\pi \theta k, \\
\left\langle \Theta_\alpha\right\rangle_{\ell,\R,\eta=-1} &\to  2\pi b^2e^{+\kappa^2 \ell} \int_{0}^{+\infty} dk e^{-\ell_{\rm JT} k^2} k\sinh 2\pi \theta k.
\end{align}

\begin{center}
\textbf{Boundary two-point function}
\end{center}
Finally we take the Schwarzian limit of the boundary two-point function expressions. Setting $\beta_M =  2b h$ and $s = 2bk_1, s' = 2bk_2$, and using \eqref{limits}, the building blocks \eqref{fukhoso} reduce to:
\begin{align}
\label{lfirst}
d(\beta|s_-, s'_{-}) &\to \Gamma(h \pm i ( k_1 - k_2) ) \Gamma(\frac{1}{2}+ h \pm i (k_1 + k_2)) , \\
\label{lsecond}
d'(\beta|s_-, s'_{-}) &\to -\Gamma(h \pm i ( k_1 + k_2) ) \Gamma(\frac{1}{2}+ h \pm i (k_1 - k_2)), \\
\label{lthird}
d(\beta|s_+, s'_{+}) &\to \Gamma(h \pm i ( k_1 + k_2) ) \Gamma(h \pm i (k_1 - k_2)), \\
\label{lfourth}
d'(\beta|s_+, s'_{+}) &\to \Gamma(\frac{1}{2}+h \pm i ( k_1 - k_2) ) \Gamma(\frac{1}{2}+ h \pm i (k_1 + k_2)),
\end{align}
and the denominators in \eqref{final1} and \eqref{final2} have the limit
\begin{align}
S_{\NS}(2\beta_M) \to \Gamma(2h).
\end{align}
Use is made of formulas contained in appendix \ref{app:spec}. The results can now be compared with the $\mathcal{N}=1$ super-Schwarzian bilocal correlators \eqref{super2pt}. The NS boundary operator insertion with boundary label $\eta=+1$ \eqref{final1} precisely limits to this super-JT correlator. The boundary with $\eta=-1$ is less interesting for our purposes and limits to a linear combination of bosonic JT results.\footnote{In \cite{Mertens:2017mtv}, the super JT bilocal correlators were constructed by considering $\mathcal{N}=1$ super-Liouville theory on a cylinder in the $\widetilde{\text{NS}}$ sector between a pair of ZZ identity branes \cite{Zamolodchikov:2001ah}. This means one uses periodic (R) boundary conditions around the circumference of the cylinder, identifying this with the $\eta = +1$ boundary condition on the Liouville supergravity disk studied in this work. From the ZZ-ZZ brane perspective, choosing anti-periodic (NS) boundary conditions around the cylinder circumference, leads to a removal of all fermions in the JT limit and one retrieves only the bosonic Schwarzian system. This is then identified with the $\eta=-1$ fermionic boundary condition on the Liouville supergravity disk. This corresponds to \eqref{lthird} and \eqref{lfourth}. The fourth line can be viewed as the reparametrized operator $1/\tau^{2h+1}$ inserted into the bosonic Schwarzian system, indeed giving just the $h \to h +1/2$ result. We will come back to this interpretation in the ZZ-ZZ system in appendix \ref{s:aside}.}

\section{Minimal superstring and matrix interpretation}
\label{s:matrix}
\begin{figure}[t!]
\centering
\begin{tikzpicture}[scale=0.7]
\draw[fill=blue!50!white,opacity=0.7] (0,0) ellipse (2 and 1);
\draw[fill] (-2,0) circle (0.08); 
\node at (-3,0) {$M_2(x_1)$};
\node at (-1.4,-1.1) {$\mu_{B3}$};
\draw[fill] (1,0.86) circle (0.08);
\node at (1.4,1.3) {$M_2(x_2)$};
\draw[fill] (1,-0.86) circle (0.08);
\node at (1.5,-1.35) {$M_2(x_3)$};
\node at (-1,1.2) {$\mu_{B1}$};
\node at (2.6,0) {$\mu_{B2}$};
\draw[-latex] (3.5,0) -- (4.5,0);
\draw[fill=blue!50!white,opacity=0.7] (7.5,0) ellipse (2 and 1);
\node at (10.5,1) {\small $\ell_1 + \ell_2 + \ell_3$};
\end{tikzpicture}
\caption{FZZT brane segments between $n$ marking operators $M_2(x_i)$ \eqref{marki2} leads upon transforming to the fixed length basis with length $\ell \equiv \sum_j \ell_j$. In the figure we show an example with $n=3$.}
\label{markingfour}
\end{figure}
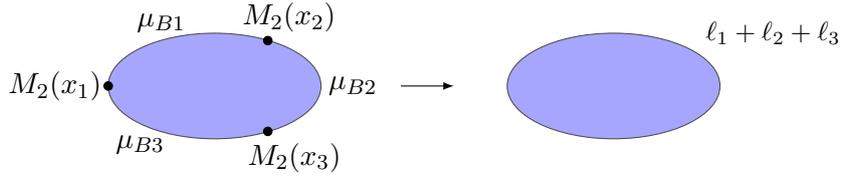

The above results hold for generic choice of matter sector. From here on, we specify to the $(p,q)$ super-minimal model briefly reviewed in section \ref{s:Lsugra}. It is known that a two-matrix description is possible for the generic choice of $p$ and $q$. Specifying further to $p=2$, we have a single-matrix description, and it is possible to describe some of the properties derived above more cleanly from the perspective of this matrix model. We first make some observations on the partition functions and marking operators, and then in subsection \ref{s:mampl} we proceed with our main goal of constructing the minimal superstring boundary two-point functions using the matrix model perspective as a guide towards writing down the analogous proposal for higher genus corrections.
\\~\\
For any random matrix integral, the resolvent is defined as $R(x) = \text{Tr}\frac{1}{x-H}$ and equals the singly marked disk amplitude.
For the $\mathcal{N}=1$ case, and in the Ramond sector, the random matrix in question is a ``supercharge'' matrix $Q$, with the Hamiltonian matrix related to it as $H=Q^2$. We define the Ramond sector resolvent by 
\begin{equation}
\mathcal{R}^{\R}(x) \equiv \frac{1}{2} \text{Tr} \frac{x}{x^2-H} = \text{Tr}\frac{1}{x-Q} + (\text{regular}),
\end{equation}
where the second way of writing it shows that it is also the usual resolvent of a random matrix model with matrix $Q$.\footnote{The regular terms do not have poles or branch cuts in the physical region of interest and are immaterial for our purposes.}
With this definition, the R-sector resolvent produces the ``charge'' spectral density by an \textbf{s}discontinuity:
\begin{equation}
\text{\textbf{s}Disc}[\mathcal{R}^{\R}(ix) ]=  \mathcal{R}^{\R}(ix+\epsilon) + \mathcal{R}^{\R}(-ix + \epsilon) = -2\pi x \rho_{\R}(-x^2).
\end{equation}
In terms of $x = \mu_B$, transforming in the R-sector to the fixed length amplitude through \eqref{Rint}, we get indeed:
\begin{equation}
\int_{-i\infty}^{+i\infty} dx e^{-x^2 \ell}  \frac{1}{2} \text{Tr} \frac{x}{x^2-H} = \int_{-i\infty}^{+i\infty} dx e^{-x^2 \ell}\text{Tr}\frac{1}{x-Q} = e^{- H \ell}.
\end{equation}
For the $(2,4\mathbf{k})$ one-matrix superminimal models with $\mathbf{k} \in \mathbb{N}$, we have an analogous story as for the bosonic $(2,2\mathfrak{m}-1)$ models. Let us be more explicit for the case $\eta=+1$, related to JT supergravity as found in \eqref{SJTpf}. The resolvent in this case is determined in terms of the uniformizing coordinate $s$ as:
\begin{equation}
\mathcal{R}^{\R}(x) = \cosh \frac{\pi}{b}s, \qquad x = \kappa \cosh \pi b s.
\end{equation}
The \textbf{s}discontinuity of the resolvent is twice its value along either side of the cut and we find for the resolvent for the matrix $H$:
\begin{equation}
R(-x \pm i \epsilon) = \frac{\mathcal{R}^{\R}(\pm i \sqrt{x} + \epsilon)}{\pm i \sqrt{x} + \epsilon} = \frac{\cosh \frac{1}{b^2}\text{arcsinh} \sqrt{x}}{i \sqrt{x}}.
\end{equation}
In the SJT limit, we set $x = 4\pi^2 b^4 x_{\rm SJT}$ as $b\to 0$, and get:
\begin{equation}
R(-x \pm i \epsilon) \sim \frac{\cosh2\pi \sqrt{x_{\rm SJT}}}{\sqrt{x_{\rm SJT}}},
\end{equation}
up to a prefactor, reproducing the super-JT spectral curve if we set $x_{\rm SJT} = -z_{\rm SJT}^2$.
\\~\\
We can write the marking two-point function, composed of the difference of the two equations in \eqref{2mark}, for either case $\eta = \pm 1$ in the following ways:
\begin{align}
\frac{\mathcal{R}^{\R}(x_{1})+\mathcal{R}^{\R}(x_{2})}{x_{1}+x_{2}} - \frac{\mathcal{R}^{\R}(x_{1}) - \mathcal{R}^{\R}(x_{2})}{x_{1} - x_{2}} 
= \text{Tr} \frac{x_{1}}{x_{1}^2-H}\, \frac{x_{2}}{x_{2}^2-H},
\end{align}
illustrating that this inverse Laplace transforms to 
\begin{equation}
e^{-(\ell_1+\ell_2)H}.
\end{equation}
This is a simple proof that marking operators act as the identity operator in the fixed-length basis.\footnote{Notice that it is only by combining $e^{\frac{b}{2}\phi}$ and $\psi e^{\frac{b}{2}\phi}$, that we get a function without any branch cuts (only a function of $x^2$).}
\\~\\
Multiple local marking operators are then readily accommodated by the expression:
\begin{equation}
\text{Tr} \frac{x_1}{x_1^2-H} \frac{x_2}{x_2^2-H} \hdots \frac{x_n}{x_n^2-H},
\end{equation}
yielding a boundary with $n$ different cosmological constants $x_i$ between all marking operators $M_2(x)$, as illustrated in figure \ref{markingfour}.
\\~\\
For a NS-sector brane boundary, the treatment is similar to the bosonic minimal string, as we discussed in \cite{Mertens:2020hbs}. Defining
\begin{equation}
\mathcal{R}^{\NS}(x) \equiv \text{Tr} \frac{1}{x^2-H},
\end{equation}
we transform to fixed length using \eqref{NSint}
\begin{equation}
\int_{-i\infty}^{+i\infty} dx^2 e^{-x^2 \ell}\text{Tr} \frac{1}{x^2-H} = e^{- H \ell},
\end{equation}
where now the spectral density is obtained by a genuine discontinuity:
\begin{equation}
\text{Disc}\left[\mathcal{R}^{\NS}(ix)\right] = \mathcal{R}^{\NS}(ix+\epsilon) - \mathcal{R}^{\NS}(-ix + \epsilon) = -2\pi i\rho_{\NS}(-x^2).
\end{equation}
This NS sector is related to the $(2,4\mathbf{k}+2)$ (one-matrix) superminimal models.\footnote{These $(2,4\mathbf{k}+2)$ models are not strictly speaking (super)minimal models, since they contain an infinite number of super-Virasoro primaries. The combined background is argued to be a RR-background, and not describable in terms of Liouville coupled to a matter model \cite{Klebanov:2003wg}. However, the simple observation we make here suggests that the NS-sector brane boundaries are related to the corresponding matrix model. It would be interesting to understand this better.} The discontinuity of the resolvent is twice its value along either side of the cut and we get e.g. for $\eta=+1$:
\begin{equation}
R(-x \pm i \epsilon) = \mathcal{R}^{\NS}(\pm i \sqrt{x} + \epsilon) =  \pm \sinh \frac{1}{b^2}\text{arcsinh} \sqrt{x},
\end{equation}
giving the spectral density of \eqref{NSspdens}. In the JT limit, we set $x = 4\pi^2 b^4 x_{\rm JT}$ as $b\to 0$, and get:
\begin{equation}
R(-x \pm i \epsilon) = \sinh2\pi \sqrt{x_{\rm JT}},
\end{equation}
 reproducing the bosonic JT spectral curve if we set $x_{\rm SJT} = -z_{\rm SJT}^2$.

\subsection{Minimal superstring: boundary tachyon correlators}
\label{s:mampl}
Next, we consider how we would insert the tachyon boundary vertex operators \eqref{MM1} in the minimal superstring. The main goal here is to get intuition into formulating a proposal for higher genus corrections to these kinds of boundary correlators. We follow a strategy very similar to the bosonic case in section \ref{s:hightop}, where we write the Liouville amplitudes in terms of solely the matrix resolvents.
\begin{center}
$\bm{\eta=+1}$
\end{center}
For $j \in \mathbb{N}= 1,2,\hdots $, the super-Liouville boundary two-point expressions \eqref{fukhoso} simplify to:\footnote{We neglect overall normalization factors here, but restore them in the end in \eqref{mssdegcor} to match with the SJT limit of degenerate bilocal operators. }
\begin{align}
d(b+bj|s_{1+},s_{2+}) &= \frac{\cosh \frac{\pi}{b}s_1 + (-)^{j} \cosh \frac{\pi}{b}s_2}{\prod_{m=-j}^{j}(\cosh \pi b s_1 + (-)^{j-m}\cosh \pi b (s_2+imb) )}, \\
d'(b+bj|s_{1+},s_{2+}) &= \frac{\cosh \frac{\pi}{b}s_1 - (-)^{j} \cosh \frac{\pi}{b}s_2}{\prod_{m=-j}^{j}(\cosh \pi b s_1 - (-)^{j-m}\cosh \pi b (s_2+imb) )}.
\end{align}
Transforming $d(b+bj|s_{1+},s_{2+}) - d'(b+bj|s_{1+},s_{2+})$ to the fixed-length basis, we will contour transform both $\mu_B$ integrals. The integrand contains both poles and branch cuts. However, just like in the bosonic case in section \ref{s:hightop}, we first do one of the integrals by evaluating the residues. Deforming the second contour, we effectively set $s_i \to s_i \pm i \frac{1}{2b}$, and we obtain:
\begin{align}
\label{mssdegcor}
\left\langle  \mathcal{B}_{1,2j+1}\mathcal{B}_{1,2j+1} \right\rangle_{\ell_1,\ell_2} =&\int_{0}^{+\infty} ds_1 \cosh \pi  b s\,  \text{\textbf{s}Disc}[\mathcal{R}^{\R}_{\eta=+1}(s_1)] e^{-\kappa^2\ell_1 \sinh^2 \pi b s_1} \\
&\times\sum_{n=-j}^{j} \frac{j!(-1)^{j}(-1)^n e^{-\kappa^2\ell_2 \sinh^2 \pi b (s_1+inb)}}{\prod_{\stackrel{m=-j}{m \neq n}}^{j}(\sinh \pi b (s_1+inb) - (-)^{n-m}\sinh \pi b (s_1+imb) )} + (\ell_1 \leftrightarrow \ell_2), \nonumber
\end{align}
where $\rho^{\R}_{\eta=+1}(s) = \cosh \pi  b s \,\text{\textbf{s}Disc}[\mathcal{R}^{\R}_{\eta=+1}(s)] = \cosh \pi  b s \cosh \frac{\pi}{b}s$.
The numerical prefactors are chosen such that the UV limit becomes
\begin{equation}
\label{UVdeg}
\lb \mathcal{B}_{1,2j+1}\mathcal{B}_{1,2j+1} \rb \to \ell_2^{j}, \qquad \ell_2 \approx 0.
\end{equation}
The JT limit is found by letting $s=2bk$ with $b\to0$ and $k$ kept finite. One can check that one reproduces the weight $h=-j/2$ super-Schwarzian degenerate bilocal correlator. Very explicitly, for $j=1$ and $j=2$, we reproduce the $-1/2$ and $-1$ super-Schwarzian results, given by \eqref{halfSexact} and \eqref{oneSexact} respectively. This generalizes a conclusion made above to the supersymmetric case: the \emph{minimal string boundary tachyon correlators have as their JT limit precisely the degenerate operator insertions} studied throughout this work.
\\~\\
Armed with these expressions, we can now make a similar proposal for how higher genus effects are treated. Identifying again the genus zero resolvents and spectral density in the above expressions, we propose to simply replace that by the all-genus random matrix result. This in paritcular implies one only has the same kind of higher topology as in the partition function, just like we found in the bosonic theory in section \ref{s:hightop}.
 In the JT limit, the spectral curve is $\frac{\cos 2 \pi z}{z}$; its higher genus effects were studied in \cite{Stanford:2019vob}, and the non-perturbative answer was computed numerically in \cite{Johnson:2020heh}. This realizes our original goal of structurally understanding higher genus effects to degenerate super JT bilocal correlators. We postpone a deeper study of this proposal to future work.

\begin{center}
$\bm{\eta=-1}$
\end{center}
For completeness, we analyze the $\eta=-1$ boundary as well, even though it is less important for our purposes. For $j \in \mathbb{N}= 1,2,\hdots $, we have analogously:
\begin{align}
d(b+bj|s_{1-},s_{2-}) &= \frac{\sinh \frac{\pi}{b}s_1 +(-)^{j} \sinh \frac{\pi}{b}s_2}{\prod_{m=-j}^{j}(\sinh \pi b s_1 + (-)^{j-m}\sinh \pi b (s_2+imb) )},
\end{align}
\begin{align}
d'(b+bj|s_{1-},s_{2-}) &= \frac{\sinh \frac{\pi}{b}s_1 - (-)^{j} \sinh \frac{\pi}{b}s_2}{\prod_{m=-j}^{j}(\sinh \pi b s_1 - (-)^{j-m}\sinh \pi b (s_2+imb) )},
\end{align}
leading to an almost identical expression:
\begin{align}
\left\langle  \mathcal{B}_{1,2j+1}\mathcal{B}_{1,2j+1} \right\rangle_{\ell_1,\ell_2} &=\int_{0}^{+\infty} ds_1 \sinh \pi b s \, \text{\textbf{s}Disc}[\mathcal{R}^{\R}_{\eta=-1}(s_1)] e^{-\kappa^2\ell_1 \cosh^2 \pi b s_1} \\
&\times \sum_{n=-j}^{j} \frac{j!(-1)^{j}(-1)^ne^{-\kappa^2\ell_2 \cosh^2 \pi b (s_1+inb)}}{\prod_{\stackrel{m=-j}{m \neq n}}^{j}(\cosh \pi b (s_1+inb) - (-)^{n-m}\cosh \pi b (s_1+imb) )} + (\ell_1 \leftrightarrow \ell_2), \nonumber
\end{align}
where $\rho^{\R}_{\eta=-1}(s) =  \sinh \pi  b s \, \text{\textbf{s}Disc}[\mathcal{R}^{\R}_{\eta=-1}(s)] = \sinh \pi  b s \sinh \frac{\pi}{b}s$.
The JT limit can again be explicitly checked, reproducing the sum of the weight $h=-j/2$ and $h=-j/2+1/2$ \emph{bosonic} Schwarzian degenerate expressions. We checked this explicitly for $j=1$ and $j=2$, reproducing a combination of the $-1/2$ and $0$, and the $-1$ and $-1/2$ bosonic Schwarzian degenerate expressions respectively, given by \eqref{halfexact} and \eqref{oneexact}.
\\~\\
Minimal superstring Ramond boundary insertions can be treated analogously. Since we do not need these explicitly, we present the results in appendix \ref{s:ramms}.

\section{Concluding remarks}
\label{s:conc}

In this work, we have considered the JT bilocal correlators of a special integrable class of operators with weight $h \in - \mathbb{N}/2$, corresponding to degenerate Virasoro representations. 
We have shown that their structure is simpler than that of the generic $h$ correlator and have exploited this to understand more deeply some aspects of the $1/C$ and small $\tau$ perturbation series on the disk that would otherwise be hard to distill. We have also exploited its minimal string embedding to gain an understanding on how higher topology interplays with this class of correlators, illustrating that no handles crossing the bilocal line are generated in this case. \\
We generalized most of these statements to JT supergravity and its degenerate bilocal correlators.  In order to analyze the minimal superstring embedding here, we first analyzed the more general Liouville supergravities and how JT limits are obtained in those amplitudes. These results are of interest in their own right. The structure of the degenerate correlators and higher genus corrections mirrors that of the bosonic case in the end. An identical small $\tau$ expansion structure was found with no additional cancellations in the series expansion. It would be interesting to understand higher supersymmetric versions such as $\mathcal{N}=2$ and to see whether one satisfies perturbative non-renormalization theorems in those cases, simplifying the expansion. We also analyzed how Ramond boundary operators work in this case, and how they change fermionic boundary conditions. In the JT limit, this corresponds to sectors of the diagrams changing between the supersymmetric and the non-supersymmetric model.
\\~\\
We end with some remarks and points that deserve further study that we did not mention up to this point.

\subsubsection*{Massive bulk fields and HKLL}
Whereas the degenerate bilocals are largely part of a structural discussion, we should point out a genuine appearance of degenerate bilocals in a direct physical context. When constructing bulk observables and their correlators associated with mass $m$ scalar fields in JT gravity in a diff-invariant way, the Schwarzian path integral can be performed exactly in terms of multiple bilocal operator insertions \cite{Blommaert:2019hjr,Mertens:2019bvy}, interpretable as a product of HKLL kernels \cite{Hamilton:2005ju,Lowe:2008ra}. Several of these are of negative weight $1-\Delta$ where $m^2=\Delta(\Delta-1)$, and for $\Delta = 3/2, 2, 5/2 \hdots $ these are degenerate bilocals for which the results in this work have to be applied. It would be interesting to understand how they alter the deep IR bulk physics of \cite{Blommaert:2020yeo,Blommaert:2020seb}.

\subsubsection*{Beyond boundary gravitons}
We have established in appendix \ref{s:proof} that the perturbative $1/C$ disk expansion is only asymptotic for $h\notin -\mathbb{N}/2$. This means that it contains non-perturbative physics even on the disk topology. We already know the full answer \eqref{sch2pt}, but can we get a more physical picture on precisely what it contains that goes beyond boundary gravitons?

\subsubsection*{Ground ring, enhanced symmetries and higher-point degenerate correlators}
The $c<1$ model contains next to the tachyon vertex operators, also an interesting set of operators with dimension 0 and ghost number 0 \cite{Kutasov:1991qx,Seiberg:2003nm}. This is the ground ring \cite{Witten:1991zd}. For the $c=1$ model, it plays an important role in making explicit the $W_\infty$ symmetry that strongly constrains correlators \cite{Moore:1991ag}. For the $c<1$ models, similar constraints on correlators can be written down. We leave it as an interesting open question to see whether one can exploit this to make progress on fixed-length higher-point functions in the minimal string, and whether it contains non-trivial information in the JT limit on degenerate higher-point functions. See also \cite{Zamolodchikov:2003yb,Belavin:2006ex,Belavin:2009cb} for explicit calculations in this direction.

\subsubsection*{Other Liouville supergravity disk amplitudes}
In \cite{Mertens:2020hbs}, we additionally determined the bosonic fixed length amplitudes for the bulk-boundary two point function and the boundary three-point function. It would be interesting to provide these also for the supergravities studied here. However, this requires knowledge of these amplitudes in $\mathcal{N}=1$ super-Liouville CFT, which as far as we know, have not been rigourously established.

\subsubsection*{Quantum group interpretation}
The authors of \cite{Hadasz:2013bwa,Pawelkiewicz:2013wga} constructed a set of self-dual representations of the modular double of $\mathcal{U}_q (\mathfrak{osp}(1|2))$ with Casimir, and supercharge:
\begin{equation}
C = \frac{\sinh^2\pi b s}{\sin^2 \pi b^2}, \qquad Q = \pm \frac{\cosh \pi b s}{\sin \pi b^2},
\end{equation}
related as $C = Q^2 -1$. This Casimir matches structurally with the Hamiltonian in the propagation amplitudes of e.g. \eqref{inttodo} and subsequent equations (up to shifts and rescalings). Likewise, the spectral densities $\rho(s)$ are to be identified as super-Virasoro vacuum modular S-matrices and presumably Plancherel measures on these representations. It would be interesting to understand this group theoretic perspective better.
For the bosonic Liouville gravity models, we showed in \cite{Mertens:2020hbs} that the vertex functions $\frac{S_b(\beta_M \pm i s_1 \pm is_2)}{S_b(2\beta_M)}$ can be identified as the square of a 3j symbol of the quantum group associated to the modular double of $\mathcal{U}_q( \mathfrak{sl}(2,\mathbb{R}))$, where two entries are mixed parabolic matrix elements or Whittaker functions in the principal series representation, and the last entry is a discrete representation insertion. It would be interesting to perform the analogous computation with the Whittaker function of (the modular double of) $\mathcal{U}_q (\mathfrak{osp}(1|2))$ and interpret the vertex functions in the second lines of \eqref{final1} and \eqref{final2} in this same manner. 

On a related front, the authors of \cite{Berkooz:2020xne} found a solution to the double-scaled supersymmetric SYK model, related to structure found in a certain quantum group. In the bosonic case, it is known that the double-scaled SYK model and the Liouville gravities are associated to different quantum deformations of SL$(2,\mathbb{R})$, namely SU$_q(1,1)$ and SL$_q(2,\mathbb{R})$ respectively. It would be interesting to learn whether this distinction persists in the supersymmetric case.

\subsubsection*{Expansion of macroscopic loops in local operators?}
Within the context of the bosonic $c<1$ and $c=1$ models, suggestive local expansions were written down for a macroscopic loop operator in \cite{Moore:1991ir,Moore:1991ag} as:
\begin{equation}
W(\ell) = \sum_n (-)^{n+1} \frac{\ell^{n+1/2}}{n!} \sigma_n = \sum_{j=0}^{+\infty} \hat{\sigma}_j (-)^j  (2j+1) \frac{I_{j+1/2}(\kappa\ell)}{\kappa^{j+1/2}},
\end{equation}
where $\kappa \sim \sqrt{\mu}$, the bulk Liouville cosmological constant. The first expansion is a Laurent expansion, and the second expansion for the $(2,2\mathfrak{m}-1)$ minimal string corresponds to the local operators $\hat{\sigma}_j$ which are to be identified with the minimal string bulk tachyon operator insertions \cite{Moore:1991ir} for $j=0\hdots \mathfrak{m}-2$. It seems such expansions have no sensible JT limit, since e.g. the loop amplitude itself in the JT limit:
\begin{equation}
\left\langle W(\ell) \right\rangle \sim \ell^{-1} K_{\mathfrak{m}-1/2}(\ell) \quad  \to \quad \ell_{\rm JT}^{-3/2} e^{\frac{\pi^2}{\ell_{\rm JT}}},
\end{equation}
is not Laurent expandable around $\ell_{\rm JT} =0$.\footnote{For the $(2,2\mathfrak{m}-1)$, the dominant piece is  $\left\langle W(\ell)\right\rangle \sim \ell^{-\mathfrak{m}-1/2}$ and hence in the JT limit where $\mathfrak{m}\to \infty$, this becomes an essential singularity.} \\
The expansion in terms of the $\hat{\sigma}_j$ on the other hand corresponds to picking up poles in the complex energy $E$-plane when writing the double trumpet amplitude in terms of boundary wavefunctions and propagator \cite{Moore:1991ag}, as
\begin{equation}
\left\langle W(\ell_1) W(\ell_2)\right\rangle = \int_{-\infty}^{+\infty} dE \, G(E)\,  \psi_E(\ell_1) \psi_E(\ell_2).
\end{equation}
For the minimal string, it was shown that the $(2,2\mathfrak{m}-1)$ propagator $G(E) \sim 1/\cosh \pi E$ has the JT limit $\sim 1$ \cite{Mertens:2020hbs} and hence all of the complex poles shift away to infinity and are beyond the JT regime. \\
An expansion that does make sense in the JT regime is in terms of the physical intermediate set of macroscopic Liouville operators. For the bosonic case, the relation of this expansion to its JT limit was studied in \cite{Mertens:2020hbs}. We present the supersymmetric analogue below. 

\subsubsection*{Double trumpet amplitudes for the $(2,q)$ minimal superstring}
Given the recently understood relevance of higher topology \cite{Saad:2019lba,Marolf:2020xie}, including multi-boundary amplitudes (see also \cite{Blommaert:2018iqz,Blommaert:2019wfy}), it is of interest to understand the structure of the multi-boundary and higher genus amplitudes in Liouville supergravity.\footnote{Some results for bosonic Liouville gravity in this framework were studied in \cite{Mertens:2020hbs}.} Restricting to the $(p,q)$ minimal superstring, we have a matrix approach at our disposal as well. For the $(p,q)$ minimal superstring, annulus amplitudes were extensively discussed from the continuum approach in \cite{Irie:2007mp,Okuyama:2005rn}. The annulus amplitude for two $\NS$ branes with fermionic labels $\eta$ and $\eta'$, and with matter vacuum branes, is given by:
\begin{align}
Z_{\NS\NS}^{\eta,\eta'}(s,s') = \frac{\eta \eta'}{2}\int_{-\infty}^{+\infty} \frac{d \lambda}{\lambda} \frac{\cos 2\pi \frac{\lambda s}{\sqrt{pq}}\cos 2\pi \frac{\lambda s'}{\sqrt{pq}}}{\sinh \pi \lambda}\frac{\sinh \frac{p-1}{p} \pi \lambda }{\sinh \pi \lambda/p}.
\end{align}
Restricting to $p=2$, and transforming both loops to the length basis in units where $\kappa=1$, we have:
\begin{align}
\label{gluens}
Z_{\NS\NS}^{\eta,\eta'}(\ell_1, \ell_2) = \eta \eta' \int_{0}^{+\infty} d\lambda \lambda \tanh \pi \lambda K_{i\lambda} (\ell_1) K_{i\lambda} (\ell_2) ,
\end{align}
with the same gluing measure $d\lambda \lambda \tanh \pi \lambda$ as the bosonic minimal string.\footnote{This parameter $\lambda$ plays the role of the energy variable $E$ from the previous paragraph.}
\\~\\
In case of Ramond boundaries, we have the following expressions \cite{Irie:2007mp}:\footnote{This is for the case where $p$ and $q$ are both even, which is the relevant one for the $(2,q)$ models we consider here. We have chosen the matter identity brane, in the notation of \cite{Irie:2007mp} this is $k=l=k'=l'=1$. Notice that the $\eta=\eta'=+1$ expression looks divergent due to the $\lambda \approx 0$ region. Just as in the bosonic case, we expect this can be suitably regulated and interpreted \cite{Kutasov:2004fg}.}
\begin{align}
Z_{\R\R}^{\eta=+1,\eta'=+1}(s,s') &= \frac{1}{2}\int_{-\infty}^{+\infty} \frac{d \lambda}{\lambda} \frac{\cos 2\pi \frac{\lambda s}{\sqrt{pq}}\cos 2\pi \frac{\lambda s'}{\sqrt{pq}}}{\sinh \pi \lambda} \frac{\cosh \frac{p-1}{p}\pi \lambda}{\cosh \pi\lambda /p}, \\
Z_{\R\R}^{\eta=-1,\eta'=-1}(s,s') &= \frac{1}{2}\int_{-\infty}^{+\infty} \frac{d \lambda}{\lambda} \frac{\sin 2\pi \frac{\lambda s}{\sqrt{pq}}\sin 2\pi \frac{\lambda s'}{\sqrt{pq}}}{\sinh \pi \lambda} \frac{\cosh \frac{p-1}{p}\pi \lambda}{\cosh \pi\lambda /p}.
\end{align}
Transforming these to the length basis (without any additional marking in the Ramond sector as explained in appendix \ref{app:rampf}) and setting $p=2$, we get the results:
\begin{align}
\label{gluer}
Z_{\R\R}^{\eta=+1,\eta'=+1}(\ell_1,\ell_2) &= \int_{0}^{+\infty} \frac{d\lambda}{\lambda } \coth \pi \lambda (K_{1/2+i\lambda}(\ell_1) + K_{1/2-i\lambda}(\ell_1))(K_{1/2+i\lambda}(\ell_2) + K_{1/2-i\lambda}(\ell_2)), \nonumber \\
Z_{\R\R}^{\eta=-1,\eta'=-1}(\ell_1,\ell_2) &= \int_{0}^{+\infty} \frac{d\lambda}{\lambda} \coth \pi \lambda (K_{1/2+i\lambda}(\ell_1) - K_{1/2-i\lambda}(\ell_1))(K_{1/2+i\lambda}(\ell_2) - K_{1/2-i\lambda}(\ell_2)).
\end{align}
Both \eqref{gluens} and \eqref{gluer} are interpretable as summing over all intermediate labels of two bulk one-point functions \eqref{bonep} or \eqref{bonep1} and \eqref{bonep2} where $\alpha = Q/2 + i b\lambda$, with a certain gluing measure $d\lambda \rho_{\text{glue}}(\lambda)$ that can be read from these equations. Diagrammatically, we read this as
\begin{align}
Z_{\NS\NS}^{\eta,\eta'}(\ell_1,\ell_2) &= \int_0^{+\infty} d\lambda \rho^{\NS}_{\text{glue}}(\lambda) \, \quad \raisebox{-8mm}{\includegraphics[width=0.15\textwidth]{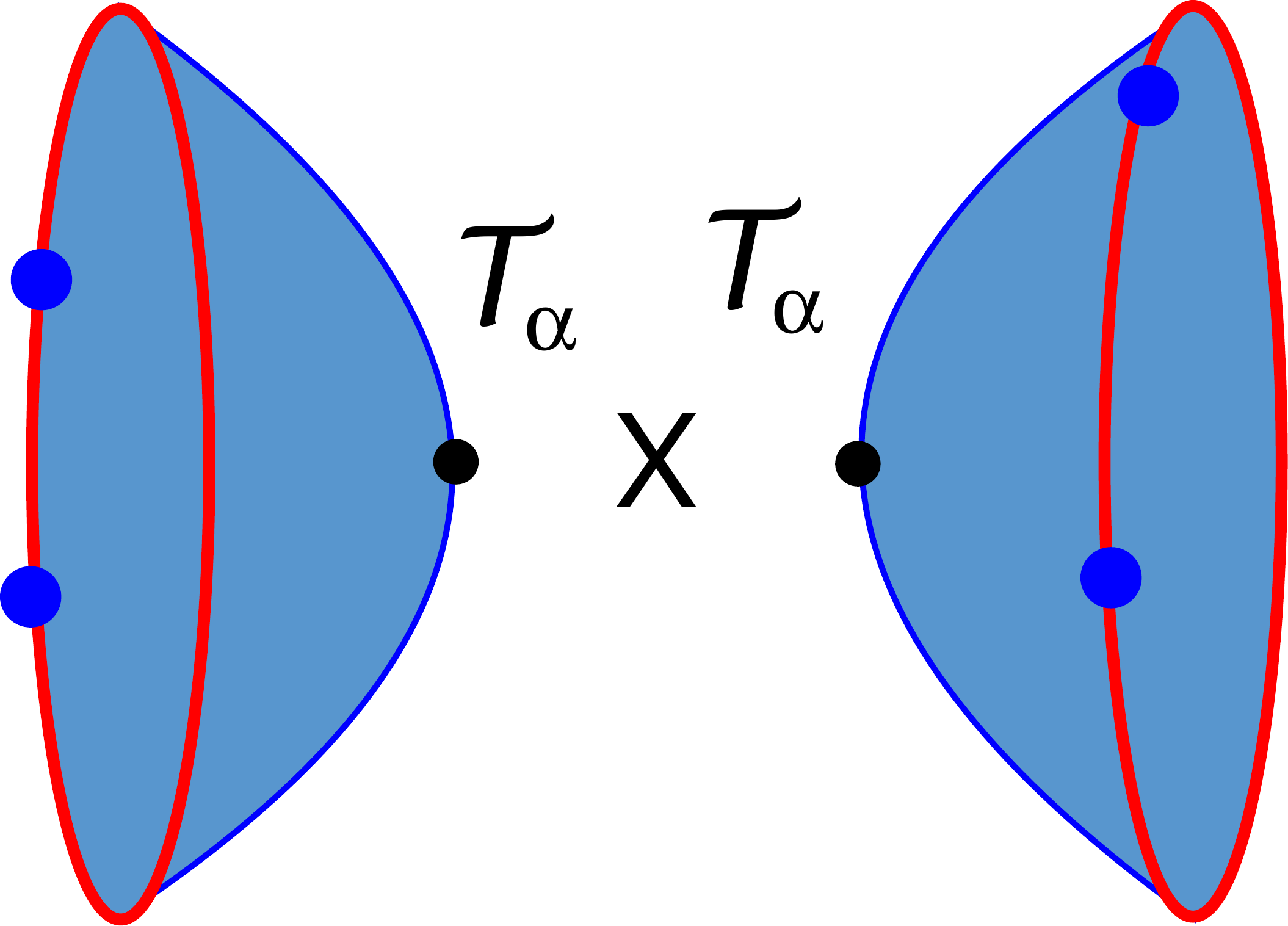}}
\quad = \quad  \raisebox{-8mm}{\includegraphics[width=0.2\textwidth]{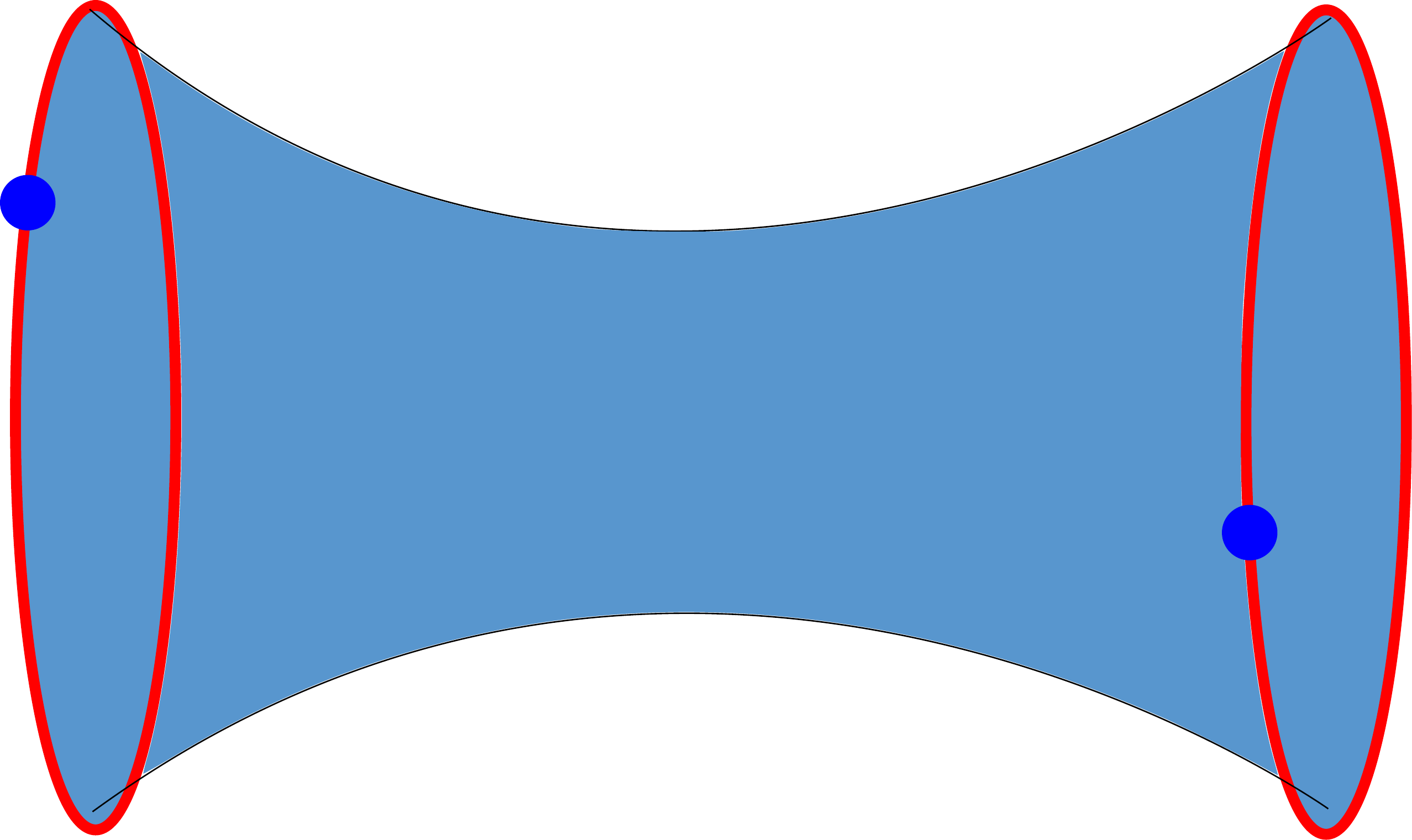}} \\
Z_{\R\R}^{\eta,\eta'}(\ell_1,\ell_2) &= \int_0^{+\infty} d\lambda \rho^{\R}_{\text{glue}}(\lambda) \, \quad \raisebox{-8mm}{\includegraphics[width=0.15\textwidth]{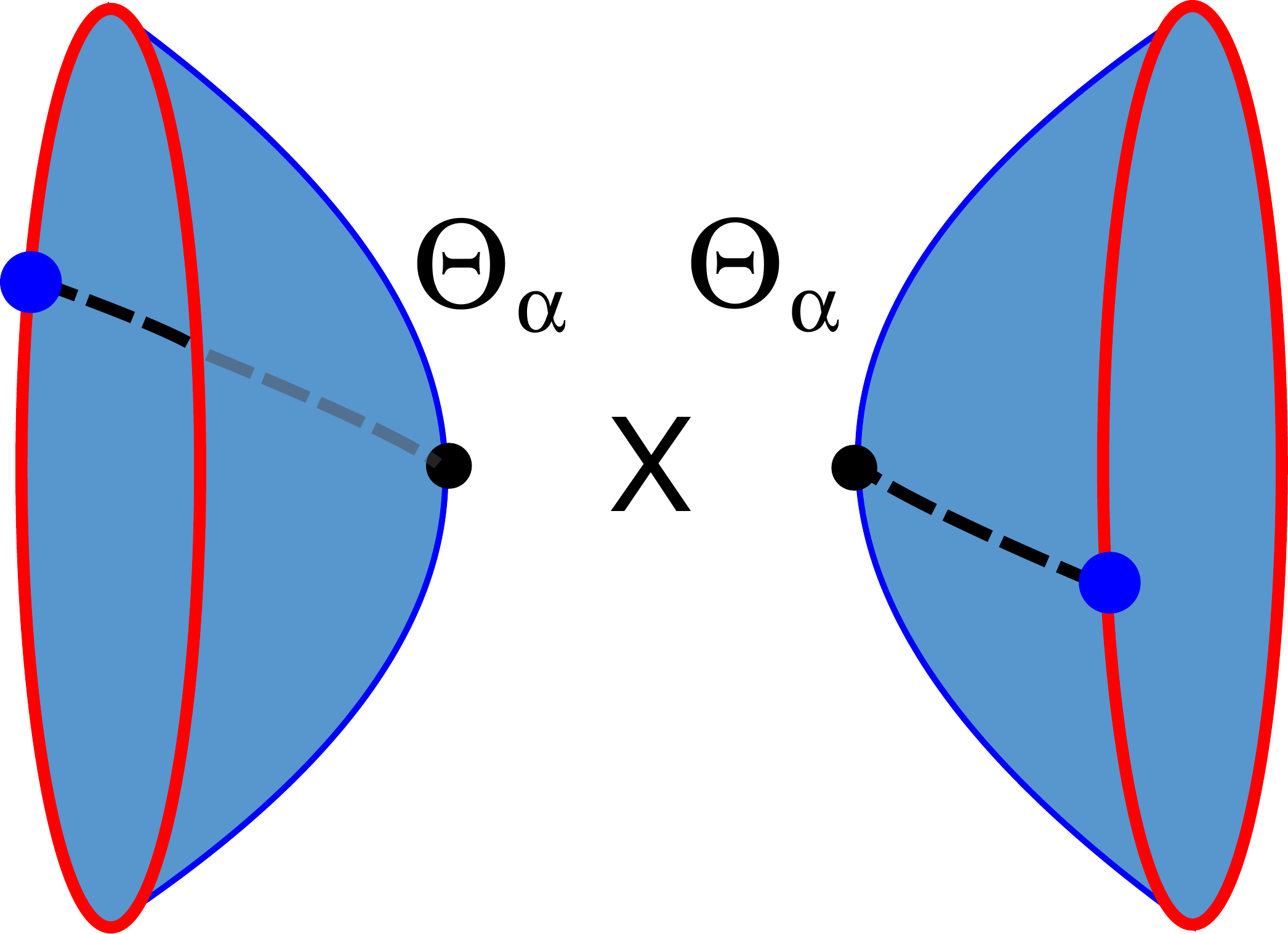}}
\quad = \quad  \raisebox{-8mm}{\includegraphics[width=0.2\textwidth]{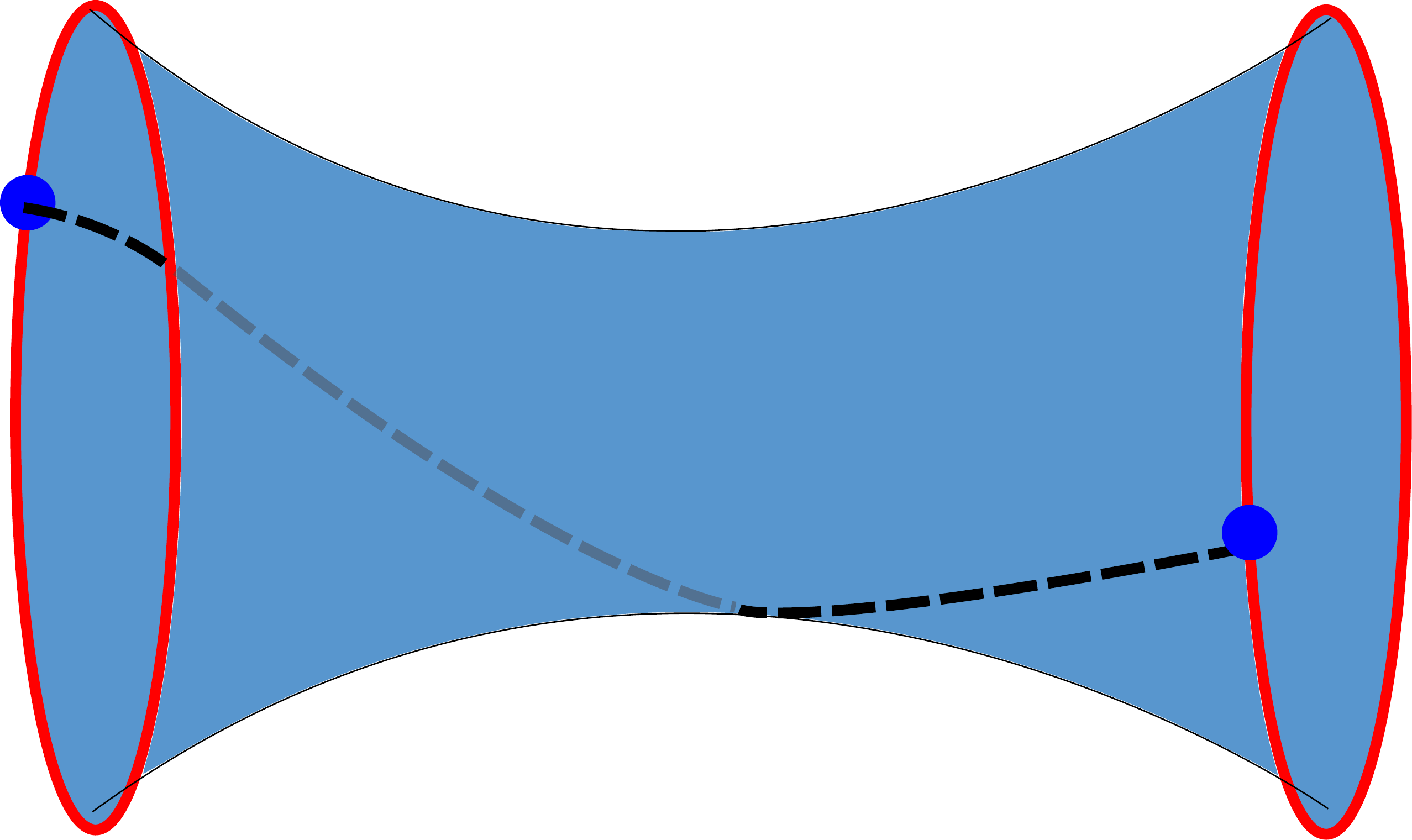}}
\end{align}
where we drew the bulk one-point function diagrams from figure \ref{onepointSum}, and where the branch cuts from the spin fields (drawn as dashed lines) are glued together by this procedure. 

It would be interesting to understand this structure better for more involved topology, and to tie it to some of the known results in JT supergravity and its super Weil-Petersson volumes \cite{Stanford:2019vob}. In order to achieve this, we need to get a better grasp on the underlying matrix model \cite{Klebanov:2003wg,Johnson:2003hy}. Suppose we start with the genus zero density of states \eqref{dosR} for the $(2,4\mathbf{k})$ minimal superstring. Then the string equation at genus zero of the resulting matrix model is written in terms of an auxiliary quantity $f(u)$, defined as:
\begin{equation}
f(u) = -\frac{\partial x}{\partial u}, \qquad \rho_0(E) =  \frac{1}{2\pi}  \int_{0}^{E} du \frac{f(u)}{\sqrt{E-u}}.
\end{equation}
We can find $f(u)$ by expanding $\rho_0(E)$ as a polynomial in $\sqrt{E}$ as:
\begin{align}
\frac{\cosh 2 \mathbf{k} \,\text{arcsinh}\pi \sqrt{E}}{\pi\sqrt{E}} = \frac{1}{\pi\sqrt{E}} + \sum_{j=1}^{\mathbf{k}} \pi^{2j-1}\frac{4^j (\mathbf{k}+j-1)! \mathbf{k}}{(2j)! (\mathbf{k}-j)!} (\sqrt{E})^{2j-1},
\end{align}
from which, following the steps outlined in e.g. \cite{Johnson:2019eik,Johnson:2020heh,Johnson:2020exp}, we can find 
\begin{equation}
\label{fu}
f(u) = 2\pi^2 \, {}_2F_1(1+\mathbf{k},1-\mathbf{k};2;-\pi^2 u/\mathbf{k}^2) = \frac{2\pi^2}{\mathbf{k}} P^{(1,0)}_{\mathbf{k}-1}(1+2\pi^2 u/\mathbf{k}^2),
\end{equation}
where  $P^{(1,0)}_{\mathbf{k}-1}$ is a Jacobi polynomial of order $\mathbf{k}-1$.\footnote{In the bosonic minimal string, this was a Legendre polynomial $P_n \equiv P_n^{(0,0)}$ \cite{Seiberg:2003nm,Mertens:2020hbs}.} This indeed satisfies the relation:
\begin{align}
\frac{1}{2\pi}  \int_{0}^{E} du \frac{f(u)}{\sqrt{E-u}} + \frac{1}{\pi \sqrt{E}} = \frac{\cosh 2 \mathbf{k}\, \text{arcsinh} \pi \sqrt{E}}{\pi \sqrt{E}}.
\end{align}
The string equation is then found by integrating \eqref{fu} and taking care of the $E^{-1/2}$ contribution, leading to:
\begin{equation}
\label{streq}
\boxed{
x = 2 - 2 P^{(0,-1)}_{\mathbf{k}-1}(1+2\pi^2 u/\mathbf{k}^2)}.
\end{equation}
At large $\mathbf{k}$, these become the super JT result derived in \cite{Johnson:2020heh}\footnote{To find both equations \eqref{l1}, one can use the Mehler-Heine relation
\begin{equation}
\lim_{n\to +\infty}n^{-\alpha}P^{(\alpha,\beta)}_n( \cos \frac{z}{n}) = \left(\frac{z}{2}\right)^{-\alpha}J_\alpha(z).
\end{equation}
}
\begin{equation}
\label{l1}
f(u) \,\, \to \,\, 2\pi \frac{I_1(2\pi \sqrt{u})}{\sqrt{u}}, \qquad x = 2 - 2 I_0(2\pi \sqrt{u}).
\end{equation}
The multi-loop genus zero amplitudes are then fully determined by the knowledge of the relation $u(x)$ from inverting the string equation \eqref{streq}, with a general formula written in e.g. \cite{Moore:1991ir}. We leave a more elaborate study to future work.

\paragraph{Acknowledgements} 

I thank A. Blommaert, Y. Fan and especially G. J. Turiaci for useful discussions. TM gratefully acknowledges financial support from Research Foundation Flanders (FWO Vlaanderen).

\appendix

\section{Degenerate braiding and OTOCs}
\label{s:degfus}
In this appendix, we examine the crossings of higher-point functions where at least one of the bilocal lines has a label in the degenerate Virasoro representation $j$. \\
We will see that the amplitudes for both uncrossed and crossed diagrams can be accommodated by the following vertex function rule:
\bea
\label{vertex}
\begin{tikzpicture}[scale=0.7, baseline={([yshift=-0.1cm]current bounding box.center)}]
\draw[thick] (-.2,.9) arc (25:-25:2.2);
\draw[fill,black] (0,0) circle (0.08);
 \draw[thick](-1.5,0) -- (0,0);
\draw (.3,-0.95) node {\footnotesize $\textcolor{black}{k_2}$};
\draw (.3,0.95) node {\footnotesize$\textcolor{black}{k_1}$};
\draw (-1,.3) node {\footnotesize$h$};
\draw (2.5,0.1) node {$\mbox{$\ =\  \, \frac{\Gamma(h \pm i k_1 \pm ik_2)^{1/2}}{\Gamma(2h)^{1/2}},$}$}; \end{tikzpicture}
\qquad \qquad
\begin{tikzpicture}[scale=0.7, baseline={([yshift=-0.1cm]current bounding box.center)}]
\draw[thick] (-.2,.9) arc (25:-25:2.2);
\draw[fill,black] (0,0) circle (0.08);
 \draw[thick,dashed](-1.5,0) -- (0,0);
\draw (.3,-0.95) node {\footnotesize $\textcolor{black}{k_2}$};
\draw (.3,0.95) node {\footnotesize$\textcolor{black}{k_1}$};
\draw (-1,.3) node {\footnotesize$j$};
\draw (5.5,0.1) node {$\mbox{$\ =\  \, \sum_{m=-j}^{+j} c_m^{j}(k_1,k_2) \delta (k_1 - k_2 + i m) ,$}$}; \end{tikzpicture}
\eea
where we add that the last rule is only written once for each combination of $k_1$ and $k_2$.\footnote{Likewise, in case of a doubly degenerate bulk vertex (see below in subsection \ref{s:dd}, only three such contributions are written, since the fourth one would be superfluous again. E.g. the two-point function has two such vertices but we only write the contribution once. We will see that this indeed gives the correct bookkeeping also for crossed lines.} As earlier, we denote a non-degenerate line by a solid line, and a degenerate line by a dashed line. \\
We supplement these diagrammatic rules with the correct crossing diagram: \eqref{vertex6} for a degenerate/non-degenerate crossing or \eqref{vertex7} for a degenerate/degenerate crossing (keeping the discussion around \eqref{slo} in mind).

\subsection{Degenerate and non-degenerate crossing}
We examine a crossed four-point function with one degenerate and one non-degenerate bilocal line:
\begin{align}
\begin{tikzpicture}[scale=0.65, baseline={([yshift=0cm]current bounding box.center)}]
\draw[thick] (0,0) circle (1.5);
\draw[thick] (0.7,-1.3) -- (-0.7,1.3);
\draw[dashed] (-0.7,-1.3) -- (0.7,1.3);
\draw[fill,black] (0.7,-1.3) circle (0.1);
\draw (-1,0) node {\small $k_1$};
\draw (0,-0.9) node {\small $k_s$};
\draw (0,0.85) node {\small $k_t$};
\draw (1,0) node {\small $k_2$};
\draw[fill,black] (-0.7,1.3) circle (0.1);
\draw[fill,black] (-0.7,-1.3) circle (0.1);
\draw[fill,black] (0.7,1.3) circle (0.1);
\draw (0.8,-0.7) node {\small $h$};
\draw (0.8,0.7) node {\small $j$};
\end{tikzpicture}
\end{align}
The case where both bilocals are degenerate, is treated further below in subsection \ref{s:dd}. For a bulk crossing between a non-degenerate and degenerate line, we use the degenerate 6j-symbol:
\bea
\label{vertex6}
\begin{tikzpicture}[scale=0.7, baseline={([yshift=-0.1cm]current bounding box.center)}]
\draw[thick] (-0.85,0.85) -- (0.85,-0.85);
\draw[dashed,thick] (-0.85,-0.85) -- (0.85,0.85);
\draw[dotted,thick] (-0.85,-0.85) -- (-1.25,-1.25);
\draw[dotted,thick] (0.85,0.85) -- (1.25,1.25);
\draw[dotted,thick] (-0.85,0.85) -- (-1.25,1.25);
\draw[dotted,thick] (0.85,-0.85) -- (1.25,-1.25);
\draw (1.5,0) node {\scriptsize $k_2$};
\draw (-1.5,0) node {\scriptsize $k_1$};
\draw (-.75,.33) node {\scriptsize $h$};
\draw (.78,.33) node {\scriptsize $j$};
\draw (0,1.5) node {\scriptsize  $k_t$};
\draw (0,-1.5) node {\scriptsize $k_s$};
\draw (4.5,0.1) node {$\mbox{$\ =\  \, \sixj{-j}{h}{k_1}{k_2}{k_s(k_1)}{k_{t}(k_2)}_{\text{deg}},$}$}; \end{tikzpicture}
\eea
An explicit expression is written in \eqref{deg6j3}, with the particular case of $j=1/2$ in \eqref{sixj}. We will determine these using two complementary methods: the first takes a Schwarzian limit from the degenerate braiding matrix in 2d Virasoro CFT, while the second method deforms the contour of the SL$(2,\mathbb{R})$ 6j symbols directly. The second method is more powerful but it is useful to have both perspectives.

\subsection*{Method 1: Virasoro CFT}
It is known how the crossed four-point function can be obtained by applying the Virasoro braiding kernel within the Schwarzian limit to the uncrossed case \cite{Mertens:2017mtv}. The braiding kernel $R$ realizes the following diagrammatic operation:
\begin{align}
\label{diagrfus}
\begin{tikzpicture}[scale=0.65, baseline={([yshift=0cm]current bounding box.center)}]
\draw[thick] (0,0) circle (1.5);
\draw[thick] (0.7,1.3) arc (150:210:2.6);
\draw[dashed] (-0.7,1.3) arc (30:-30:2.6);
\draw[fill,black] (0.7,-1.3) circle (0.1);
\draw (-1,0) node {\small $k_1$};
\draw (0,0) node {\small $k_s$};
\draw (1,0) node {\small $k_2$};
\draw[fill,black] (-0.7,1.3) circle (0.1);
\draw[fill,black] (-0.7,-1.3) circle (0.1);
\draw[fill,black] (0.7,1.3) circle (0.1);
\draw (0.8,0.7) node {\small $h$};
\end{tikzpicture}
\quad \stackrel{R}{\longrightarrow} \quad 
\begin{tikzpicture}[scale=0.65, baseline={([yshift=0cm]current bounding box.center)}]
\draw[thick] (0,0) circle (1.5);
\draw[thick] (0.7,-1.3) -- (-0.7,1.3);
\draw[dashed] (-0.7,-1.3) -- (0.7,1.3);
\draw[fill,black] (0.7,-1.3) circle (0.1);
\draw (-1,0) node {\small $k_1$};
\draw (0,-0.9) node {\small $k_s$};
\draw (0,0.85) node {\small $k_t$};
\draw (1,0) node {\small $k_2$};
\draw[fill,black] (-0.7,1.3) circle (0.1);
\draw[fill,black] (-0.7,-1.3) circle (0.1);
\draw[fill,black] (0.7,1.3) circle (0.1);
\draw (0.8,-0.7) node {\small $h$};
\end{tikzpicture}
\end{align}
It is an object defined in 2d CFT by the following operation on conformal blocks:
\begin{align}
\label{fusblock}
\begin{tikzpicture}[scale=1, baseline={([yshift=-0.2cm]current bounding box.center)}]
\draw[thick] (-1.5,0) -- (1.5,0);
\draw[dashed] (-0.5,0) -- (-0.5,1);
\draw[thick] (0.5,0) -- (0.5,1);
\draw (-1.25,-0.25) node {\small $\alpha_1$};
\draw (1.25,-0.25) node {\small $\alpha_4$};
\draw (-0.4,1.2) node {\small $\alpha_2$};
\draw (0.4,1.2) node {\small $\alpha_3$};
\draw (0,0.2) node {\small $\alpha_s$};
\end{tikzpicture} 
= R_{\alpha_s \alpha_t}\left[\, {}^{\alpha_2}_{\alpha_1} \;{}^{\alpha_3}_{\alpha_4} \, \right] 
\begin{tikzpicture}[scale=1, baseline={([yshift=-0.2cm]current bounding box.center)}]
\draw[thick] (-1.5,0) -- (1.5,0);
\draw[thick] (-0.5,0) -- (-0.5,1);
\draw[dashed] (0.5,0) -- (0.5,1);
\draw (-1.25,-0.25) node {\small $\alpha_1$};
\draw (1.25,-0.25) node {\small $\alpha_4$};
\draw (-0.4,1.2) node {\small $\alpha_3$};
\draw (0.4,1.2) node {\small $\alpha_2$};
\draw (0,0.2) node {\small $\alpha_t$};
\end{tikzpicture} 
\end{align}
The braiding matrix $R$ itself is related to the fusion matrix $F$ by
\begin{align}
\label{FR}
 F_{\alpha_s\alpha_t}\left[\, {}^{\alpha_2}_{\alpha_1} \;{}^{\alpha_4}_{\alpha_3} \, \right]  =  e^{-i\epsilon \pi (\Delta_{\alpha_1}+\Delta_{\alpha_4}- \Delta_{\alpha_s} - \Delta_{\alpha_t})}R_{\alpha_s \alpha_t}\left[\, {}^{\alpha_2}_{\alpha_1} \;{}^{\alpha_3}_{\alpha_4} \, \right],
\end{align}
where $\epsilon$ is the sign of the braiding. To be explicit, we focus here on the first degenerate primary with $j=1/2$. For a degenerate primary $\alpha_2=-b/2$ (denoted by a dashed line in \eqref{fusblock}), the fusion rules require $\alpha_s = \alpha_1 \pm b/2$ and $\alpha_t = \alpha_3 \pm b/2$. Denoting this choice of sign in the subscript of the fusion matrix, the degenerate fusion matrix is of the form:\footnote{These expressions can be found in many works. Some particularly convenient ones are \cite{Schomerus:2005aq,Alday:2009fs,Ribault:2014hia}.}
\begin{align}
 F_{--}\left[\, {}^{-b/2}_{\alpha_1} \;{}^{\alpha_4}_{\alpha_3} \, \right] &= \frac{\Gamma((2\alpha_1-b)b)\Gamma((Q-2\alpha_4)b)}{\Gamma((\alpha_1-\alpha_4+\alpha_3-b/2)b)\Gamma(1-(\alpha_1-\alpha_4-\alpha_3)b+b^2/2)}, \\
 F_{-+}\left[\, {}^{-b/2}_{\alpha_1} \;{}^{\alpha_4}_{\alpha_3} \, \right] &= \frac{\Gamma((2\alpha_1-b)b)\Gamma((2\alpha_4-Q)b)}{\Gamma((\alpha_1+\alpha_4-\alpha_3-b/2)b)\Gamma((\alpha_1+\alpha_4+\alpha_3-b/2-Q)b)}, \nonumber \\
 F_{+-}\left[\, {}^{-b/2}_{\alpha_1} \;{}^{\alpha_4}_{\alpha_3} \, \right] &= \frac{\Gamma(1+(Q-2\alpha_1)b)\Gamma((Q-2\alpha_4)b)}{\Gamma(1-(\alpha_1+\alpha_4-\alpha_3-b/2)b)\Gamma(1-(\alpha_1+\alpha_4+\alpha_3-b/2-Q)b)}, \nonumber \\
 F_{++}\left[\, {}^{-b/2}_{\alpha_1} \;{}^{\alpha_4}_{\alpha_3} \, \right] &= \frac{\Gamma(1+(Q-2\alpha_1)b)\Gamma((2\alpha_4-Q)b)}{\Gamma((-\alpha_1+\alpha_4+\alpha_3-b/2)b)\Gamma(1-(\alpha_1-\alpha_4+\alpha_3)b+b^2/2)}. \nonumber 
\end{align}
Parametrizing $\alpha_1 = Q/2 + i b k_1$, $\alpha_4 = Q/2 + ib k_2$, $\alpha_2= -b/2$ and $\alpha_3=bh$, and taking the Schwarzian limit where $b\to 0$, we obtain:\footnote{The phase factor in \eqref{FR} evaluates to $(-)^{\epsilon}$, and is an \emph{overall} sign factor which is ignored.}
\begin{align}
\label{fusionlim}
 F_{--}\left[\, {}^{-b/2}_{\alpha_1} \;{}^{\alpha_4}_{\alpha_3} \, \right] &= \frac{i}{2k_2}(h + i k_1 - ik_2 - \frac{1}{2}), \qquad  F_{-+}\left[\, {}^{-b/2}_{\alpha_1} \;{}^{\alpha_4}_{\alpha_3} \, \right] = -\frac{i}{2k_2}(h + i k_1 + ik_2 - \frac{1}{2}), \nonumber \\
 F_{+-}\left[\, {}^{-b/2}_{\alpha_1} \;{}^{\alpha_4}_{\alpha_3} \, \right] &= \frac{i}{2k_2}(h - i k_1 - ik_2 - \frac{1}{2}), \qquad  F_{++}\left[\, {}^{-b/2}_{\alpha_1} \;{}^{\alpha_4}_{\alpha_3} \, \right] = -\frac{i}{2k_2}(h - i k_1 + ik_2 - \frac{1}{2}).
\end{align}
Let us work this out for the $++$ case as an example, for which $ik_s = ik_1 + 1/2$ and $ik_t = ik_2 + 1/2$. Following the logic of \cite{Mertens:2017mtv}, the braiding procedure results in the following combination of Schwarzian vertex functions and braiding matrix:
\begin{equation}
\frac{\Gamma(h \pm (ik_1 + 1/2) \pm ik_2)}{\Gamma(2h)} \frac{1}{k_1}F_{++}\left[\, {}^{-b/2}_{\alpha_1} \;{}^{\alpha_4}_{\alpha_3} \, \right] ,
\end{equation}
where we take the vertex functions from the uncrossed diagram (the left diagram of \eqref{diagrfus}). This expression can be rewritten suggestively as:
\begin{equation}
\frac{\left(\Gamma(h \pm i k_1 \pm i k_t)\Gamma(h \pm i k_2 \pm i k_s)\right)^{1/2}}{\Gamma(2h)}\frac{1}{2ik_1k_2} \left((h -\frac{1}{2})^2 + (k_1-k_2)^2\right)^{1/2},
\end{equation}
where the first factors are the diagrammatic rule for the vertex functions \eqref{vertex} of the rightmost diagram of \eqref{diagrfus}, already evaluating the delta-functions and cancelling the measures for $k_s$ and $k_t$. The last factor can be interpreted as the degenerate 6j-symbol. \\
The other three cases $-+$, $+-$ and $--$ are worked out analogously, resulting in the 6j-symbols:\footnote{There is a technicality about the sign factors. Starting with \eqref{vertex} for $j=1/2$, one can write the resulting contributions schematically as $(+) - (-)$, being the difference between the $ik_s = ik_1 + 1/2$ and $ik_s=ik_1 -1/2$ contributions. Including now the overall sign factors of \eqref{fusionlim}, we find that the resulting four terms have relative signs as $(++) - (+-) - (-+) + (--)$, which is in accord with the signs of applying the degenerate vertex function \eqref{vertex} twice in the rightmost diagram of \eqref{diagrfus}. What is left after these signs are removed, is \eqref{sixj}.}
\begin{align}
\label{sixj}
\sixj{\hspace{-0.2cm}-1/2}{h}{k_1}{k_2}{k_s(-)}{k_t(-)}_{\text{deg}} \hspace{-0.2cm}&= \left( (h-\frac{1}{2})^2 + (k_1-k_2)^2\right)^{\frac{1}{2}}\hspace{-0.2cm}, \quad  \sixj{\hspace{-0.2cm}-1/2}{h}{k_1}{k_2}{k_s(-)}{k_t(+)}_{\text{deg}}  \hspace{-0.2cm}=\left( (h-\frac{1}{2})^2 + (k_1+k_2)^2\right)^{\frac{1}{2}}, \nonumber \\
\sixj{\hspace{-0.2cm}-1/2}{h}{k_1}{k_2}{k_s(+)}{k_t(-)}_{\text{deg}}  \hspace{-0.2cm}&= \left(( h-\frac{1}{2})^2 + (k_1+k_2)^2\right)^{\frac{1}{2}}\hspace{-0.2cm}, \quad \sixj{\hspace{-0.2cm}-1/2}{h}{k_1}{k_2}{k_s(+)}{k_t(+)}_{\text{deg}} \hspace{-0.2cm} = \left(( h-\frac{1}{2})^2 + (k_1-k_2)^2\right)^{\frac{1}{2}}.
\end{align}
Denoting the boundary lengths of the four segments $\ell_{{\rm JT},i}$, where $i=1,2,3,4$, we can write the full amplitude for the crossed four-point function as:
\begin{align}
\label{degotoc}
\frac{1}{Z} \sum_{\epsilon,\epsilon'=\pm} \int_0^{+\infty} dk_1^2 \sinh 2 \pi k_1 &dk_2^2 \sinh 2 \pi k_2 e^{-\ell_{{\rm JT},1} k_s(\epsilon)^2 - \ell_{{\rm JT},2} k_1^2 - \ell_{{\rm JT},3} k_t(\epsilon')^2 - \ell_{{\rm JT},4} k_2^2} \\
&\times \frac{\left(\Gamma(h \pm i k_1 \pm i k_s(\epsilon))\Gamma(h \pm i k_2 \pm i k_t(\epsilon'))\right)^{1/2}}{\Gamma(2h)} \epsilon\epsilon'\, \sixj{-1/2}{h}{k_1}{k_2}{k_s(\epsilon)}{k_t(\epsilon')}_{\text{deg}}. \nonumber
\end{align}

\subsection*{Method 2: contour pinching of non-degenerate 6j symbols}
\begin{figure}[H]
\centering
\includegraphics[width=0.8\textwidth]{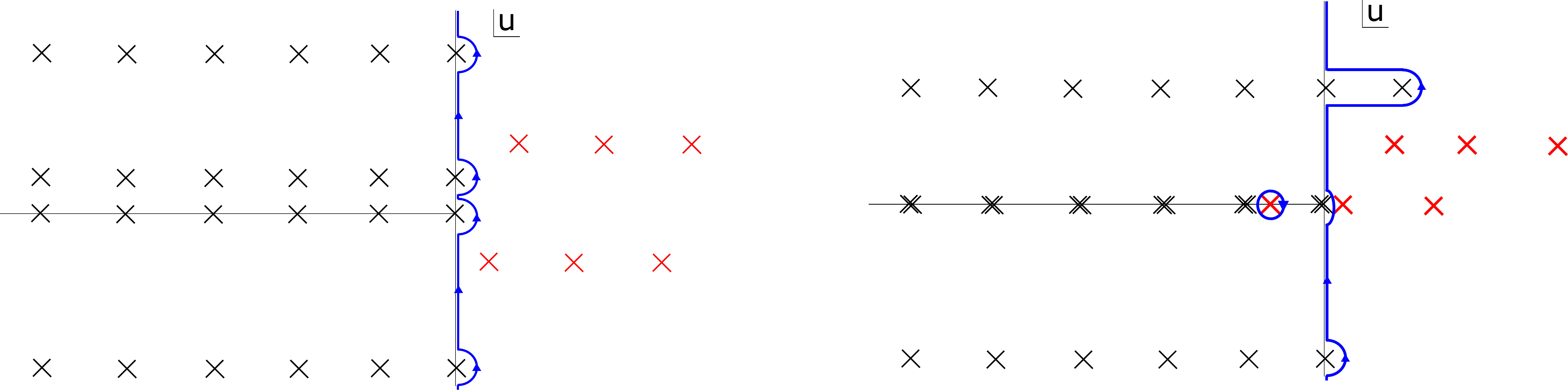}
\caption{Pole series from the 6 Gamma-functions in the numerator of \eqref{appfinal}, with the poles on the rhs of the contour colored in red. Left: starting situation where $h_{1,2} > 0$ and $k_i \in \mathbb{R}$. Right: As one changes $h_1 \to -1/2$ and $k_{s,t} \in \mathbb{R} \pm i/2$ to correspond to the degenerate insertion and fusion rules, some of the red poles cause the contour to pinch off. One picks up the residue from these poles which give singular contributions (due to coincidence with some of the other (black) poles).}
\label{polesDeg}
\end{figure}

From a 2d CFT point of view, higher degenerate fusion matrices get significantly more complicated. Instead, there is a second way of getting the expressions \eqref{sixj} for the 6j-symbols, without resorting to 2d CFT. For non-degenerate labels, the 6j-symbol has an integral expression written in \cite{Mertens:2017mtv}:
\begin{align}
&\sixj{h_1}{h_2}{k_1}{k_2}{k_s}{k_{t}} = \sqrt{\frac{\Gamma(h_1+ik_1 \pm ik_s)\Gamma(h_2-ik_1\pm ik_t)\Gamma(h_1-i k_2 \pm ik_t) \Gamma(h_2+ik_2\pm ik_s)}{\Gamma(h_1-ik_1 \pm ik_s)\Gamma(h_2+ik_1\pm ik_t)\Gamma(h_1+i k_2 \pm ik_t) \Gamma(h_2-ik_2\pm ik_s)}}\nonumber \\[2mm]
&\times \int\limits_{-i\infty}^{i\infty}\!\! \frac{du}{2\pi i} \, \frac{\Gamma(u)\Gamma(u\! -\! 2ik_s)\Gamma(u\! +\!  i k_{1+2-s+t})\Gamma(u\! -\! i k_{s+t-1-2})\Gamma(h_1\!  + \! i k_{s-1}\! -\! u)\Gamma(h_2\! +\! i k_{s-2}\! -\! u)}{\Gamma(u\! +\! h_1\!  - \! i k_{s-1})\Gamma(u\! +\! h_2\! -\! ik_{s-2})}, \nonumber \\[-4mm]
\label{appfinal}
\end{align}
where $k_{i-j} \equiv k_i  - k_j$ etc. This was obtained as the Schwarzian limit of the well-known fusion kernel of Virasoro CFT, obtained by Ponsot and Teschner in \cite{Ponsot:1999uf}. The expression can be viewed as an integral representation of the Wilson function of Groenevelt's work \cite{groenevelt}.
\\~\\
Suppose the label $h_1$ becomes degenerate, then the vertex function of the $h_1$ bilocal line contains the piece $1/\Gamma(2h_1)$ bringing the entire amplitude to zero, unless the 6j symbol \eqref{appfinal} contains a pole that precisely compensates the zero of the vertex function. This can happen when multiple poles coincide in the degenerate limit. A sketch of the poles of the integrand of \eqref{appfinal} is given in figure \ref{polesDeg}.
This leads to the contour getting pinched off between the poles. We can deform the contour by picking up only those residues which give divergent contributions due to multiple poles coinciding.

Within the full amplitude, this process results in only some of the residues of the $u$-integral that have to be considered. In this case, we only need the pole series where $u = h_1 + ik_s - ik_1 +n$ (from the fifth gamma in the numerator of the integrand of \ref{appfinal}). The integer $n$ ranges only over a short range to give a divergent pole contribution. In fact, we can keep the entire range $n \in \mathbb{N}$ since the residues of all other poles in this series turn out to vanish as we show below. The resulting contribution was evaluated in \cite{Mertens:2017mtv} in terms of a ${}_4F_3$ hypergeometric function and yields:
\begin{align}
\label{deg6j}
\sixj{-j}{h_2}{k_1}{k_2}{k_s(k_1)}{k_{t}(k_2)} &=   \frac{\Gamma(-j \pm i k_1 \pm ik_s)^{1/2}\Gamma(-j \pm i k_2 \pm i k_t)^{1/2}}{\Gamma(-2j)}\\
&\times \sqrt{\frac{\Gamma(h_2-ik_1\pm ik_t) \Gamma(h_2+ik_2\pm ik_s)}{\Gamma(h_2+ik_1\pm ik_t)\Gamma(h_2-ik_2\pm ik_s)}} \frac{\Gamma(h_2 + j +ik_1-ik_2)}{\Gamma(h_2- j -ik_1 +ik_2)} \nonumber \\
&\times {}_4F_3 \Big[\mbox{\small$\begin{array}{cccc}\! -j-ik_1+ik_s \!\! & \!\! -j-ik_1-ik_s \!\! &\!\! -j+ik_2+ik_t &\!\! -j+ik_2-ik_t \!\nspc \\[-1mm] \!\! -2j \!\! &\!\! -j+h_2 -ik_1+ik_2 \!\! &\!\! 1-j-h_2-ik_1+ik_2 \!\nspc \end{array}$}
;1 \Big], \nonumber
\end{align}
where the hypergeometric function is defined by the series expansion:
\beq
~_4F_3\Big[\mbox{\small$\begin{array}{cccc}\! a_1 \!\! & \!\! a_2 \!\! &\!\! a_3 &\!\! a_4 \!\nspc \\[-1mm] \!\! b_1 \!\! &\!\! b_2 \!\! &\!\! b_3 \!\nspc \end{array}$} ; z \Big] \equiv \sum_{n=0}^\infty \frac{ (a_1)_n(a_2)_n(a_3)_n(a_4)_n}{(b_1)_n(b_2)_n(b_3)_n}\frac{z^n}{n!},
\eeq
in terms of Pochhammer symbols $(a)_n \equiv a (a+1)\hdots (a+n-1)$. Combining the vertex functions for the $j$ bilocal line with the first line of \eqref{deg6j}, we have the $h_1 \to -1/2$ limit as:
\begin{align}
\frac{\Gamma(h_1 \pm i k_1 \pm ik_s)}{\Gamma(2h_1)} \frac{\Gamma(h_1 \pm i k_2 \pm i k_t)}{\Gamma(2h_1)} = \sum_{\epsilon \epsilon'=\pm} \epsilon \epsilon' \frac{\delta(k_1-k_s +\epsilon i/2)}{k_1 k_s \sinh 2\pi k_s}\frac{\delta(k_2-k_t + \epsilon' i/2)}{k_2 k_t \sinh 2\pi k_t},
\end{align}
where use is made of \eqref{verdeg} and \eqref{orthoprop}, evaluating to Dirac delta functions giving the degenerate fusion rules. Across a higher $j$ bilocal line, we get \eqref{vertfunc}, and the momentum labels are related as $ik_s = ik_1 + m$ and $ik_t = ik_2 + \tilde{m}$ where $-j \leq m,\tilde{m} \leq j$. With these values of $k_s$ and $k_t$ in terms of $k_1$ and $k_2$, we can see that the first and fourth entry of the ${}_4F_3$ in \eqref{deg6j} are negative integers. This means the ${}_4F_3$ truncates to a polynomial, that is called the Wilson polynomial $P_n(x;a,b,c,d)$. In the notation of \cite{groenevelt}, we can write:
\begin{align}
\label{deg6j2}
{}_4F_3 &\Big[\mbox{\small$\begin{array}{cccc}\! -j-ik_1+ik_s \!\! & \!\! -j-ik_1-ik_s \!\! &\!\! -j+ik_2+ik_t &\!\! -j+ik_2-ik_t \!\nspc \\[-1mm] \!\! -2j \!\! &\!\! -j+h_2 -ik_1+ik_2 \!\! &\!\! 1-j-h_2-ik_1+ik_2 \!\nspc \end{array}$}
;1 \Big] \\
&= P_{j-ik_2+ik_t}\big(k_s; -j-ik_1,-j+ik_1,h_2+ik_2,1-h_2+ik_2). \nonumber
\end{align}
With the value of $k_s$ and $k_t$ in terms of $k_1$ and $k_2$, the polynomial is of order min$(j-m,j+\tilde{m})$ in $k_1,k_2$. This leads to the final form of the degenerate 6j-symbols, where we stripped off the pieces that are associated to the vertex functions:
\begin{align}
\label{deg6j3}
\sixj{-j}{h_2}{k_1}{k_2}{k_s(k_1)}{k_{t}(k_2)}_{\text{deg}} &=  \sqrt{\frac{\Gamma(h_2-ik_1\pm ik_t) \Gamma(h_2+ik_2\pm ik_s)}{\Gamma(h_2+ik_1\pm ik_t)\Gamma(h_2-ik_2\pm ik_s)}} \frac{\Gamma(h_2 + j +ik_1-ik_2)}{\Gamma(h_2- j -ik_1 +ik_2)} \nonumber \\
&\times P_{j-ik_2+ik_t}\big(k_s; -j-ik_1,-j+ik_1,h_2+ik_2,1-h_2+ik_2).
\end{align}
Very explicitly, for $h_1 =-j \to -1/2$, we can evaluate the Wilson polynomial (or compute the relevant residues in \eqref{appfinal}),\footnote{For the sign choices $++$, $+-$ and $--$, only a single residue at $n=0$ is required. Alternatively, the Wilson polynomial is 1. For the $-+$ case, one needs both $n=0$ and $n=1$.} and we obtain for \eqref{deg6j3}
\begin{equation}
\left( (h_2 - \frac{1}{2})^2 + (k_1 \pm k_2)^2\right)^{1/2},
\end{equation}
which matches with \eqref{sixj}, with matching $\pm$ signs. 
\\~\\
As a further application, setting $j=0$ in \eqref{deg6j}, the second and third lines evaluate to 1, and we get:
\begin{align}
\sixj{0}{h_2}{k_1}{k_2}{k_s(k_1)}{k_{t}(k_2)} = \sqrt{\frac{\delta(k_1-k_s) \delta(k_2-k_t)}{k_1\sinh 2 \pi k_1 k_2 \sinh 2\pi k_2}}, \qquad \sixj{0}{h_2}{k_1}{k_2}{k_s(k_1)}{k_{t}(k_2)}_{\text{deg}} = 1,
\end{align}
providing a noncompact analogue to a well-known identity, required when checking that removing a bilocal line in the crossed four-point function gives back the two-point function.
\\~\\
Physically, the 6j-symbol for non-degenerate matter weights $h_1$ and $h_2$ contains the semi-classical information for gravitational shockwave scattering in the bulk of particles with masses $m_i^2 \sim h_i(h_i-1)$, and with four black hole regions of energies $k_i^2$ \cite{Jackson:2014nla,Mertens:2017mtv}. Since for degenerate operators, two of these $k_i$'s are complex, the black hole shockwave interpretation is not valid here. This is in accord with our interpretation in section \ref{s:deg} in terms of an integrable subclass of operators. 

\subsection{Degenerate crossing}
\label{s:dd}
Next we consider the case where both operator pairs are degenerate:
\begin{align}
\label{vertex7}
\begin{tikzpicture}[scale=0.85, baseline={([yshift=-0.1cm]current bounding box.center)}]
\draw[fill,black] (0,0) circle (0.1);
\draw[dashed,thick] (-0.85,0.85) -- (0.85,-0.85);
\draw[dashed,thick] (-0.85,-0.85) -- (0.85,0.85);
\draw[dotted,thick] (-0.85,-0.85) -- (-1.25,-1.25);
\draw[dotted,thick] (0.85,0.85) -- (1.25,1.25);
\draw[dotted,thick] (-0.85,0.85) -- (-1.25,1.25);
\draw[dotted,thick] (0.85,-0.85) -- (1.25,-1.25);
\draw (1.5,0) node {\scriptsize $k_2$};
\draw (-1.5,0) node {\scriptsize $k_1$};
\draw (-.75,.33) node {\scriptsize $j_2$};
\draw (.78,.33) node {\scriptsize $j_1$};
\draw (0,1.5) node {\scriptsize  $k_t$};
\draw (0,-1.5) node {\scriptsize $k_s$};
\draw (4.5,0.1) node {$\mbox{$\ =\  \, \sixj{-j_1}{-j_2}{k_1}{k_2}{k_s}{k_{t}},$}$};
\draw (10,0.1) node {$j_1,j_2 \in \mathbb{N}/2$.};
\end{tikzpicture}
\end{align}
This requires both (red) pole series in Figure \ref{polesDeg} getting pinched contour contributions, and leads to the rather lengthy expression:
\begin{align}
\label{sde}
\sixj{-j_1}{-j_2}{k_1}{k_2}{k_s}{k_{t}} &=   \frac{\Gamma(-j_1 \pm i k_1 \pm ik_s)^{1/2}\Gamma(-j_1 \pm i k_2 \pm i k_t)^{1/2}}{\Gamma(-2j_1)}\\
&\hspace{-1cm}\times \sqrt{\frac{\Gamma(-j_2-ik_1\pm ik_t) \Gamma(-j_2+ik_2\pm ik_s)}{\Gamma(-j_2+ik_1\pm ik_t)\Gamma(-j_2-ik_2\pm ik_s)}} \frac{\Gamma(-j_2 + j_1 +ik_1-ik_2)}{\Gamma(-j_2- j_1 -ik_1 +ik_2)} \nonumber \\
&\hspace{-1cm}\times {}_4F_3 \Big[\mbox{\small$\begin{array}{cccc}\! -j_1-ik_1+ik_s \!\! & \!\! -j_1-ik_1-ik_s \!\! &\!\! -j_1+ik_2+ik_t &\!\! -j_1+ik_2-ik_t \!\nspc \\[-1mm] \!\! -2j_1 \!\! &\!\! -j_1-j_2 -ik_1+ik_2 \!\! &\!\! 1-j_1+j_2-ik_1+ik_2 \!\nspc \end{array}$};1 \Big], \nonumber \\
 &+ (j_1 \leftrightarrow j_2, k_1 \leftrightarrow k_2), \nonumber
\end{align}
where the last line represents the contribution from the second pole series, and where we still need to explicitly deal with setting $j_1,j_2 \in \mathbb{N}/2$. We can rewrite this suggestively as follows. Denoting the degenerate 3j-symbols (squared) as:
\begin{equation}
V(k_i,k_j;j) \equiv \lim_{h \to -j}\frac{\Gamma(h \pm i k_i \pm ik_j)}{\Gamma(2h)} = \sum_{m=-j}^{+j} c_m^{j}(k_1,k_2) \delta (k_1 - k_2 + i m),
\end{equation}
we can write the doubly degenerate 6j-symbol \eqref{sde} as:
\begin{align}
\label{firsttwo}
\sixj{-j_1}{-j_2}{k_1}{k_2(k_1)}{k_s(k_1)}{k_{t}(k_1)} &= \sqrt{\frac{V(k_1,k_s;j_1)V(k_2,k_t;j_1)V(k_1,k_t;j_2)}{V(k_2,k_s;j_2)}} \\
&\hspace{-2cm}\times\frac{\Gamma(-j_2+n-m-\tilde{m})}{\Gamma(-j_2-n)} \frac{\Gamma(j_1-j_2-n+\tilde{m})}{\Gamma(-j_1-j_2+n-\tilde{m})} \frac{\Gamma(-j_2+2ik_1+n+m-\tilde{m})}{\Gamma(-j_2+2ik_1+n)}\nonumber \\
&\hspace{-2cm}\times P_{j_1+\tilde{m}}(k_1-im;-j_1-ik_1,-j_1+ik_1,-j_2+ik_1-\tilde{m},1+j_2+ik_1+n-\tilde{m}) \nonumber \\
 &+ (j_1 \leftrightarrow j_2, k_1 \leftrightarrow k_2), \nonumber
\end{align}
where we have already used the delta-functions in the numerator in the first line of \eqref{firsttwo} in terms of:
\begin{alignat}{3}
ik_s &= ik_1 + m, \qquad ik_t = ik_2 + \tilde{m}, \qquad &&-j_1 \leq m,\tilde{m} \leq j_1, \\
ik_t &= ik_1+ n, \qquad &&-j_2 \leq n \leq j_2, \nonumber
\end{alignat}
and hence 
\begin{equation}
ik_2 = ik_s + n - m -\tilde{m} \equiv ik_s + \tilde{n},
\end{equation}
where we defined the number $\tilde{n}$ in terms of the others. Graphically, the four quantum numbers $m, \tilde{m}, n$ and $\tilde{n}$ are defined as
\begin{align}
\label{gra}
\begin{tikzpicture}[scale=0.7, baseline={([yshift=-0.1cm]current bounding box.center)}]
\draw[fill,black] (0,0) circle (0.1);
\draw[dashed,thick] (-0.85,0.85) -- (0.85,-0.85);
\draw[dashed,thick] (-0.85,-0.85) -- (0.85,0.85);
\draw[dotted,thick] (-0.85,-0.85) -- (-1.25,-1.25);
\draw[dotted,thick] (0.85,0.85) -- (1.25,1.25);
\draw[dotted,thick] (-0.85,0.85) -- (-1.25,1.25);
\draw[dotted,thick] (0.85,-0.85) -- (1.25,-1.25);
\draw[thick,->] (1.2,-1.5) arc (-60:-30:1);
\draw[thick,<-] (-1.2,1.5) arc (120:150:1);
\draw[thick,<-] (1.2,1.5) arc (60:30:1);
\draw[thick,<-] (-1.2,-1.5) arc (240:210:1);
\draw (1.5,0) node {\scriptsize $ik_2$};
\draw (-1.5,0) node {\scriptsize $ik_1$};
\draw (-.75,.33) node {\scriptsize $j_2$};
\draw (.78,.33) node {\scriptsize $j_1$};
\draw (0,1.5) node {\scriptsize  $ik_t$};
\draw (0,-1.5) node {\scriptsize $ik_s$};
\draw (-1.8,-1.8) node {\small \color{red} \scriptsize $+m$};
\draw (1.8,1.8) node {\small \color{red} \scriptsize $+\tilde{m}$};
\draw (-1.8,1.8) node {\small \color{red} \scriptsize $+n$};
\draw (1.8,-1.8) node {\small \color{blue} \scriptsize $+\tilde{n}$};
\end{tikzpicture}
\end{align}
There are two features of the expression \eqref{firsttwo} that deserve special attention: the treatment of the delta-functions, and the selection rule on the quantum number $\tilde{n} \equiv n - m -\tilde{m}$ that follows from consistency. 

\subsubsection*{Delta-functions}
The presence of square roots of delta-functions in \eqref{firsttwo} is warranted since they will always combine with other such factors coming from either 3j-symbols, or other 6j-symbols for adjacent bulk crossings. For instance, when inserted in a crossed four-point function:
\begin{align}
\label{dideg}
\begin{tikzpicture}[scale=0.65, baseline={([yshift=0cm]current bounding box.center)}]
\draw[thick] (0,0) circle (1.5);
\draw[dashed] (0.7,-1.3) -- (-0.7,1.3);
\draw[dashed] (-0.7,-1.3) -- (0.7,1.3);
\draw[fill,black] (0.7,-1.3) circle (0.1);
\draw (-1,0) node {\small $k_1$};
\draw (0,-0.9) node {\small $k_s$};
\draw (0,0.85) node {\small $k_t$};
\draw (1,0) node {\small $k_2$};
\draw[fill,black] (-0.7,1.3) circle (0.1);
\draw[fill,black] (-0.7,-1.3) circle (0.1);
\draw[fill,black] (0.7,1.3) circle (0.1);
\draw (0.8,-0.7) node {\small $j_2$};
\draw (0.8,0.7) node {\small $j_1$};
\end{tikzpicture}
\end{align}
the first line in \eqref{firsttwo} can be economically combined with the vertex functions in \eqref{dideg}, yielding the momentum space amplitude
\begin{align}
\label{slo}
\mathcal{A} &= V(k_1,k_s;j_1)V(k_2,k_t;j_1)V(k_1,k_t;j_2) \\
&\times\frac{\Gamma(-j_2+n-m-\tilde{m})}{\Gamma(-j_2-n)} \frac{\Gamma(j_1-j_2-n+\tilde{m})}{\Gamma(-j_1-j_2+n-\tilde{m})} \frac{\Gamma(-j_2+2ik_1+n+m-\tilde{m})}{\Gamma(-j_2+2ik_1+n)}\nonumber \\
&\times P_{j_1+\tilde{m}}(k_1-im;-j_1-ik_1,-j_1+ik_1,-j_2+ik_1-\tilde{m},1+j_2+ik_1+n-\tilde{m}) \nonumber \\
 &+ (j_1 \leftrightarrow j_2, k_1 \leftrightarrow k_2). \nonumber
\end{align}
We notice that since there are three delta-functions, only a single momentum label remains in diagram \eqref{dideg} (which we choose as $k_1$). \\
Following \eqref{vertex} above, it is possible to redefine the normalization of the degenerate 6j-symbol by interpreting the first line of \eqref{slo} as coming entirely from the vertex functions. However, we refrain from doing that here since we believe the above form \eqref{firsttwo} is more transparant in the doubly degenerate case.

\subsubsection*{Selection rules}
The first two factors in the second line of \eqref{firsttwo} result in selection rules to which we turn next. They can be written as
\begin{equation}
\label{fac}
\prod_{s=0}^{j_1+\tilde{m}-1}(-j_2-n + s) \prod_{s=-m-1}^{-j_1}(-j_2+ n - \tilde{m} + s),
\end{equation}
and provide selection rules on the quantum numbers $m, \tilde{m}$ and $n$ in the sense that for compatibility we require that 
\begin{equation}
\label{nbound}
-j_2 \leq \tilde{n} \leq j_2,
\end{equation}
where we have the relation $\tilde{n} \equiv n-m-\tilde{m}$ because we can go in the ``opposite direction'' around the bulk crossing vertex \eqref{gra}. A priori we only have $m, \tilde{m}$ and $n$, and hence this quantum number $\tilde{n}$ is only bounded as
\begin{equation}
-2j_1-j_2 \leq n - m -\tilde{m} \leq 2j_1 + j_2.
\end{equation} 
However, using \eqref{fac}, one readily sees that the 6j-symbol expression vanishes additionally in the intervals:
\begin{align}
\label{alsozero}
\tilde{n} &\in \left\{-j_2-m-\tilde{m}, \hdots, -j_2+j_1-m-1\right\}, \\
\tilde{n} &\in \left\{j_2+1, \hdots,  j_2+j_1-m\right\}.
\end{align}
Hence the 6j-symbol \eqref{firsttwo} is zero outside of the restricted range \eqref{nbound}, where the first line of \eqref{alsozero} produces the lower bound, and the second line of \eqref{alsozero} produces the upper bound of \eqref{nbound}.

\subsubsection*{Generalization}
When going to more complicated diagrams with multiple bulk crossings, the procedure generalizes as follows. The square root of delta functions in \eqref{firsttwo} can now also combine with the other bulk 6j-symbols from the crossings. Working out a few examples immediately shows that one indeed gets the correct number of delta-functions, resulting in a single undetermined momentum label in the disk (say $k_1$), with a selection rule for the sectors around each bulk crossing vertex.

\section{Precision testing the two-point function and perturbative expansion}
\label{s:proof}
In this appendix, we wish to show that the perturbative $1/C$ expansion of the bosonic Schwarzian bilocal correlator \eqref{sch2pt} is always asymptotic except at negative half-integers $h \in - \mathbb{N}/2$, which we investigated above in section \ref{s:deg}. \\
We will start with a warm-up and consider the particular examples $h=+1/2$ and $h=+1$ for which we illustrate that the perturbative series is asymptotic for these cases, developing the technical tools we need further on. 

\subsection{Warm-up: $h=1/2,1$ case study}
The zero-temperature Schwarzian bilocal correlator is given by taking the $\beta\to+\infty$ limit of \eqref{sch2pt}, which effectively projects onto the $k_2=0$ sector:
\begin{align}
\label{schz2pt}
\left\langle \mathcal{O}_h(\tau_1,\tau_2)\right\rangle_{\beta\to \infty} = G_{h}^{\beta \to \infty}(\tau)
 &= \frac{1}{(2C)^{2h}} \int d\mu(k)  e^{ - \tau \frac{k_1^2}{2C}}\, \frac{ \Gamma( h \pm i k)^2}{{2\pi^2}\, \Gamma(2h)}.
\end{align}
In order to distill lessons on the perturbative $1/C$ expansion of this object, we focus on the cases where $h \in +\mathbb{N}/2$ first. Using the asymptotic expansions
\begin{equation}
\label{asex}
\coth(\pi k) = 1 + 2 \sum_{j=1}^{+\infty} \exp(-2j\pi k), \qquad \tanh(\pi k) = 1 - 2 \sum_{j=1}^{+\infty} (-)^{j-1}\exp(-2j\pi k),
\end{equation}
one can write a (formal) asymptotic series expansion of the $h=1$ and $h=1/2$ result in terms of Bernoulli numbers as:
\begin{align}
\label{exone}
G_{h=1}^{\beta \to \infty}(\tau) &= 2\frac{1}{(2C)^2}\int_{0}^{+\infty}dk e^{-\frac{\tau}{2C} k^2}k^3 \coth(\pi k) = \frac{1}{\tau^2} - \frac{1}{(2C)^2}\sum_{n=0}^{+\infty}B_{2n+4}\frac{2\Gamma(2n+4)}{n!(2n+4)!}\left(\frac{\tau}{2C}\right)^n \nonumber \\
&\approx \frac{1}{\tau^2} + \frac{1}{240C^2} -\frac{1}{1008} \frac{\tau}{C^3}+\frac{1}{3840}\frac{\tau^2}{C^4} + \mathcal{O}(\tau^3), \\
\label{exhalf}
G_{h=1/2}^{\beta \to \infty}(\tau) &= 2\frac{1}{2C}\int_{0}^{+\infty}dk e^{-\frac{\tau}{2C} k^2}k \tanh(\pi k) = \frac{1}{\tau} + \frac{1}{2C}\sum_{n=0}^{+\infty}B_{2n+2}\frac{2\Gamma(2n+2)}{n!(2n+2)!}\left(2^{-2n-1}-1\right)\left(\frac{\tau}{2C}\right)^n \nonumber \\
&\approx \frac{1}{\tau} - \frac{1}{24C}+ \frac{7}{1920} \frac{\tau}{C^2}-\frac{31}{64512}\frac{\tau^2}{C^3} + \mathcal{O}(\tau^3).
\end{align}
This is done in a multi-step process: first we insert the above asymptotic expansion \eqref{asex} for the hyperbolic functions. Then we Taylor-expand the Boltzmann factor $e^{-\frac{\tau}{2C} k^2}$, perform the $k$-integral and finally resum the series-expansion of the coth and tanh in terms of Bernoulli numbers $B_{2n} \equiv \frac{ (-)^{n+1}2(2n)!}{(2\pi)^{2n}} \zeta(2n)$.
\\~\\
This series expansion is asymptotic as one can check very explicitly: the coefficients of these series diverge at large orders. This means the correlators contain non-perturbative (in $G_N \sim 1/C$) contributions. We will later show that the non-perturbative contribution to the correlator we have just discussed on the one hand, and the higher-genus contributions on the other hand, are both contributing at the same order in the $G_N \sim 1/C$ expansion, both as $\sim e^{-\frac{\#}{C}}$.

\subsection{Explicit form of zero-temperature expansion}
\label{s:explzero}
Generalizing to higher value of $h \in \mathbb{N}/2$, each coefficient at any fixed order in the expansion is larger than the $h=1$ coefficient, meaning these series are even more divergent. An explicit expression can be obtained in terms of combinations of Bernoulli numbers, but is not very revealing. Instead we focus on obtaining generic expressions for the first few lowest-order terms.

The trick is to start with $h \in \mathbb{N}$, for which we can write
\begin{align}
\label{integl}
G_{h\in \mathbb{N}}^{\beta \to \infty}(\tau) &= 2\frac{1}{(2C)^{2h}\Gamma(2h)}\int_{0}^{+\infty}dk e^{-\frac{\tau}{2C} k^2}k^3 \coth(\pi k) \prod_{m=1}^{h-1}(m^2+k^2)^2.
\end{align}
We expand the product as a power series in $k$:
\begin{align}
\prod_{m=1}^{h-1}(m^2+k^2)^2 = k^{4h-4} + g_2 k^{4h-6} + g_4 k^{4h-8} + g_6 k^{4h-10} + \hdots 
\end{align}
with coefficients explicitly determined by the combinatorics as:
\begin{align}
g_2 &= 2\sum_{n=1}^{h-1} n^2, \qquad g_4 = - \sum_{n=1}^{h-1} n^4 + 2\left(\sum_{n=1}^{h-1} n^2\right)^{2}, \qquad g_6 = \frac{2}{3}\sum_{n=1}^{h-1} n^6 - 2 \sum_{n=1}^{h-1}n^4 \sum_{n=1}^{h-1} n^2 + \frac{4}{3} \left(\sum_{n=1}^{h-1}\right)^{3}.
\end{align}
These coefficients have closed expressions, and doing the $k$-integral, one can write down the expansion:
\begin{align}
\label{zerotexp}
\left\langle \mathcal{O}_{h}(\tau,0)\right\rangle_{\beta\to+\infty} = \frac{1}{\tau^{2h}}\left[ 1 + \frac{h(h-1)}{6C}\tau + \frac{h(10h^3-24h^2+14h+3)}{720}\frac{\tau^2}{C^2} \right. \nonumber \\
+\left. \frac{1}{90720}h(h-2)(70h^4-154h^3+86h^2+73h+15)\frac{\tau^3}{C^3}+ \hdots \right].
\end{align}
While derived only for integer $h$, the structure of the Schwarzian perturbation series at this order, reviewed in section \ref{s:rev}, shows that for \emph{any} $h$ the quantity $g_m$ has to be a polynomial of $m$'th order in $h$, and hence the above formula is correct for any real $h$.\footnote{Indeed, evaluating this correction for $h=1$, $h=1/2$ and $h=-1/2$ agrees with the second- and third-order terms in all of the above series expansions \eqref{exone}, \eqref{exhalf} and \eqref{halfexact}.} \\
The third term is positive for $h>0$ and the fourth for $h > 2$. Some other lessons are that all singular terms as $\tau\to0$ are always positive, and integer $h$ always has no $1/\tau$-term, reflected in the presence of $h(h-n)$ factors at the levels $\tau^\text{odd}$. The positive terms in the expansion $\sim \tau^{>0}$ are always alternating in sign, as is seen by \eqref{expanco}.\footnote{This also resolves a small puzzle we faced in section 4 of \cite{Mertens:2017mtv}. We performed the small-$\tau$ expansion of the $h=1$ bilocal and matched the first correction to the $1/\tau^2$ pole to $\frac{1}{6}\left\langle T\right\rangle$, but we didn't explain the zero-temperature additive constant. With formula \eqref{zerotexp} and our general understanding of the $\tau$-expansion, we can indeed check numerically that the zero-temperature piece is indeed captured by the third term in \eqref{zerotexp}.}

\subsection{General proof of asymptotic series}
\label{s:asymp}
In this section, we will show explicitly that the $1/C$ Schwarzian perturbative expansions are asymptotic series. We will do this by working up from the zero-temperature case, via the fixed energy eigenstate, to the thermal ensemble.

\begin{center}
\textbf{Ground state}
\end{center}
For any \emph{real} value of $h$, one can write the zero-temperature two-point function \eqref{schz2pt} as:
\begin{align}
G_{h}^{\beta \to + \infty}(\tau) = \sum_{n=-2h}^{+\infty}C_n(h) \tau^n = \frac{2}{(2C)^{2h}\Gamma(2h)}\int_{0}^{+\infty}dk e^{-\frac{\tau}{2C} k^2}k^3 \coth(\pi k) \frac{\Gamma(h \pm ik)^2}{\Gamma(1\pm ik)^2}.
\end{align}
This consists of a finite number of terms for which $e^{-\frac{\tau}{2C} k^2}$ cannot be Taylor-expanded (the negative-order $n<0$ terms $\sim \frac{1}{ \tau^{>0}}$ in the $\tau$-expansion) and an infinite number of terms ($n>0$) where it can. The latter is the important contribution to test convergence of the power series. For any fixed value of $k$, the function $f_k(h) = \frac{\Gamma(h \pm ik)^2}{\Gamma(1\pm ik)^2}$ has the following properties:
\begin{itemize}
\item
It is positive and monotonically increasing as a function of $h$.
\item
For any real value of $h$, this function is bounded by a polynomial in $k$; this is because for $h \in \mathbb{N}$ it is polynomial. Since any $h$ has an integer on top of it, and since the function is monotonically increasing, this result follows.
\end{itemize}
Hence, the coefficient at a fixed positive order $n$ in the $\tau$-expansion, $\sum_n C_n(h) \tau^n$, in the expansion is given by:
\begin{equation}
\label{expanco}
C_{n}(h) = \frac{4}{(2C)^{2h}\Gamma(2h)} \sum_{j=1}^{+\infty}  \int_{0}^{+\infty}dk \frac{1}{n!}\left(-\frac{1}{2C} k^2\right)^n k^3 \exp(-2j\pi k) f_k(h).
\end{equation}
The integral is convergent by the polynomial boundedness of $f_k$. \\
The series itself is an alternating series with the following important property. If the series is asymptotic for some value $h^*$, then it is automatically so for higher values of $h > h^*$; indeed, due to the monotonicity properties of $f_k(h)$, the size of the coefficient $\left|C_n\right|$ increases for higher values of $h$. \\
We hence investigate the limiting case where $h \to - \infty$, for which we can use the asymptotic Stirling result:
\begin{equation}
\Gamma(h\pm ik) \sim 2\pi\left|h\right|^{2h-1}e^{-2\pi k}e^{-2h}.
\end{equation}
The coefficient $C_n$ in this limit becomes:\footnote{There is no need to expand the coth-function in this case, as the gamma-functions are already contributing decaying expontial factors to make the $k$-integral convergent.}
\begin{equation}
\label{coef}
C_{n,}(h\to-\infty) \sim \int_{0}^{+\infty}dk \frac{(-)^n}{n!}\left(\frac{1}{2C} k^2\right)^n k \sinh(2\pi k) e^{-4\pi k}.
\end{equation}
One checks explicitly that this diverges as $C_n \sim n$ as $n \to +\infty$, meaning even the $h \to - \infty$ series is asymptotic. By the above argument, this then happens for any $h \in \mathbb{R}$.

\begin{center}
\textbf{Energy eigenstate}
\end{center}
For an energy eigenstate with $k_2 = M$, one only replaces the function $f_k(h)$ by:
\begin{equation}
f_{k,M}(h) = \frac{\Gamma(h \pm ik \pm iM)}{\Gamma(1\pm ik \pm i M)},
\end{equation}
which is likewise positive, monotonically increasing as a function of $h$ and polynomially bounded in $k$. The expansion coefficient is now:
\begin{equation}
\label{expancoM}
C_{n}(h,M) = \frac{4}{(2C)^{2h}\Gamma(2h)}\sum_{j=1}^{+\infty}  \int_{0}^{+\infty}dk \frac{1}{n!}(-)^n \left(\frac{1}{2C} k^2\right)^n k(k^2-M^2) \exp(-2j\pi k) \cosh 2 \pi j M f_{k,M}(h),
\end{equation}
with $C_{n}(h,M=0) \equiv C_n(h)$. As before, it suffices to look at the relevant $h \to - \infty$ asymptotics:
\begin{equation}
\Gamma(h \pm ik \pm iM) \sim \frac{4\pi^2}{\left|h\right|^2}\left|h\right|^{4h}e^{-4\pi \text{max}(k,M)}e^{-4 h}.
\end{equation}
Using this, one obtains the $h \to - \infty$ expansion coefficient:\footnote{Notice the presence of the $1/C^n$ coefficient which has to be here for dimensional reasons.}
\begin{equation}
\label{coefM}
C_{n}(h\to-\infty,M) \sim \int_{0}^{+\infty}dk \frac{(-)^n}{n!}\left(\frac{1}{2C} (k^2-M^2)\right)^n k \sinh(2\pi k) e^{-4\pi \text{ max}(k,M)}.
\end{equation}
It is readily seen that $C_n(h\to-\infty,M)$ decreases for increasing $M$, but increases for increasing $n$ sufficiently large.\footnote{This last property can be checked numerically.} This behavior with $n$ immediately shows that the resulting $\tau/C$ expansion is again asymptotic, for any value of $h$.

\begin{center}
\textbf{Thermal ensemble}
\end{center}
Finally, for the thermal ensemble, one has to take the convergent Laplace $M$-integral of the previous answer:
\begin{equation}
\frac{1}{Z}\int_{0}^{+\infty}d M M \sinh(2\pi M) e^{-\frac{\beta}{2C} M^2}\left( \sum_{n=-2h}^{+\infty} \tau^n C_n(h,M)\right) \equiv \sum_{n=-2h}^{+\infty} D_n\left(h,\frac{\beta}{C}\right) \frac{\tau^n}{C^n},
\end{equation}
where we defined the new coefficients $D_n$ as
\begin{equation}
\label{laplc}
D_n\left(h,\frac{\beta}{C}\right) = \frac{C^n}{Z}\int_{0}^{+\infty}d M M \sinh(2\pi M) e^{-\frac{\beta}{2C} M^2} C_n(h,M).
\end{equation}
The $D_n$ still increase as $n$ is sufficiently large and increasing, simply because if the $C_n(h,M) > C_m(h,M)$ for any $M$ and for $n>m$ and both sufficiently large, then the integral transform in \eqref{laplc} with its positive factors respects this inequality.
\\~\\
This implies the small $\tau$-series expansion
\begin{equation}
\sum_{n=-2h}^{+\infty} D_n\left(h,\frac{\beta}{C}\right)  \frac{\tau^n}{C^n}
\end{equation}
has larger coefficients $D_n\left(h,\frac{\beta}{C}\right) $ for $n$ sufficiently large, and is thus asymptotic only. This is a property of the series expansion \eqref{smaltau} written in the introduction.\footnote{From the perspective of this section it seems harder to make explicit the form of the coefficients $D_n\left(h,\frac{\beta}{C}\right) $ in terms of multi-stress tensor correlators $\left\langle T^n\right\rangle$, which is something where the degenerate cases $h \in - \mathbb{N}/2$ are more convenient for.}\\
If one imagines creating the thermal background in the bulk through gravitational effects, then it is natural to take $\beta \sim C$. The above result then shows that if we think of the coherent background as created through gravitational fluctuations, then that perturbative series is asymptotic.
\\~\\
Now let's not assume $\beta$ proportional to $C$, and interpret $\beta$ as an independent length scale. One can rearrange the series into a power series in $1/C$, by writing:
\begin{equation}
\label{serienew}
\sum_{n=-2h}^{+\infty}  D_n\left(h,\frac{\beta}{C}\right) \frac{\tau^n}{C^n}\equiv \sum_{p=-2h}^{+\infty} \tilde{D}_p \left(h,\frac{\tau}{\beta}\right) \frac{\tau^p}{C^p},
\end{equation}
in term of new coefficients $ \tilde{D}_p \left(h,\frac{\tau}{\beta}\right)$, and this series must hence also be divergent. \\
We now prove that the coefficients $ \tilde{D}_p \left(h,\frac{\tau}{\beta}\right)$ appearing in the $1/C$ expansion \eqref{serienew} are well-defined, and that the divergence of the series on the r.h.s. of \eqref{serienew} only comes from the summation over $p$, making the $1/C$ expansion asymptotic as well. \\
The proof uses the fact that the coefficient $ \tilde{D}_p \left(h,\frac{\tau}{\beta}\right)$ in this series can be found in the Schwarzian $1/C$ perturbation series that we reviewed in the main text. In particular, its computation requires using the higher vertices of the Schwarzian action \eqref{expa} (which only contribute a $\tau/\beta$-independent factor), the propagator \eqref{propa} and the expansion of the bilocal itself \eqref{opexpa}. A crucial property is that the coefficient  $\tilde{D}_p$ has no singularity as $\tau/\beta \to 0$, due to the existence of a well-defined zero-temperature limit. The coefficient $ \tilde{D}_p \left(h,\frac{\tau}{\beta}\right)$ at fixed order $p$ is a (complicated) analytic function of $\tau/\beta$ within its convergence radius $0 \leq \tau < \beta$, with a convergent Taylor series expansion within that radius. An example is the first-order result \eqref{msy}. The divergence of the series on the r.h.s. of \eqref{serienew} can then only arise from the summation over $p$, meaning the $1/C$ Schwarzian perturbation series is an asymptotic series as well. \\
We have hence proved rather generically that the series expansion is asymptotic both at zero temperature, for an energy eigenstate, and at finite temperature.
\\~\\
However, there is an exception to this analysis: if $\Gamma(2 h) = \infty$, and hence $h = 0,-1/2,-1 \hdots$, the above $k$-integrals diverge from the $k_1-k_2\approx im$ region (for a (half)integer $m$) but this is cancelled by a similar divergence from $\Gamma(2 h)$. Hence this case requires separate treatment, and this is our main story in section \ref{s:deg}.

\begin{center}
\textbf{Non-perturbative corrections}
\end{center}
It is interesting to assess the size of the non-perturbative contributions, that goes beyond boundary gravitons, by looking at the growth of the coefficients of this asymptotic series. We will only discuss the zero-temperature case, as we don't expect finite temperature to change this. For both $h=1$ and $h =1/2$ the coefficients at large $n$ grow as\footnote{In the $h \to -\infty$ regime, the coefficients grow as slow as $C_n \sim n$, but of course still divergent as we used above.}
\begin{equation}
C_n \sim \frac{n!}{\sqrt{n}(2\pi^2)^n}.
\end{equation}
To estimate the size of non-perturbative corrections, a useful diagnostic is to Borel transform this series as \cite{Shenker:1990uf}:
\begin{equation}
B(t) = \sum_n t^n \frac{C_n}{n!}, \qquad F(\tau/C) = \int_0^{+\infty} dt e^{-t\frac{C}{\tau}}B(t),
\end{equation}
where the pole in the Borel $t$-plane closest to the origin signals the size of the leading non-perturbative corrections. Since this Borel series has its convergence properties from the geometric series, we find the radius of convergence $t_* \sim 2\pi$, leading to non-perturbative effects of order 
\begin{equation}
\mathcal{A}_{\text{non-pt}}\sim e^{-\frac{C}{\tau}}.
\end{equation}
It is interesting to note that this is of the same order in $G_N \sim 1/C$ as the \emph{perturbative} genus expansion of the matrix integral in powers of $e^{- \# S_0} = e^{- \# C}$.

\section{Expansion of classical bilocal in SL$(2,\mathbb{R})$ invariants}
\label{s:paradox}
Starting with a classical bilocal \eqref{bil}, one can perform a small $\tau$-expansion with coefficients that have to be, due to local $\sltr$ invariance, composed of the Schwarzian derivative, its derivatives and its powers. Can we understand such small time expansions within correlation functions,
\begin{align}
\left\langle \frac{\dot{F}_1\dot{F}_2}{(F_1-F_2)^2}\right\rangle \stackrel{?}{=} \sum_{n=-2}^{+\infty} \left\langle \,  C_n\left(\left\{F,\tau_2\right\}^{m}, \left\{F,\tau_2\right\}' \hdots \right)\, \right\rangle \tau^{2n} .
\end{align}
We will take up the simplest degenerate correlator with $h=-1/2$ as a case study. Denoting the Schwarzian derivative $T(\tau) \equiv \left\{F,\tau\right\}$, the classical bilocal \eqref{bil} for $h=-1/2$ has the small time expansion:
\begin{align}
\label{pt}
\mathcal{O}_{\ell=-1/2}(\tau,0) &=  \frac{F_1-F_2}{\sqrt{F_1'F_2'}} = \tau - \frac{1}{12} \left.T\right|_0\tau^3 - \frac{1}{24}\left.T'\right|_0\tau^4 - \frac{1}{1920}\left[24 T'' - 4 T^2\right]_0 \tau^5 \\
&- \frac{1}{2880}\left[8T''' - 6 T'T\right]_0 \tau^6  - \frac{1}{322560}\left[160 T'''' -160 (T')^2 -208 T''T+ 8 T^3 \right]_0 \tau^7 + \hdots \nonumber
\end{align}
Comparing \eqref{pt} and \eqref{precisionhalf}, several terms match, such as the $\left\langle T\right\rangle$ contribution at $\tau^3$ and the $\left\langle T^2\right\rangle$ at $\tau^5$. 
However, other terms are more puzzling: such as the $\tau^2$ term and $\tau^4$ terms.
\\~\\
In order to properly identify contributions, a crucial point is the renormalization of the composite stress tensors, which start at $\mathcal{O}(\tau^5)$.\footnote{The subtleties associated to composite multi-Schwarzian derivatives were studied recently in \cite{Iliesiu:2020zld} in a different context.} The Schwarzian derivative two-point function is:
\begin{equation}
\left\langle T(\tau)T(0)\right\rangle = -\frac{1}{C}\delta''(\tau) - \frac{2}{C}\left\langle T(0)\right\rangle\delta(\tau) + \frac{4\pi^4}{\beta^4}+\frac{10\pi^2}{\beta^3 C} + \frac{15}{4\beta^2 C^2} + \frac{\#}{C^4},
\end{equation}
and diverges at $\tau=0$ due to the contact terms. The last term written here is a zero-temperature contribution that depends on precisely how one defines the operator. Note that even renormalizing by subtracting the zero-temperature answer still isn't enough to renormalize this correlator. 

These terms actually cause a breakdown of the perturbative expansion \eqref{pt}. Nonetheless, we would like to find out which renormalization is implicitly used in the correlation functions \eqref{precisionhalf}. We propose to replace the contact terms by a pole instead using the rule 
\begin{equation}
\boxed{
\delta^{(n)}(\tau) \to 1/\tau^{n+1}},
\end{equation} 
replacing the $n$-fold derivative of the delta-function by a pole. We were not able to determine the precise coefficient, but will check now that this rule explains the general structure in \eqref{precisionhalf}.\footnote{This interpretation is inspired by the 2d CFT origin of the Schwarzian theory \cite{Mertens:2017mtv}, where the singular contact terms are coming from the singular terms in the OPE expansion.} Using this rule, these contribute to the perturbative expansion at \emph{lower} orders in the $\tau$ expansion. This is dimensionally correct, and we illustrate here how the previous missing terms and the particular structure in \eqref{precisionhalf} can be explained.
\\~\\
The $\tau^6$-term only contributes
\begin{equation}
\left\langle T(\tau)'T(0)\right\rangle = -\frac{1}{C}\delta'''(\tau) - \frac{2}{C}\left\langle T(0)\right\rangle\delta'(\tau),
\end{equation}
giving a zero-temperature $\tau^2$-term and a $\left\langle T\right\rangle$ $\tau^4$-term.
\\~\\
At order $\tau^7$, we need the following expressions:
\begin{align}
\left\langle T(\tau)'T(0)'\right\rangle &= \frac{1}{C}\delta''''(\tau) + \frac{2}{C}\left\langle T(0)\right\rangle\delta''(\tau), \\
\left\langle T(\tau)''T(0)\right\rangle &= -\frac{1}{C}\delta''''(\tau) - \frac{2}{C}\left\langle T(0)\right\rangle\delta''(\tau), \\
\label{threeT}
\left\langle T(\tau_1)T(\tau_2)T(\tau_3)\right\rangle &= \left(\frac{3}{C^2} \delta_{12}'' \delta_{23} - \frac{2}{C^2} \left\langle T(\tau_1)\right\rangle \delta_{12} \delta_{23} - \frac{2}{C}\left\langle T(\tau_1)T(\tau_2)\right\rangle \delta_{23}\right) + (\text{perm}) \nonumber \\
&+ 
\frac{8\pi^6}{\beta^6}+\frac{42\pi^4}{\beta^5 C} + \frac{105 \pi^2}{2\beta^4 C^2} + \frac{105}{8 \beta^3 C^3} + \frac{\#}{C^6},
\end{align}
contributing a zero-temperature term at $\tau^2$, $\left\langle T\right\rangle$ at $\tau^4$, a zero-temperature term at $\tau^3$, $\left\langle T\right\rangle$ at $\tau^5$, $\left\langle T^2\right\rangle$ at $\tau^6$ (containing a further zero-temperature term for $\tau^3$ and $\left\langle T\right\rangle$ for $\tau^5$) and finally the $\left\langle T^3\right\rangle$ at $\tau^7$ itself. \\
Equation \eqref{threeT} can be found by generalizing the arguments of Stanford and Witten \cite{Stanford:2017thb}.\footnote{The constant term is found by integrating this equation thrice and relating it to derivatives w.r.t. $C$ of the partition function $Z$. The first line should be summed over all 3 cyclic permutations of the indices.}

\subsection{Aside: normal-ordered bilocal operators}
Within the gravitational quantum theory, one can alternatively define the normal-ordered bilocal operator by entirely removing all singular contact term contributions studied above. Since these contact terms arise from poles in the OPE when coming from 2d CFT \cite{Mertens:2017mtv}, this is the same definition of normal-ordering of an interacting 2d CFT \cite{DiFrancesco:1997nk}. We will denote this bilocal operator as:
\begin{equation}
:\mathcal{O}_1\mathcal{O}_2: \,\, \equiv \,\, :\left(\frac{F'_1F'_2}{(F_1-F_2)^2}\right)^{h}:
\end{equation}
For this case of $h =-1/2$ \eqref{halfexact}, one has a simple adjustment of the bilocal operator that sets to zero all contact term contributions, by setting:
\begin{equation}
\left( \frac{\beta}{\pi} \sin \frac{\pi}{\beta} \tau \right)e^{\frac{\tau}{8C}(1-\frac{\tau}{\beta})} \qquad \to \qquad :\mathcal{O}_1\mathcal{O}_2:\,\, \equiv\,\, \left( \frac{\beta}{\pi} \sin \frac{\pi}{\beta} \tau \right)e^{-\frac{\tau^2}{8C\beta}}.
\end{equation}
One checks explicitly that indeed stripping off this factor of $e^{\frac{\tau}{8C}}$, one finds solely the $\left\langle T^n\right\rangle$ contributions to the series expansion. We will call this the \emph{normal-ordered} bilocal operator. Explicitly
\begin{equation}
\label{nohalf}
\left( \frac{\beta}{\pi} \sin \frac{\pi}{\beta} \tau \right)e^{-\frac{\tau^2}{8C\beta}} = \sum_{n=1}^{+\infty} \frac{(-)^{n+1}}{\Gamma(2n)2^{n-1}}\left\langle T^{n-1}\right\rangle \tau^{2n-1}.
\end{equation}
This means one has a simple procedure for computing an arbitrary $n$-point correlator of Schwarzian derivatives, by considering the normal-ordered $h=-1/2$ bilocal as a generating function. The generalization of the correlator to arbitrary $h \in \mathbb{R}$, including the supersymmetric $\mathcal{N}=1$ case, is readily made:
\begin{equation}
\left\langle :\left(\frac{F'_1F'_2}{(F_1-F_2)^2}\right)^{h}:\right\rangle = \frac{1}{Z} \int d\mu(k) e^{-\frac{\beta}{2C} k^2} \left( \frac{k/2C}{\sin \left( \frac{k \tau}{2C}\right)}\right)^{2h},
\end{equation}
where one has $d\mu(k) = dk k \sinh (2\pi k)$ for the bosonic theory, and $d\mu(k) = dk 2\cosh(2\pi k)$ for the $\mathcal{N}=1$ case. This has a small $\tau$ expansion of the same structure as \eqref{nohalf}.

\section{Special Functions}
\label{app:spec}
The $b$-deformed Gamma-function is defined by
\beq
\Gamma_b(x) \equiv \frac{\Gamma_2(x | b, b^{-1})}{\Gamma_2 (Q/2|b,b^{-1})},
\eeq
where $\Gamma_2(z|\epsilon_1,\epsilon_2)$ is the Barnes double gamma function. We collect several useful formulae.
\\~\\
The b-deformed gamma function $\Gamma_b(x)$ has the shift properties:
\begin{align}
\Gamma_b(x+b) = \frac{\sqrt{2\pi}b^{bx-1/2}}{\Gamma(bx)}\Gamma_b(x), \qquad \Gamma_b(x+1/b) = \frac{\sqrt{2\pi}b^{-x/b+1/2}}{\Gamma(x/b)}\Gamma_b(x). 
\end{align}
One defines the double sine function $S_b(x)$ as
\begin{equation}
S_b(x) = \frac{\Gamma_b(x)}{\Gamma_b(Q-x)},
\end{equation}
which has the properties:
\begin{align}
S_b(Q-x) &= 1/ S_b(x), \qquad S_b(x+b) &= 2 \sin \pi  b x \, S_b(x), \qquad S_b\left(x+\frac{1}{b}\right) = 2 \sin \frac{\pi}{b} x \,  S_b(x),
\end{align}
and the small $b$-limits:
\begin{equation}
\label{limits}
S_b(bx) \to (2\pi b^2)^{x-1/2}\Gamma(x), \qquad  S_b\left(\frac{1}{2b}+bx\right) \to 2^{x-1/2}.
\end{equation}
The supersymmetric extensions are defined as:
\begin{align}
S_{\NS}(x) &= S_b\left(\frac{x}{2}\right) S_b\left(\frac{x}{2} + \frac{Q}{2}\right), \qquad S_{\R}(x) = S_b\left(\frac{x}{2} + \frac{b}{2}\right) S_b\left(\frac{x}{2} + \frac{1}{2b}\right) \\
\Gamma_{\NS}(x) &= \Gamma_b\left(\frac{x}{2}\right) \Gamma_b\left(\frac{x}{2} + \frac{Q}{2}\right), \qquad \Gamma_{\R}(x) = \Gamma_b\left(\frac{x}{2} + \frac{b}{2}\right) \Gamma_b\left(\frac{x}{2} + \frac{1}{2b}\right),
\end{align}
and have the properties:
\begin{equation}
S_{\NS}(Q-x) = 1/S_{\NS}(x), \qquad S_{\R}(Q-x) = 1/S_{\R}(x),
\end{equation}
\begin{alignat}{2}
\label{susyshift}
S_{\NS}(x+b) && = 2 \cos \left(\frac{\pi b x}{2} \right) S_{\R}(x), \qquad S_{\NS}\left(x+\frac{1}{b}\right) &= 2 \cos \left(\frac{\pi x}{2b} \right) S_{\R}(x), \\
S_{\R}(x+b) && = 2 \sin \left(\frac{\pi b x}{2} \right) S_{\NS}(x), \qquad S_{\R}\left(x+\frac{1}{b}\right) &= 2 \sin \left(\frac{\pi x}{2b} \right) S_{\NS}(x),  \nonumber
\end{alignat}
and
\begin{align}
\label{gammashift}
\Gamma_{\NS}(x+b) &= \frac{\sqrt{2\pi} b^{bx/2}}{\Gamma(bx/2+1/2)} \Gamma_{\R}(x), \qquad \Gamma_{\NS}(x+1/b) = \frac{\sqrt{2\pi} b^{-x/2b}}{\Gamma(x/2b+1/2)} \Gamma_{\R}(x), \nonumber \\
\Gamma_{\R}(x+b) &= \frac{\sqrt{2\pi} b^{bx/2-1/2}}{\Gamma(bx/2)} \Gamma_{\NS}(x), \qquad \Gamma_{\R}(x+1/b) = \frac{\sqrt{2\pi} b^{-x/2b+1/2}}{\Gamma(x/2b)} \Gamma_{\NS}(x).
\end{align}

\section{Details on $\mathcal{N}=1$ super-Schwarzian boundary correlators}
This appendix contains some technical details on the $\mathcal{N}=1$ boundary two-point functions on the disk.

\subsection{Bilocal correlators}
\label{a:corr}
The elementary $h = 1/2$ superspace bilocal operator is
\begin{equation}
\frac{D_1 \theta'_1 D_2 \theta'_2}{\tau_1'-\tau_2'-\theta_1'\theta_2'}.
\end{equation}
This can be expanded in the $\theta$'s and gives two bosons ($\sim 1$ and $\sim \theta_1\theta_2$) and two fermions ($\sim \theta_1$ and $\sim \theta_2$). When considering a two-point correlator, the fermionic pieces vanish by fermion number conservation. Explicitly, the bottom component $(\sim 1$), denoted $\mathcal{B}$, is
\begin{align}
\label{bottom}
\mathcal{B} 
&= \frac{\sqrt{f'_1f'_2}}{f_1-f_2}\left[1 + \frac{\eta_1\eta'_1}{2}\right]\left[1 + \frac{\eta_2\eta'_2}{2}\right] + \frac{\eta_1\eta_2f'_1f'_2}{(f_1-f_2)^2},
\end{align}
and the top component ($\sim \theta_1\theta_2$), denoted $\mathcal{T}$, is 
\begin{align}
\label{top}
&\mathcal{T} = - \frac{\sqrt{f'_1f'_2}}{f_1-f_2}\left[\frac{1}{4}\eta_1\eta_2 \frac{f''_1}{f'_1}\frac{f''_2}{f'_2} - \frac{1}{2}\eta_2\eta'_1 \frac{f''_2}{f'_2} + \frac{1}{2}\eta_1\eta'_2 \frac{f''_1}{f'_1} + \eta'_1\eta'_2\right] \nonumber \\
&+\frac{f'_1f'_2}{(f_1-f_2)^2}\left[1 + \frac{1}{2}\eta_1\eta_2 f''_2 \sqrt{\frac{f'_1}{f'_2}} - \frac{1}{2}\eta_1\eta_2 f''_1 \sqrt{\frac{f'_2}{f'_1}} + 2\eta_1\eta_2 \frac{\sqrt{f'_1 f'_2}}{f_1-f_2} + \eta_1\eta'_2 \sqrt{\frac{f'_1}{f'_2}} + \eta_2\eta'_1 \sqrt{\frac{f'_2}{f'_1}} \right].
\end{align}
For higher $h$, the classical bilocal operator is of the form:
\begin{equation}
\label{susyhigher2}
\mathcal{O}_{h}(\tau_1,\tau_2, \theta_1,\theta_2) \equiv \left( \frac{D_1 \theta'_1 D_2 \theta'_2}{\tau_1'-\tau_2'-\theta_1'\theta_2'}\right)^{2h}.
\end{equation}
The supersymmetric correlators were given in equations \eqref{super2pt} and \eqref{ssuper2pt}. Let us mention some properties.
\begin{itemize}
\item
In the semi-classical large $C$ regime, one finds the limits
\begin{align}
\label{slargeC}
G_h^{\mathcal{B}}(\tau) \to \frac{1}{\left(\frac{\beta}{\pi}\sin \frac{\pi}{\beta}\tau \right)^{2h}}, \qquad G_h^{\mathcal{T}}(\tau) \to 2h\frac{1}{\left(\frac{\beta}{\pi}\sin \frac{\pi}{\beta}\tau \right)^{2h+1}}.
\end{align}
This is obtained by using similar tricks as in \cite{Lam:2018pvp}. For $G_h^{\mathcal{B}}(\tau)$, the second term in \eqref{super2pt} dominates, whereas in $G_h^{\mathcal{T}}(\tau)$ the first term in \eqref{ssuper2pt} dominates, which is why we call the second term the bottom piece and the first term the top piece in the main text. \\
Going to real time $\tau \to \pm it + \epsilon$, quasi-normal modes can be readily read off from these expressions by Fourier-transforming the $1/\sinh$ real-time correlator, or by looking at the poles of the Gamma-functions of \eqref{super2pt} and \eqref{ssuper2pt}:
\begin{alignat}{2}
\omega_n &= - \frac{2\pi}{\beta}i \left( n + h\right), \quad &&\text{bosonic}, \\
\omega_n &= - \frac{2\pi}{\beta}i \left( n + h + \frac{1}{2}\right), \quad &&\text{superpartner},
\end{alignat}
with the $1/2$ shift corresponding to the shift in conformal weight of the superpartner.

\item
Bilocal correlation functions satisfy the following recursion relations:
\begin{align}
G^{\mathcal{B}}_h &= \frac{1}{2h-1}\left[\frac{1}{2C}\left(h - 1/2 \right) ^2G^{\mathcal{B}}_{h-1/2} + G^{\mathcal{T}}_{h-1/2}\right], \\
\label{suprec}
G^{\mathcal{T}}_h &= \frac{1}{2h-1}\left[\frac{1}{2C}\left(h - 1/2 \right)^2\left(-G^{\mathcal{T}}_{h-1/2} + 2 \partial_\tau G^{\mathcal{T}}_{h-1/2} + 4 \frac{\partial_\beta(G^{\mathcal{B}}_{h-1/2}Z)}{Z}\right) + \partial_\tau^2 G^{\mathcal{B}}_{h-1/2}\right].
\end{align}
These are readily derived explicitly using \eqref{super2pt} and \eqref{ssuper2pt}, and checked in the large $C$-regime \eqref{slargeC}.\footnote{For the superpartner semi-classical limit, one uses $\partial_\beta \ln Z \to  - \frac{2\pi^2 C}{\beta^2}$ and only the $\beta$-derivative and the last term contribute in the rhs of \eqref{suprec}.}
At zero temperature $\beta \to +\infty$, one has $G_h^{\mathcal{B}}(\tau) = 1/\tau^{2h}$, $G_h^{\mathcal{T}}(\tau) = (2h)/\tau^{2h+1}$, which satisfy these relations as well. The bosonic bilocal correlator \eqref{sch2pt} satisfies an analogous recursion relation relating $h$ to $h +1$ though it is not very illuminating.
\end{itemize}

\subsection{Degenerate two-point function}
\label{a:super1}
Compared to the bosonic theory in section \ref{s:deg}, one only replaces the continuous irrep wavefunctions of SL$(2,\mathbb{R})$ by those of OSp$(1|2)$ \cite{Blommaert:2018oro}.  Equivalently, we employ the $\mathcal{N}=1$ super-Liouville minisuperspace wavefunctions \cite{Mertens:2017mtv}. We define the supersymmetric wavefunctions as:
\begin{equation}
\psi_{k}(x) \equiv K_{\frac{1}{2}+2ik}(x) + K_{\frac{1}{2}-2ik}(x), \qquad x=e^{\phi} >0.
\end{equation}
The vertex functions for JT supergravity in \eqref{super2pt} are then found by performing the auxiliary integral:
\begin{align}
\int_0^{+\infty} dx \psi_{k_1}(x) \psi_{k_2}(x) x^{2h} = 4^{h-1}\frac{\left(\Gamma\bigl(\textstyle \frac{1}{2}+h \pm i(k_1-k_2)\bigr)\, \Gamma\bigl(h \pm i(k_1+k_2)\bigr) + (k_2 \to -k_2)\right)}{\Gamma(2h)}.
\end{align}
Mirroring the bosonic case in section \ref{s:deg}, to determine this expression for $h\in - \mathbb{N}/2$, we combine the following normalization constraint \cite{Hikida:2007sz}:
\begin{align}
\int_{0}^{+\infty} dx \psi_{k_1}(x) \psi_{k_2}(x) = \frac{\pi^2}{\cosh 2 \pi  k_1} \delta(k_1-k_2),
\end{align}
with the 1d fusion property of the modified Bessel function \eqref{besselprop}, leading to e.g. the relation:
\begin{align}
\int_{0}^{+\infty} \frac{dx}{x}\,  &\psi_{k_1}(x) \psi_{k_2}(x) = \int_{0}^{+\infty} dx \, \psi_{k_1}(x) \frac{ \psi_{k_2-\frac{i}{2}}(x) -  \psi_{k_2}(x)}{\frac{1}{2}+2ik_2}+ \frac{\psi_{k_2+\frac{i}{2}}(x) - \psi_{k_2}(x)}{\frac{1}{2}-2ik_2} \nonumber \\
&= \frac{\pi^2}{\cosh 2 \pi  k_1} \left(\frac{\delta(k_1-k_2+\frac{i}{2})}{\frac{1}{2}+2ik_2} + \frac{\delta(k_1-k_2-\frac{i}{2})}{\frac{1}{2}-2ik_2} - \frac{\delta(k_1-k_2)}{(\frac{1}{2}-2ik_2)(\frac{1}{2}+2ik_2)}\right),
\end{align}
playing the same role as \eqref{orthoprop} above. In the last equality, the Dirac delta functions are to be interpreted in $k$-integrals where one move the contour from the real axis to $\pm i/2$. Applying this relation consecutively, one arrives at the supersymmetric degenerate vertex functions written in the main text in \eqref{vertfuncS}.
\\~\\
Let us give some examples. For $h = -1/2$, one has
\begin{align}
\label{halfSexact}
G_{-1/2}^{\mathcal{B}}(\tau) =  \frac{(2C)}{Z} \int dk &\cosh 2\pi k e^{- \beta k^2 }\left[e^{\frac{\tau}{8 C}}\left(\frac{e^{i k \tau/2C}}{\frac{1}{2}+2ik} - \frac{e^{-i k \tau/2C}}{-\frac{1}{2}+2ik}\right)  + \frac{1}{(\frac{1}{2}+2ik)(-\frac{1}{2}+2ik)}\right].
\end{align}
For $h=-1$, one has
\begin{align}
\label{oneSexact}
G_{-1}^{\mathcal{B}}(\tau) =  \frac{(2C)^{2}}{Z} \int &dk \cosh 2\pi k e^{- \beta k^2 }\left[e^{\frac{\tau}{2 C}}\left(\frac{e^{i k \tau/C}}{(\frac{3}{2}+2ik)(\frac{1}{2}+2ik)} + \frac{e^{-i k \tau/C}}{(-\frac{3}{2}+2ik)(-\frac{1}{2}+2ik)}\right)  \right. \nonumber \\
&\left.- 2e^{\frac{\tau}{8 C}}\left(\frac{e^{\frac{i k \tau}{2C}}}{(\frac{3}{2}+2ik)(\frac{1}{2}+2ik)(-\frac{1}{2}+2ik)} - \frac{e^{-\frac{i k \tau}{2C}}}{(-\frac{3}{2}+2ik)(\frac{1}{2}+2ik)(-\frac{1}{2}+2ik)}\right) \right. \nonumber \\
&\left.- \frac{2}{(\frac{1}{2}+2ik)(-\frac{1}{2}+2ik)}\right] .
\end{align}
At zero temperature $\beta \to \infty$, one finds the simple results:
\begin{align}
\frac{1}{(2C)}G^{\mathcal{B}}_{h=-1/2}(\tau) &= 4e^{\frac{\tau}{8C}} - 4 = \frac{\tau}{2C} + \mathcal{O}(\tau^2), \\
\frac{1}{(2C)^{2}}G^{\mathcal{B}}_{h =-1}(\tau) &= \frac{8}{3}e^{\frac{\tau}{2C}} - \frac{32}{3}e^{\frac{\tau}{8C}} + 8 = \frac{\tau^2}{(2C)^2} + \mathcal{O}(\tau^3).
\end{align}

\subsection{Degenerate superpartner two-point function}
\label{a:super}
The superpartner two-point function is obtained by multiplying $(k_1-k_2)^2$ for the bottom piece and $(k_1+k_2)^2$ for the top piece:
\begin{align}
c_{mB}^j (k_1,k_2) &= \frac{1}{\cosh 2\pi k_1} (-)^{m+j}(im)^2\left(\begin{array}{c}
2j \\
m+j \\
\end{array}\right) \frac{1}{(2ik_2 - j +\frac{1}{2} + m)_{2j}}, \\
c_{mT}^j (k_1,k_2) &=  \frac{2j}{\cosh 2\pi k_1} (-)^{m+j-1/2}(2k_2-i m)^2\left(\begin{array}{c}
2j-1 \\
m+j-1/2 \\
\end{array}\right) \frac{1}{(2ik_2 - j + m)_{2j+1}}.
\end{align}
Since this makes the top piece dominate at large $k$, the top part $c^j_{mT}$ is giving the contribution to the semiclassical limit $- 2j \left(\frac{\beta}{\pi}\sin \frac{\pi}{\beta}\tau \right)^{2j-1}$. \\
As an example of this, the super-partner two-point function for $j = 1$ is readily obtained as:
\begin{align}
\frac{(2C)^{2}}{Z} \int &dk \cosh 2\pi k e^{- \beta k^2 }\left[-e^{\frac{\tau}{2 C}}\left(\frac{e^{i k \tau/C}}{(\frac{3}{2}+2ik)(\frac{1}{2}+2ik)} + \frac{e^{-i k \tau/C}}{(-\frac{3}{2}+2ik)(-\frac{1}{2}+2ik)}\right)  \right. \\
&\left.- 2e^{\frac{\tau}{8 C}}\left(\frac{(2k-i/2)^2e^{\frac{i k \tau}{2C}}}{(\frac{3}{2}+2ik)(\frac{1}{2}+2ik)(-\frac{1}{2}+2ik)} - \frac{(2k+i/2)^2e^{-\frac{i k \tau}{2C}}}{(-\frac{3}{2}+2ik)(\frac{1}{2}+2ik)(-\frac{1}{2}+2ik)}\right) \right], \nonumber
\end{align}
with zero-temperature limit and small $\tau$-expansion:
\begin{align}
\frac{1}{(2C)^{2}}G^{\mathcal{T}}_{h=-1}(\tau) = -\frac{8}{3}e^{\frac{\tau}{2C}} + \frac{8}{3}e^{\frac{\tau}{8C}} = -2\frac{\tau}{(2C)} + \mathcal{O}(\tau^2).
\end{align}

\section{Details on $\mathcal{N}=1$ super-Liouville model with boundaries}
\label{s:detliou}

\subsection{Lagrangian approach}
\label{s:lagr}
\begin{figure}[h]
\centering
\includegraphics[width=0.3\textwidth]{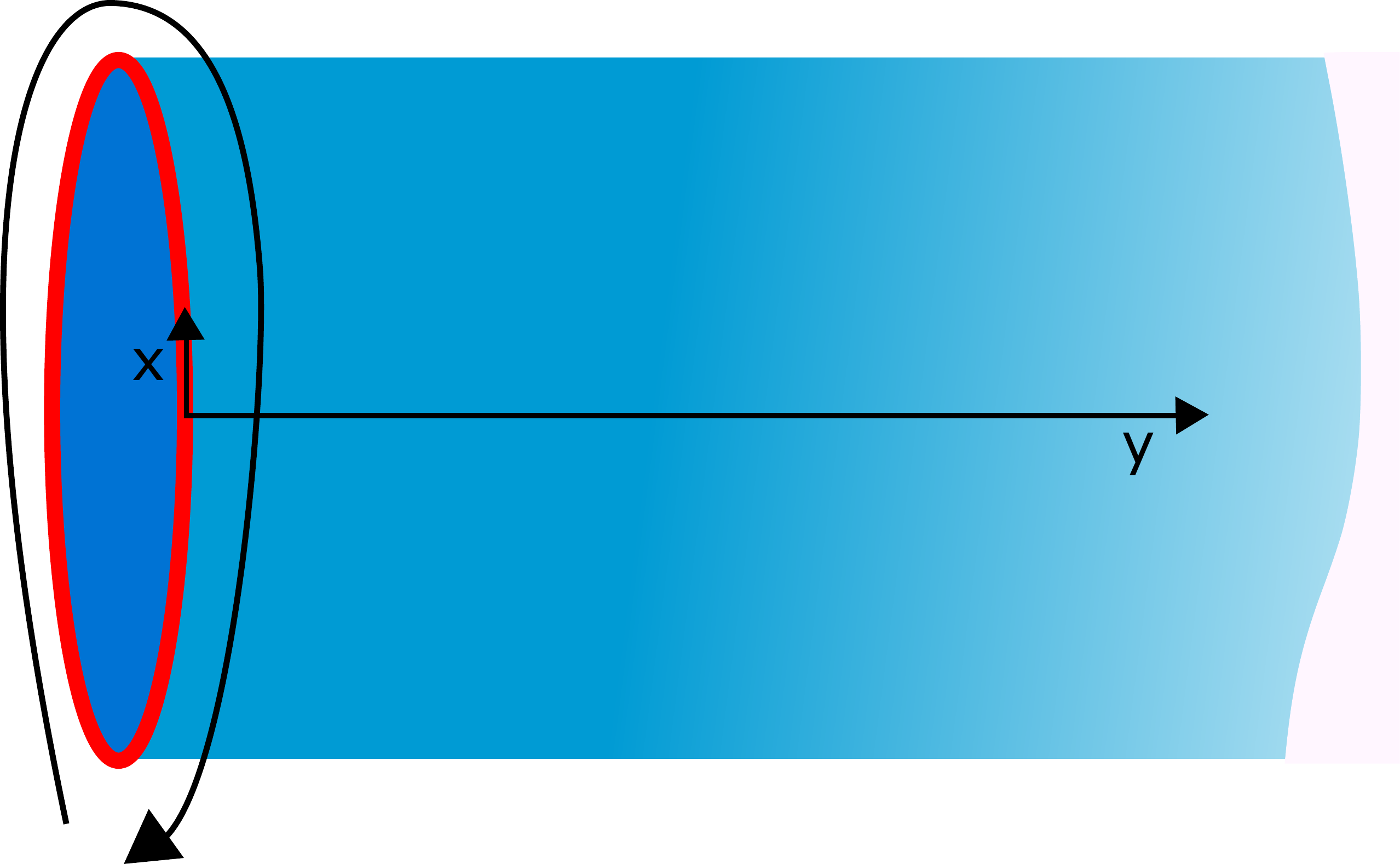}
\caption{Boundary in $\mathcal{N} =1$ Liouville theory. Preserving half of supersymmetry requires left- and right movers to be related as $T_F(z) +\eta \bar{T}_F(\bar{z}) = 0$. One also has a global boundary condition $T_F(z+2\pi) = \mp T_F(z)$ labeling the boundary state as NS or R respectively.}
\label{SUSYbrane}
\end{figure}

In order to motivate the transformation formulas to the fixed length basis mentioned in section \ref{s:Lsugra}, and to develop intuition, we provide here a discussion of the classical Lagrangian of $\mathcal{N}=1$ Liouville theory, its boundary terms and the transition to fixed length amplitudes. This story is well-known in the literature \cite{Fukuda:2002bv,Ahn:2002ev,Douglas:2003up}, but it is not that simple to relate the different works, which we try to remedy here.
\\~\\
We study $\mathcal{N}=1$ Liouville theory on a manifold with a circular boundary. We set the coordinates as $z=x+iy$, with the boundary at $y=0$, and $x \sim x+2\pi$ periodically identified. \\ 
To preserve half of the supersymmetry, one has the superconformal boundary conditions on the UHP: $T(z) = \bar{T}(\bar{z})$ and $T_F(z) +\eta \bar{T}_F(\bar{z}) = 0$ which allows a sign-factor $\eta = \pm 1$ and hence gives two types of branes. Next to this, the fermionic fields can be either periodic (R) or antiperiodic (NS) upon circling the boundary: $T_F(z+2\pi) = \pm T_F(z)$ (see Figure \ref{SUSYbrane}).  
For most of the discussion in this work, we will work with fixed fermionic boundary condition. When making contact with higher topology and with matrix model techniques to compute minimal superstring amplitudes, one has to sum over different spin structures, but this will not be important for our purposes and is deferred to future work.
\\~\\
Defining a scalar superfield $\Phi$ in terms of the scalar fields $\phi$ and $F$ and the two Majorana-Weyl fermions $\psi$ and $\bar{\psi}$ as
\begin{equation}
\Phi = \phi + i \theta \psi + i \bar{\theta} \bar{\psi} + \theta \bar{\theta} F,
\end{equation}
the supersymmetric Liouville model, on a manifold with a boundary that preserves half of the supersymmetry (labeled by $\eta)$, is given in terms of the superspace Lagrangian
\begin{equation}
\mathcal{L} = \frac{1}{4\pi}\left[D\Phi \bar{D}\Phi  + 2 i\mu_0e^{b\Phi}\right], \qquad D=\partial_\theta + \theta \partial, \bar{D} = \partial_{\bar{\theta}} + \bar{\theta} \bar{\partial},
\end{equation}
as
\begin{align}
S_L + S_L^{\partial}&= \int d^2z d^2\theta \mathcal{L} + i\eta  \oint dx \mathcal{L}(\theta=\bar{\theta}=0) \\
&= \frac{1}{4\pi} \int d^2z \left( \partial \phi \bar{\partial} \phi + \psi \bar{\partial}\psi + \bar{\psi}\partial \bar{\psi} -F^2 - 2 \mu b e^{b\phi}F + 2 i \mu_0 b^2 e^{b\phi} \psi \bar{\psi}\right) \nonumber \\
&{} \qquad \qquad \qquad \qquad - \oint dx\left( \frac{i}{2}\eta \psi \bar{\psi} + \eta \mu_0 e^{b\phi}\right).
\end{align}
We follow the notations and conventions of \cite{Douglas:2003up} in this subsection. Enforcing the boundary variation to vanish, we can distinguish two boundary conditions.
\begin{itemize}
\item
One can find a classical ZZ-brane solution with the Dirichlet boundary conditions 
\begin{equation}
\left.\psi = -\eta \bar{\psi}\right|_{y=0}, \qquad \left.\phi\right|_{y=0} = +\infty.
\end{equation}
\item
Imposing Neumann (free) boundary conditions is done by imposing
\begin{equation}
\left.\psi = \eta \bar{\psi}\right|_{y=0}, \qquad \partial_y \phi + 2\eta\mu_0 b e^{b\phi} =0.
\end{equation}
These are FZZT branes, with boundary cosmological constant $\eta \mu_0$ fixed by supersymmetry.
\end{itemize}
These cases can be neatly summarized in superspace as
\begin{align}
\left.D_t \Phi\right|_{y=0} &= 0, \qquad \text{Dirichlet}, \qquad D_t = D+ \eta \bar{D}, \\
\label{bdyN}
\left.D_n \Phi\right|_{y=0} &= 0, \qquad \text{Neumann}, \qquad D_n = D- \eta \bar{D}.
\end{align}

We can still preserve supersymmetry by also adding the following boundary term in terms of the fermionic boundary superfield $\Gamma$ \cite{Ghoshal:1993tm,Nepomechie:2001qr,Douglas:2003up,Ahn:2003wy}:
\begin{align}
 S_{\text{FZZT}} &=\frac{1}{2\pi} \oint dx d\theta_t \left( \Gamma D_t \Gamma + 2 i \mu_B b \Gamma e^{\frac{b}{2}\Phi} \right), \qquad \Gamma = \gamma + i \theta_t f, \nonumber \\
\label{susybdy}
&= \frac{1}{2\pi} \oint dx \left( \gamma \partial_x \gamma -f^2 - 2 \mu_B f e^{b/2\phi} - \mu_B b \gamma (\psi + \eta \bar{\psi}) e^{b/2\phi} \right).
\end{align}
The fields $\gamma$ and $f$ are boundary degrees of freedom that need to be quantized or integrated out to obtain the theory in its original variables, see also \cite{Ghoshal:1993tm,Nepomechie:2001qr,Ahn:2003wy}.
The addition of this term $ S_{\text{FZZT}}$ is necessary to find regular classical solutions with generalized Neumann boundary conditions. \\
Including the term $ S_{\text{FZZT}}$, the Neumann boundary condition \eqref{bdyN} is adjusted into the superspace expression:
\begin{equation}
\left.D_n \Phi + 2\mu_B \Gamma b e^{\frac{b}{2}\Phi}\right|_{z=\bar{z}, \theta=\eta \bar{\theta}}= 0,
\end{equation}
which resembles the bosonic Liouville FZZT boundary condition. In components:
\begin{align}
\label{bdyco1}
&\left.i(\psi - \eta \bar{\psi}) + 2\mu_B b \gamma e^{\frac{b}{2}\phi}\right|_{y=0}=0, \\
\label{bdyco2}
&\left.\partial_y \phi + 2\eta\mu_0 b e^{b\phi} - 2\mu_B b f e^{\frac{b}{2}\phi} - 2\mu_B b \gamma (\psi + \eta \bar{\psi}) e^{\frac{b}{2} \phi}\right|_{y=0} =0.
\end{align}
The boundary equations of motion derived from \eqref{susybdy} for $\Gamma \equiv \gamma + i \theta f$, are
\begin{align}
f &= \mu_B e^{\frac{b}{2}\phi}, \\
\partial_x \gamma &= \frac{1}{2}\mu_B b (\psi + \eta \psi) e^{\frac{b}{2}\phi}.
\end{align}
Plugging these back into the boundary conditions \eqref{bdyco1},\eqref{bdyco2}, we get:
\begin{align}
&\left.i\frac{\partial}{\partial x}(\psi - \eta \bar{\psi}) + \mu_B^2 \frac{b^2}{2}(\psi+\eta \bar{\psi}) e^{b\phi} -i \frac{b}{2}(\psi-\eta \bar{\psi})\partial_x \phi \right|_{y=0}= 0, \\
&\left.\partial_y \phi + 2b(\eta\mu_0 -\mu_B^2) e^{b\phi} +2 i \eta \psi \bar{\psi}\right|_{y=0}=0.
\end{align}
The first equation shows that if $\mu_B=0$ we return to Neumann boundary conditions for the fermions $\left.\psi - \eta \bar{\psi}\right|_{y=0} = 0$, whereas when $\mu_B\to\infty$, we get Dirichlet boundary conditions $\left.\psi + \eta \bar{\psi}\right|_{y=0} = 0$.
\\~\\
Beyond classical field theory, we integrate out the boundary fields $\gamma$ and $f$. The path integral over $f$ yields:
\begin{equation}
\label{susybdy2}
S_{\text{FZZT}}= \frac{1}{2\pi} \oint dx \left( \gamma \partial_x \gamma - \mu_B b \gamma (\psi + \eta \bar{\psi}) e^{\frac{b}{2}\phi} + \mu_B^2 : e^{\frac{b}{2}\phi}:^2 \right).
\end{equation}
The final term is a contact contribution within correlation functions that can be neglected. It will be important however when transforming to the fixed length basis. \\

The boundary (Grassmann) fermion $\gamma(x)$ can be canonically quantized by the Clifford algebra $\left\{\gamma,\gamma\right\}=1$, leading to a description in terms of a 2-dimensional Pauli matrix $\gamma = \frac{1}{\sqrt{2}} \sigma$. This doubles the dimension of the boundary state Hilbert space. This is the usual description used in the super-Liouville literature, where $\gamma^2=1$.
Within the path integral, the boundary fermion field can have periodic (Ramond) boundary conditions: $\gamma(x+2\pi)=\gamma(x)$ or antiperiodic (Neveu-Schwarz) boundary conditions: $\gamma(x+2\pi)= - \gamma(x)$. Since the Lagrangian needs to be single-valued, the last term of \eqref{susybdy} shows that this boundary condition has to match with that of the field $\psi$ and hence with the label NS or R of the boundary state under consideration. 
\\~\\
The boundary fermion field $\gamma$ needs to be path-integrated out in the Liouville amplitude, much like the boson $f$. This can either be done by its equation of motion
\begin{equation}
\label{eqmog}
\partial_x \gamma = \frac{1}{2}\mu_B b  (\psi+\eta \bar{\psi}) e^{\frac{b}{2} \phi},
\end{equation}
solvable as
\begin{equation}
\gamma(x) = \int^x dx' \frac{1}{2} \mu_B b (\psi+\eta \bar{\psi}) e^{\frac{b}{2} \phi} +\gamma_0,
\end{equation}
where the zero-mode $\gamma_0$ has been extracted; or by finding the Green's function of $\partial_x G(x|x') = \delta(x-x')$ as $G(x|x') = \frac{1}{2}\text{sign}(x-x')$, upon first removing the zero-mode $\gamma_0$ again. Using this expression \eqref{eqmog} together with the boundary conditions \eqref{bdyco1}, we find
\begin{align}
\gamma \partial_x \gamma = - i \frac{\eta}{2} \psi \bar{\psi}, \qquad -\mu_B b \gamma(\psi+\eta\bar{\psi})e^{\frac{b}{2}\phi} = i \eta \psi \bar{\psi}.
\end{align}
This results in the total boundary action
\begin{align}
\label{susybdye}
S_L^{\partial} + S_{\text{FZZT}} =  \frac{1}{2\pi} \oint dx \left( ( \mu_B^2 -\eta \mu_0 ) e^{b\phi}  -  \mu_B b  \gamma_0 (\psi+\eta \bar{\psi}) e^{\frac{b}{2}\phi} \right).
\end{align}

Within a canonical framework, the fermionic zero-mode satisfies $\gamma_0^2=1$.\footnote{This follows from $\left\{\gamma(x),\gamma(x)\right\}=1$, combined with the Hamiltonian evolution equation \eqref{eqmog}, which together show that $\left\{\gamma(x),\gamma(x')\right\}=1$ for any $x$ and $x'$. Integrating twice along the entire boundary circle, gives the result.} This reduces the description to the boundary action of \cite{Fukuda:2002bv,Ahn:2002ev} used in the (super)conformal bootstrap method used to determine the correlation functions.
\\~\\
Taking a derivative of this action with respect to $\mu_B$, brings down a factor of 
\begin{equation}
\label{marki1}
M_{1} \equiv \frac{1}{2\pi}b \gamma_0 \oint dx (\psi+\eta \bar{\psi}) e^{\frac{b}{2}\phi},
\end{equation}
which we call the global marking operator $M_1$. We will verify this explicitly in section \ref{s:mark} further on. Notice that no contribution $\sim \mu_B e^{b\phi}$ is written as such a term comes from the auxiliary field $f$ which has a contact-term propagator $\left\langle f(x)f(x')\right\rangle \sim \delta(x-x')$. The resulting marking operator would have a second piece which contributes only contact terms to quantum correlation functions and is usually neglected by suitable choice of external operator dimensions.
\\~\\ 
Next to this, we will need a second local marking operator that combines bosonic and fermionic contributions as
\begin{equation}
\label{marki2}
M_{2}(x) \equiv \left( \frac{b}{2}\gamma_0 (\psi(x)+\eta \bar{\psi}(x)) + i \right) e^{\frac{b}{2}\phi(x)}.
\end{equation}
This second (local) marking operator $M_2(x)$ will play the role of no insertion at all, i.e. the $h= 0$ limit of the boundary two-point function, much like the bosonic marking operators of \cite{Mertens:2020hbs}.
\\~\\
Within the path integral, $\gamma_0$ is a Grassmann number. Performing the $\gamma_0$-integral by bringing down a factor of $\gamma_0 \mu_B \times \frac{1}{2\pi} \oint (\psi+\eta \bar{\psi}) e^{\frac{b}{2}\phi}$, and then putting it back in the exponent by inserting a dummy $\gamma_0$-integral again, gives the suggestive way of writing the path integrand as
\begin{equation}
\label{susybdyt}
\mu_B e^{-\frac{1}{2\pi} \oint dx \left(  (\mu_B^2-\eta \mu_0 ) e^{b\phi} -  \gamma_0 (\psi+\eta \bar{\psi}) e^{\frac{b}{2}\phi} \right)}.
\end{equation}

Transforming to the length basis is done by the integral transform
\begin{equation}
\label{R:integ}
\boxed{
-i\int_{\mathcal{C}} d\mu_B e^{\mu_B^2\ell} \hdots, \qquad \mathcal{C}={\mu_B^2-\eta\mu_0 :-i\infty \to+i\infty} },
\end{equation}
along the hyperboloid $\mathcal{C}$ where $\mu_B^2-\eta\mu_0 :-i\infty \to+i\infty$. Inserting \eqref{susybdyt}, we can evaluate the $\mu_B$-integral into the fixed-length amplitude:
\begin{equation}
\left\langle \hdots\right\rangle_{\ell} = \int \left[\mathcal{D}\phi\right] d\gamma_0 \hdots \delta\left( \ell - \frac{1}{2\pi}\oint e^{b\phi}\right) e^{-\frac{1}{2\pi} \oint dx \left(  -\eta \mu_0 e^{b\phi} -  \gamma_0 (\psi+\eta \bar{\psi}) e^{\frac{b}{2}\phi} \right)}e^{-S_L}.
\end{equation}

For NS-branes, the zero-mode $\gamma_0$ does not exist. Instead one finds the total boundary action
\begin{align}
\label{susybdyens}
S_L^{\partial} + S_{\text{FZZT}} =  \frac{1}{2\pi} \oint dx \left( (\mu_B^2 -\eta \mu_0 ) e^{b\phi}  \right),
\end{align}
leading to the fixed length prescription
\begin{equation}
\label{NSprescri}
-i\int_{\mathcal{C}} d\mu_B^2 e^{\mu_B^2\ell} \hdots, \qquad \mathcal{C}={\mu_B^2-\eta\mu_0 :-i\infty \to+i\infty},
\end{equation}
and correlator:
\begin{equation}
\left\langle \hdots\right\rangle_{\ell} = \int \left[\mathcal{D}\phi\right] \hdots \delta\left( \ell - \frac{1}{2\pi}\oint e^{b\phi}\right) e^{-\frac{1}{2\pi} \oint dx \left(  -\eta \mu_0  e^{b\phi} \right)}e^{-S_L}.
\end{equation}
For the NS-brane, one can mark the boundary by differentiating an amplitude w.r.t. $\mu_B^2$, bringing down $\oint e^{b \phi}$, which is of the same structure as in the bosonic Liouville CFT.
\\~\\
If one consider only brane segments, connected through boundary operator insertions, each brane segment has its own $\mu_{Bi}$ and a different fermion zero-mode $\gamma_{0i}$ (which can be absent). Transforming to the length basis is then done by individually transforming each segment to the length basis, using either \eqref{R:integ} or \eqref{NSprescri} depending on whether the fermion zero-mode is present or not in the segment of interest.

\subsection{Super-Liouville amplitudes with FZZT brane boundaries}
\label{s:liou}
In this appendix, we summarize the super-Liouville amplitudes we will need.
\\~\\
As always in a supersymmetric worldsheet theory, bulk and boundary vertex operators fall in Ramond and Neveu-Schwarz types. \\
The $\mathcal{N}=1$ Liouville bulk one-point functions have been determined through the conformal bootstrap in \cite{Fukuda:2002bv,Ahn:2002ev} as:
\begin{align}
\label{NS:1}
\left\langle V_\alpha \right\rangle_{s_\eta}  &= \Big(\mu \pi \gamma\left( bQ/2 \right) \Big)^{\frac{Q-2\alpha}{2b}} \Gamma\Big( b(\alpha-\frac{Q}{2})\Big) \Gamma\Big(1+\frac{1}{b}(\alpha-Q/2)\Big) \cosh \pi (2\alpha-Q) s, \\
\label{R:1}
\left\langle \Theta^{\epsilon,\bar{\epsilon}}_\alpha \right\rangle_{s_+}&= \Big(\mu \pi \gamma\left(bQ/2 \right) \Big)^{\frac{Q-2\alpha}{2b}} \Gamma\Big(\frac{1}{2} + b(\alpha-\frac{Q}{2})\Big) \Gamma\Big(\frac{1}{2}+\frac{1}{b}(\alpha-\frac{Q}{2})\Big) \cosh \pi (2\alpha-Q) s, \\
\label{R:2}
\left\langle \Theta^{\epsilon,\bar{\epsilon}}_\alpha \right\rangle_{s_-} &= \delta_{\epsilon,\bar{\epsilon}}\epsilon \Big(\mu \pi \gamma\left( bQ/2 \right) \Big)^{\frac{Q-2\alpha}{2b}} \Gamma\Big(\frac{1}{2} + b(\alpha-\frac{Q}{2})\Big) \Gamma\Big(\frac{1}{2}+\frac{1}{b}(\alpha-\frac{Q}{2})\Big) \sinh \pi (2\alpha-Q) s,
\end{align}
where the parameter $s$ is related to the boundary cosmological constant $\mu_B$ by the relations
\begin{equation}
\label{muba}
\mu_B = \kappa \begin{cases}
\cosh \pi  b s ,\qquad \eta=+1,\\
\sinh \pi b s, \qquad \eta=-1,
\end{cases}
\qquad \kappa= \sqrt{\frac{2\mu}{\cos \frac{\pi b^2}{2}}}.
\end{equation}
We leave implicit the worldsheet coordinate dependence of this correlator. It will disappear in any case in the end when we include the (super)ghost and matter contributions. \\
We used the notation $V_\alpha = e^{\alpha \phi}$ for the Neveu-Schwarz bulk operator insertion. The Ramond puncture contains a spin field insertion:
\begin{equation}
\Theta^{\epsilon,\bar{\epsilon}}_\alpha = \sigma^{\epsilon \bar{\epsilon}}e^{\alpha \phi}.
\end{equation}
The NS (R) sector one-point function is only non-zero provided the fermions satisfy antiperiodic (NS), respectively periodic (R) boundary conditions around the boundary circle, directly relating the fermionic sector of the boundary to the type of bulk insertion.
\\~\\
The boundary two-point function for two vertex operators $V_\beta$ in $\mathcal{N}=1$ super-Liouville CFT is of the form:
\begin{equation}
\left\langle V_{\beta_1}(x)V_{\beta_1}(0)\right\rangle = \frac{\delta(\beta_1+\beta_2-Q) + \mathbf{d}(\beta|s_1s_2) \delta(\beta_1-\beta_2)}{\left|x\right|^{2\Delta_{\beta_1}}},
\end{equation}
where the reflection coefficient $\mathbf{d}(\beta|s_1s_2)$ is the dynamical information that will be discussed below. It multiplies a $\delta(0)$ factor corresponding to the infinite Liouville volume. Within the full theory, we mod by the (super)conformal Killing group of the twice-punctured disk and get a finite factor from this. This is discussed in the main text. Moreover, the worldsheet coordinate dependence will be cancelled by similar factors coming from the matter- and superghost boundary two-point functions. First, we focus solely on the super-Liouville amplitudes themselves.
\\~\\
The NS-sector primary boundary vertex operators are denoted by
\begin{equation}
\label{defNs}
B^\beta \equiv e^{\frac{\beta}{2}\phi}, \qquad \Lambda^\beta \equiv \frac{\beta}{2}(\psi +\eta \bar{\psi})B^\beta.
\end{equation}
The boundary two-point functions with FZZT-boundary parameters $s_\eta$ and $s'_{\eta'}$ are denoted as:
\begin{align}
\begin{tikzpicture}[scale=1, baseline={([yshift=0cm]current bounding box.center)}]
\draw[thick] (0,0) circle (0.6);
\draw[fill,black] (-0.6,0) circle (0.1);
\draw[fill,black] (0.6,0) circle (0.1);
\draw (0,0.8) node {\small \color{red}$s_\eta$};
\draw (0,-0.8) node {\small \color{red}$s'_{\eta'}$};
\draw (-1.3,0) node {\small $B^\beta_{s_\eta s'_{\eta'}}$};
\draw (1.3,0) node {\small $B^\beta_{s_\eta s'_{\eta'}}$};
\end{tikzpicture} = d(\beta|s_\eta, s'_{\eta'}) \qquad 
\begin{tikzpicture}[scale=1, baseline={([yshift=0cm]current bounding box.center)}]
\draw[thick] (0,0) circle (0.6);
\draw[fill,black] (-0.6,0) circle (0.1);
\draw[fill,black] (0.6,0) circle (0.1);
\draw (0,0.8) node {\small \color{red}$s_\eta$};
\draw (0,-0.8) node {\small \color{red}$s'_{\eta'}$};
\draw (-1.3,0) node {\small $ \Lambda^\beta_{s_\eta s'_{\eta'}}$};
\draw (1.3,0) node {\small $ \Lambda^\beta_{s_\eta s'_{\eta'}}$};
\end{tikzpicture} = d'(\beta|s_\eta, s'_{\eta'})
\end{align}
and are characterized by a (super)Virasoro representation label $s$ and the sign $\eta$. To find non-zero, one needs $\eta$ and $\eta'$ of equal sign. 
\\~\\
The two R-sector vertex operators are constructed by applying the spin field $\sigma^{\epsilon}$:
\begin{equation}
\label{defR}
\Theta^{\epsilon \beta} \equiv \sigma^\epsilon B^{\beta}, \qquad \epsilon = \pm,
\end{equation}
and the R-sector two-point function is:
\begin{align}
\begin{tikzpicture}[scale=1, baseline={([yshift=0cm]current bounding box.center)}]
\draw[thick] (0,0) circle (0.6);
\draw[fill,black] (-0.6,0) circle (0.1);
\draw[fill,black] (0.6,0) circle (0.1);
\draw (0,0.8) node {\small \color{red}$s_\eta$};
\draw (0,-0.8) node {\small \color{red}$s'_{\eta'}$};
\draw (-1.3,0) node {\small $\Theta^{\epsilon\beta}_{s_\eta s'_{\eta'}}$};
\draw (1.3,0) node {\small $\Theta^{\epsilon\beta}_{s_\eta s'_{\eta'}}$};
\end{tikzpicture} = \tilde{d}(\beta, \epsilon |s_\eta, s'_{\eta'})
\end{align}
One needs $\eta$ and $\eta'$ to be opposite sign to find non-zero: R-sector boundary operators change the chirality of the boundary.

With these definitions, Fukuda and Hosomichi found the following boundary two-point functions \cite{Fukuda:2002bv}:\footnote{Their definition of fermionic boundary condition is called $\zeta$ which is related as $\zeta = - \eta$. We have absorbed some overall minus signs into the definitions of these amplitudes.}
\begin{align}
\label{fukhoso}
d(\beta|s_+, s'_{+}) &= \frac{(\pi \mu \gamma(bQ/2)b^{1-b^2})^{\frac{Q-2\beta}{2b}}\Gamma_{\NS}(2\beta-Q)\Gamma_{\NS}(Q-2\beta)^{-1}}{S_{\NS}(\beta + i s +is') S_{\NS}(\beta-is+is')S_{\NS}(\beta+is-is')S_{\NS}(\beta-is-is')}, \\
d'(\beta|s_+, s'_{+}) &= \frac{(\pi \mu \gamma(bQ/2)b^{1-b^2})^{\frac{Q-2\beta}{2b}}\Gamma_{\NS}(2\beta-Q)\Gamma_{\NS}(Q-2\beta)^{-1}}{S_{\R}(\beta + i s +is') S_{\R}(\beta-is+is')S_{\R}(\beta+is-is')S_{\R}(\beta-is-is')}, \nonumber\\
d(\beta|s_-, s'_{-}) &= \frac{(\pi \mu \gamma(bQ/2)b^{1-b^2})^{\frac{Q-2\beta}{2b}}\Gamma_{\NS}(2\beta-Q)\Gamma_{\NS}(Q-2\beta)^{-1}}{S_{\R}(\beta + i s +is') S_{\NS}(\beta-is+is')S_{\NS}(\beta+is-is')S_{\R}(\beta-is-is')},\nonumber \\
d'(\beta|s_-, s'_{-}) &= \frac{(\pi \mu \gamma(bQ/2)b^{1-b^2})^{\frac{Q-2\beta}{2b}}\Gamma_{\NS}(2\beta-Q)\Gamma_{\NS}(Q-2\beta)^{-1}}{S_{\NS}(\beta + i s +is') S_{\R}(\beta-is+is')S_{\R}(\beta+is-is')S_{\NS}(\beta-is-is')}, \nonumber
\end{align}
and the Ramond boundary two-point function:
\begin{align}
\label{fukhoso2}
\tilde{d}(\beta, \epsilon |s_+, s'_-) &= i\frac{(\pi \mu \gamma(bQ/2)b^{1-b^2})^{\frac{Q-2\beta}{2b}}\Gamma_{\R}(2\beta-Q)\Gamma_{\R}(Q-2\beta)^{-1}}{S_{\NS}(\beta + i s +i \epsilon s') S_{\NS}(\beta-is+i\epsilon s')S_{\R}(\beta+is-i\epsilon s')S_{\R}(\beta-is-i\epsilon s')},
\end{align}
where the special functions are defined in Appendix \ref{app:spec}.

\subsection{A comment on the Ramond partition function}
\label{app:rampf}
\begin{figure}[h]
\centering
\includegraphics[width=0.9\textwidth]{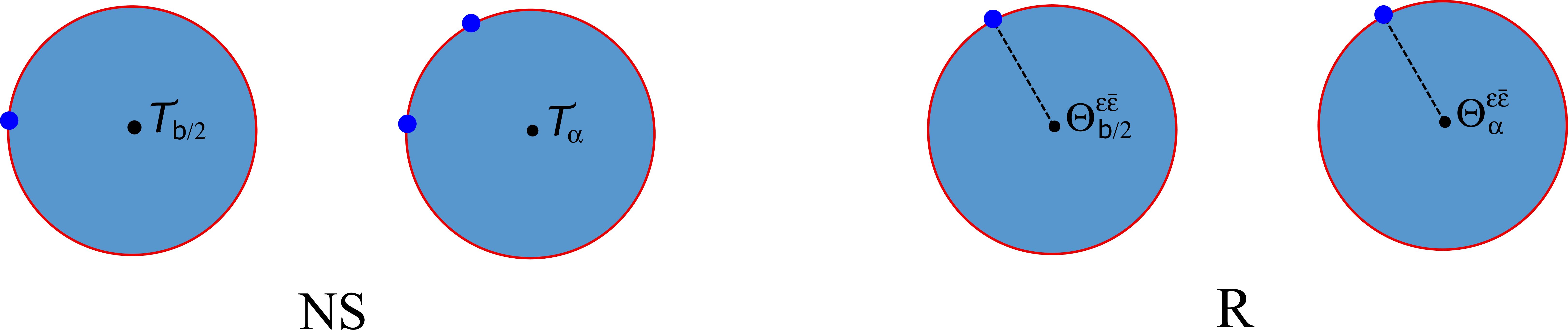}
\caption{Left diagrams: partition function $Z_{\NS}$ and bulk one-point functions $\left\langle \mathcal{T}_\alpha \right\rangle$. The blue dots on the boundary represent the markings. Right diagrams: $Z_{\R}$ and $\left\langle \Theta_\alpha\right\rangle$. The NS partition function is once-marked, whereas the NS bulk one-point function is twice marked. All R-sector cases have a single marking due to the branch cut of the spin field.
}
\label{onepointSum}
\end{figure}

We can get the Ramond brane partition functions also by applying a single bulk spin field $\Theta^{\epsilon,\bar{\epsilon}}_\alpha$ with $\theta=1$ or $\alpha=b/2$. Indeed, inserting this value in \eqref{RR1} and \eqref{RR2}, we reproduce the fixed length partition functions \eqref{inttodo} and \eqref{inttodo2}. This means this bulk operator is not really doing anything besides marking the boundary: the amplitudes are marked since the branch cut from the spin field hits the boundary somewhere. These results illustrate the equivalence:
\begin{equation}
\Theta^{\epsilon,\bar{\epsilon}}_{\frac{b}{2}} \sim \sigma^{\epsilon \bar{\epsilon}}e^{\frac{b}{2} \phi} \qquad\to \qquad M_1(x) = \gamma_0(\psi + \eta \bar{\psi})e^{\frac{b}{2} \phi}.
\end{equation}
This also shows that applying a single boundary marking operator $M_1$ is effectively changing the fermionic boundary condition from NS to R.
\\~\\
The full situation is summarized in figure \ref{onepointSum}.

In the bosonic case, this same mnemonic applies to the bulk one-point function $\sim \cosh \pi (2\alpha -Q)s$, where indeed setting $\alpha = b/2$ in $e^{2\alpha \phi}$ gives the $s$-dependence of the marked partition function.

\subsection{Consistency check on marking operators}
\label{s:markapp}
As a check on the expressions \eqref{2mark} within $\mathcal{N}=1$ Liouville supergravity, take the $s_1 \to s_2$ limit of the second and fourth expressions:
\begin{align}
d'(b|s_+s_+) = \frac{\sinh\frac{\pi}{b}s_1}{b^2\kappa\sinh \pi b s_1}, \qquad d'(b|s_-s_-) = \frac{\cosh\frac{\pi}{b}s_1}{b^2 \kappa \cosh \pi b s_1}.
\end{align}
This amplitude corresponds to two insertions of \eqref{marki1}\footnote{The $\gamma_0$ insertions in $M_1$ cancel pairwise due to $\gamma_0^2=1$.}
\begin{equation}
M_1 = \oint dx \Lambda^b(x) = \left(c e^{-\varphi}\right)\Lambda^b(0), \qquad \Lambda^b \equiv \frac{b}{2}(\psi +\eta \bar{\psi})e^{\frac{b}{2}\phi},
\end{equation}
where we have written the operator both in its gauge-invariant and in its gauxe-fixed form, including the superghosts. Since insertions of $M_1$ correspond to marking the boundary by taking derivatives w.r.t. $\mu_B$, this is to be compared to the twice-marked Ramond boundary amplitudes, obtained by marking \eqref{mark1} and \eqref{mark2} once more:
\begin{align}
Z(\mu_B)^{\M\M}_{\eta=+1} &\equiv \partial_{\mu_B}Z(\mu_B)^{\M}_{\eta=+1} = \frac{\sinh \frac{1}{b^2}\text{arccosh} \mu_B}{b^2\sqrt{\mu_B^2-\kappa^2}} = \frac{\sinh \frac{\pi}{b}s}{b^2\kappa \sinh \pi b s}, \\
Z(\mu_B)^{\M\M}_{\eta=-1} &\equiv \partial_{\mu_B}Z(\mu_B)^{\M}_{\eta=-1} = \frac{\cosh \frac{1}{b^2}\text{arcsinh} \mu_B}{b^2\sqrt{\kappa^2+\mu_B^2}} = \frac{\cosh \frac{\pi}{b}s}{b^2 \kappa \cosh \pi b s},
\end{align}
which indeed agree.

\section{Ramond operators}
\label{s:ramond}
In this section, we consider Ramond boundary operators. They cause a change in fermionic sector between $\eta=+1$ to $\eta=-1$. Morever they cause also an effective change from NS to R sector in terms of how the transform to the fixed length basis should be implemented, \eqref{Rint} or \eqref{NSint}, in the different segments. We first discuss the generic weight Ramond insertion from the continuum approach. Then we discuss the special class of operator insertions corresponding to the minimal superstring Ramond operators. Finally, we present a way of getting them starting with super-Liouville CFT between a pair of identity ZZ branes.

\subsection{Continuum approach}
We consider $\left\langle \Sigma^\beta  \Sigma^\beta  \right\rangle_{+-}$ on a boundary that changes fermionic boundary conditions between $\eta=\pm 1$. We combine the super-Liouville two-point functions from \eqref{fukhoso2} as 
\begin{equation}
D^{\beta}_{s,s'} \equiv \tilde{d}(\beta,+|s_+s_-') + \tilde{d}(\beta,-|s_+s_-'),
\end{equation}
and take as the \textbf{s}discontinuity a combination of plus and minus signs:
\begin{align}
\label{Discdefspin}
\text{sDisc } D^{\beta}_{s,s'} \equiv D^{\beta}_{s+\frac{i}{2b},s'+\frac{i}{2b}} + D^{\beta}_{s-\frac{i}{2b},s'+\frac{i}{2b}} - D^{\beta}_{s+\frac{i}{2b},s'-\frac{i}{2b}} - D^{\beta}_{s-\frac{i}{2b},s'-\frac{i}{2b}}.
\end{align}
This corresponds to picking the brane with label $s$ to be in the R-sector, but the $s'$-brane is in the NS-sector. Using this, one computes:
\begin{align}
\text{sDisc } D^{\beta}_{s,s'} = &\left[-16i \sin \frac{\pi}{\beta}\left(\beta-\frac{1}{2b}\right)  \cos \frac{\pi \beta}{b}\right] \nonumber \\
&\hspace{-1cm}\times\cosh \frac{\pi s}{b} \sinh \frac{\pi s'}{b} \left( \tilde{d}(\beta + \frac{1}{b},+|s_-s_+') +  \tilde{d}(\beta + \frac{1}{b},-|s_-s_+')  \right),
\end{align}
where one sees that the $s$ and $s$' branes generate a different spectral factor, signaling indeed that the Ramond vertex operators cause transitions between them. Note that the discontinuity has minus signs for the $s'$-parameter, which signals a spin field insertion is causing a transition from a R boundary brane segment (having a fermionic zero-mode) into a NS segment (without a fermionic zero-mode), recalling the bosonic Liouville theory has minus signs in the discontinuity.
\\~\\
Within Liouville supergravity, the Ramond boundary tachyon vertex operators are defined as
\begin{align}
\mathcal{B}^{+}_{\epsilon \beta_M} &= (\pi \mu \gamma(bQ/2))^{\frac{2\beta-Q}{4b}}\Gamma(\frac{b}{2}(Q-2\beta)+1/2) \,\, \left(c e^{-\varphi/2}\right) \left[\sigma^{\epsilon} e^{\frac{\beta}{2} \phi} e^{\beta_M \chi}+ (\text{superpartner})\right], \\
\mathcal{B}^{-}_{\epsilon \beta_M} &= (\pi \mu \gamma(bQ/2))^{\frac{2\beta-Q}{4b}}\Gamma(\frac{1}{2b}(Q-2\beta)+1/2) \,\, \left(c e^{-\varphi/2}\right) \left[\sigma^{\epsilon} e^{\frac{\beta}{2} \phi} e^{(q-\beta_M) \chi} + (\text{superpartner})\right],
\end{align}
which include explicit Ramond-sector legpole factors. The insertion of a pair of these Ramond operators leads, after an analogous computation as for the NS operators of subsection \ref{s:twop}, to the answer:
\begin{align}
\label{final3}
&\left\langle \mathcal{B}^{+}_{\epsilon \beta_M} \mathcal{B}^{-}_{\epsilon \beta_M}\right\rangle_{\ell_1,\ell_2} = \int ds_1ds_2 \rho_{\R,\eta=+1}(s_1) \rho_{\NS,\eta=-1}(s_2) e^{-\mu_B(s_1) \ell_1} e^{-\mu_B(s_2) \ell_2}\nonumber \\
&\times\left[ \frac{S_{\NS}(\beta_M + i s_1 \pm i s_2)S_{\R}(\beta_M  - i s_1 \pm  i s_2)}{S_{\R}(2\beta_M)} + \frac{S_{\R}(\beta_M + i s_1 \pm i s_2)S_{\NS}(\beta_M  - i s_1 \pm  i s_2)}{S_{\R}(2\beta_M)}\right],
\end{align}
where $\rho^{\NS}_{\eta=-1}(s) = \sinh 2 \pi  b s \sinh \frac{\pi}{b}s$. 
\\~\\
Within the super-JT limit, we can use
\begin{align}
\label{llast}
\tilde{d}(\beta,\epsilon |s_+, s'_-) \to \Gamma(h +i\epsilon k_2 \pm i k_1 ) \Gamma(\frac{1}{2}+ h - i \epsilon k_2 \pm i k_1), \qquad S_{\R}(2\beta_M) \to \Gamma(2h + \frac{1}{2}).
\end{align}
In the JT limit, one changes between the super JT and the bosonic JT theory upon crossing the bilocal line. It would be interesting to understand whether this result can be understood directly in the JT theory, without going through its embedding in Liouville supergravity. \\
For $h \in - \mathbb{N}/2 - \frac{1}{4}$, the Ramond sector bilocal \eqref{final3} becomes degenerate. These degenerate values of $h$ need to be considered separately. Within the framework of the minimal superstring, we consider them next.

\subsection{Minimal superstring Ramond operators}
\label{s:ramms}
Minimal superstring Ramond operators are located at $\beta_M = 2bh = -bj$ where $j \in \mathbb{N}+\frac{1}{2} = \frac{1}{2}, \frac{3}{2} \hdots $. In this case, we can evaluate explicitly:
\begin{align}
\label{gendegSUSY}
\tilde{d}\left(b+bj,+|s_{1+},s_{2-} \right) &= \frac{\cosh \frac{\pi}{b}s_1 + (-)^{j+1/2} i \sinh \frac{\pi}{b}s_2}{\prod_{m=-j}^{j} \left( \sinh \pi b s_2 - (-)^{j-m}\sinh \pi b (s_1+ imb)\right)} \\
&= \frac{\cosh \frac{\pi}{b}s_1 + (-)^{j+1/2} i \sinh \frac{\pi}{b}s_2}{\prod_{m=-j}^{j} \left( \cosh \pi b s_1 + (-)^{j-m}\cosh \pi b (s_2 + imb)\right)}, \nonumber \\
\tilde{d}\left(b+bj,-|s_{1+},s_{2-} \right) &= \frac{\cosh \frac{\pi}{b}s_1 - (-)^{j+1/2} i \sinh \frac{\pi}{b}s_2}{\prod_{m=-j}^{j} \left( \sinh \pi b s_2 + (-)^{j-m}\sinh \pi b (s_1+ imb)\right)},
\end{align}
which, after picking up the residues and the discontinuity, leads to
\begin{align}
\left\langle  \mathcal{B}_{1,2j+1}\mathcal{B}_{1,2j+1} \right\rangle_{\ell_1,\ell_2} &= \int_{0}^{+\infty} ds_1 \cosh \pi b s_1 \, \text{\textbf{s}Disc}[\mathcal{R}^{\R}_{\eta=+1}(s_1)] \, e^{-\kappa^2\ell_1 \sinh^2 \pi b s_1} \\
&\times \sum_{n=-j}^{j} \frac{(j+1/2)!(-1)^{j}(-)^{n-\frac{1}{2}} e^{-\kappa^2\ell_2 \cosh^2 \pi b (s_1+inb)}}{\prod_{\stackrel{m=-j}{m\neq n}}^{j} (\cosh \pi b (s_1+i nb) - (-)^{n-m}\cosh \pi b (s_1+imb))} \nonumber \\
&-i \int_{0}^{+\infty} ds_2  \sinh 2\pi b s_2\,  \text{Disc}[\mathcal{R}^{\NS}_{\eta=-1}(s_2)] \, e^{-\kappa^2\ell_2 \cosh^2 \pi b s_2} \nonumber \\
&\times \sum_{n=-j}^{j} \frac{(j+1/2)!(-1)^{j}(-)^{n-\frac{1}{2}} e^{-\kappa^2\ell_1 \sinh^2 \pi b (s_1+inb)}}{\prod_{\stackrel{m=-j}{m\neq n}}^{j} (\sinh \pi b (s_2+i nb) - (-)^{n-m}\sinh \pi b (s_2+imb))} ,\nonumber
\end{align}
in terms of the densities of states:
\begin{align}
\rho^{\R}_{\eta=+1}(s) &= \cosh \pi  b s \, \text{\textbf{s}Disc}[\mathcal{R}^{\R}_{\eta=+1}(s)] = \cosh \pi  b s \cosh \frac{\pi}{b}s, \\
\rho^{\NS}_{\eta=-1}(s) &= \sinh 2 \pi b s \, \text{Disc}[\mathcal{R}^{\NS}_{\eta=-1}(s)] = \sinh 2 \pi  b s \sinh \frac{\pi}{b}s,
\end{align}
which describes a transition from an R-sector to an NS-sector. We indeed recognize the NS density of states from \eqref{NSpf}.

\subsection{Spin fields from the ZZ-ZZ brane perspective}
\label{s:aside}
\begin{figure}[h]
\centering
\includegraphics[width=0.3\textwidth]{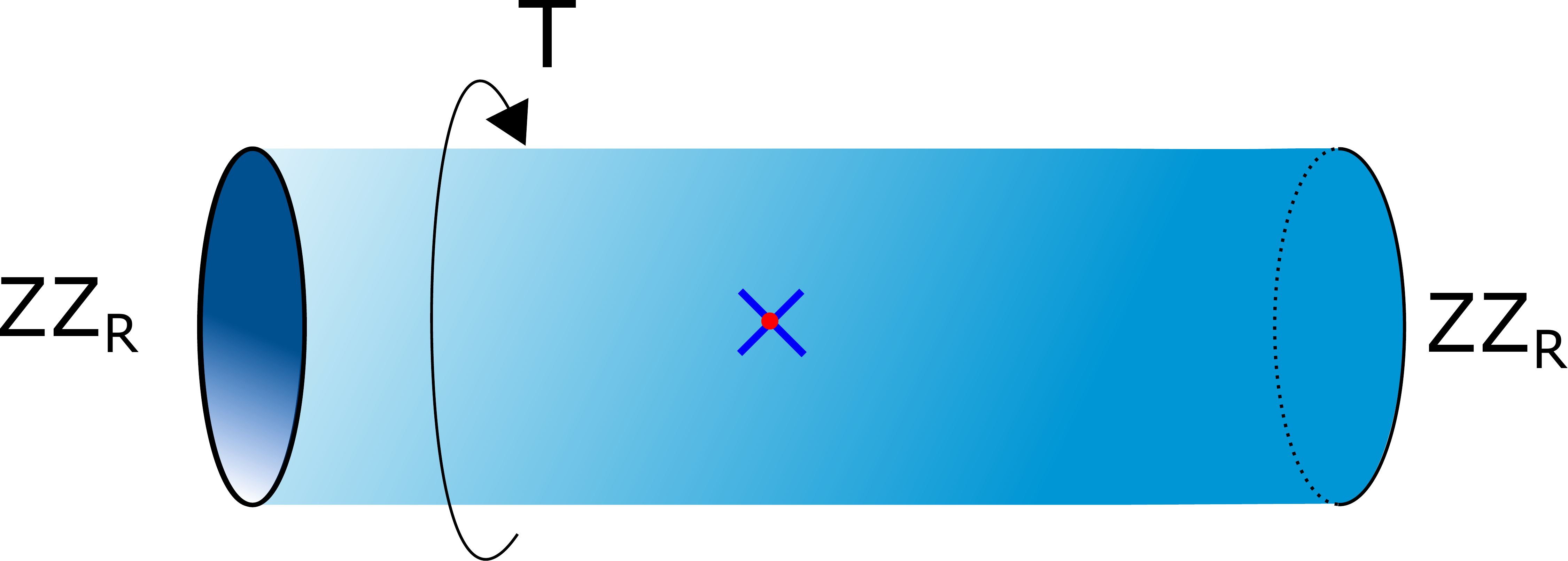}
\caption{Super-Liouville primary field insertion $e^{\alpha \phi}$ between two vacuum $R$-sector branes. The Schwarzian bilocal correlators are obtained in the double scaling limit where the circumference of the cylinder $T \to0$ while $c\to +\infty$ keeping the product $\frac{cT}{24\pi} = C$ fixed as the Schwarzian coupling constant.}
\label{superLiouvilleZZ}
\end{figure}

\begin{figure}[h]
\centering
\includegraphics[width=0.8\textwidth]{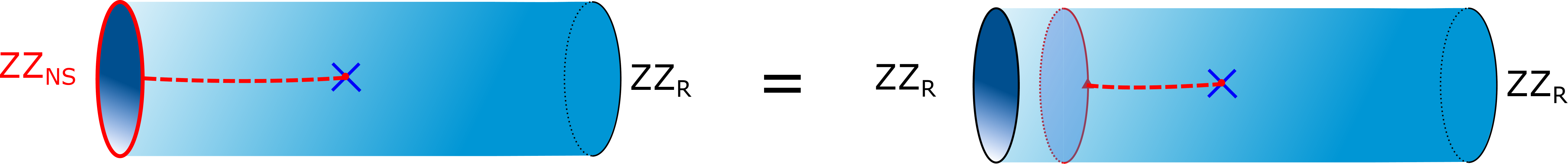}
\caption{Left: Insertion of spin operator between a ZZ$_{\NS}$ and a ZZ$_{\R}$ brane. Right: Insertion of Verlinde loop (topological defect) allows one to view this as a single defect consisting of the combination $\mathcal{W} \Theta^{\pm\pm}_{\alpha_2}$ .}
\label{branchcut}
\end{figure}

\begin{figure}[H]
\centering
\includegraphics[width=0.95\textwidth]{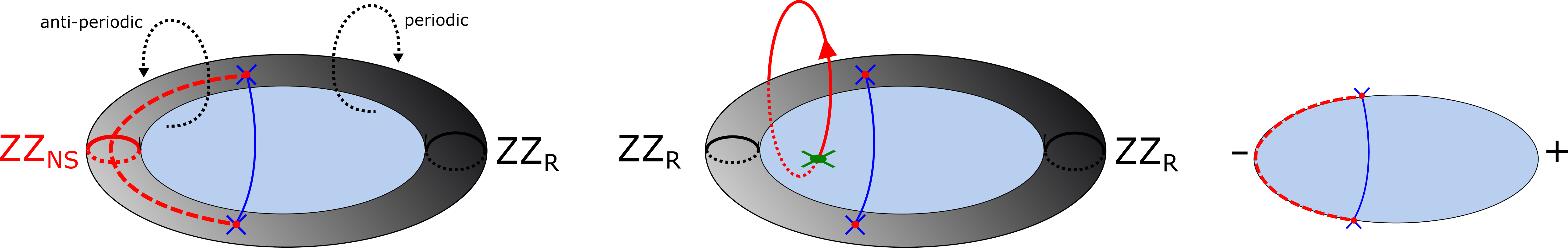}
\caption{Left: Doubling the ZZ-ZZ cylinder into a chiral torus with branch cut (red dashed) between the operator and its image. Fermions have anti-periodic (NS) boundary conditions around the small circle between the ends of the bilocal, and periodic (R) boundary conditions elsewhere. Middle: Verlinde loop (= Chern-Simons Wilson loop) linked with the bilocal Wilson line, and soaking up the branch cut. Right: JT limit where the branch cut becomes the FZZT boundary condition with $\eta = -1$, and the usual $\eta=+1$ elsewhere.}
\label{branchcut2}
\end{figure}

In this complementary section, we elaborate on a perspective that was developed in \cite{Mertens:2017mtv} and \cite{Mertens:2019tcm} on Schwarzian amplitudes. In that work, the strategy was to consider the cylinder amplitude between a pair of vacuum branes in solely the Liouville CFT. Within that construction, a double-scaling limit where the central charge $c\to + \infty$, the amplitude reduces to the Schwarzian partition function. \\
The extension to Schwarzian bilocal operators is made by taking the same double-scaling limit of the cylinder amplitude with a single Liouville primary vertex operator inserted in the middle. One of the main benefits of this approach is that several generalizations are readily studied, exploiting the many investigations of Liouville and super-Liouville CFT done during the past decades. Indeed, we computed the super-Schwarzian bilocal correlators by inserting a super-Liouville primary vertex operator $e^{\alpha \phi}$ between two Ramond identity branes ZZ$_{\R}$ (Figure \ref{superLiouvilleZZ}).
In this work, we have seen that there is also a boundary spin field insertion in Liouville supergravity, and the resulting bilocal correlator \eqref{final3} was not obtained before in \cite{Mertens:2017mtv}. In this section we remedy this and study spin field insertions in the super-Liouville CFT between vacuum branes and match these to expressions we found in this work using the Liouville supergravity approach.
\\~\\
We compute the amplitude with the insertion of a super-Liouville spin field operator between one ZZ$_{\NS}$ brane and one ZZ$_{\R}$ brane (Figure \ref{branchcut} left). 
Within the minisuperspace limit, we expand the boundary states as
\begin{equation}
\left|\text{ZZ}_{\NS}\right\rangle = \int dP \Psi_{\NS}(P)\left|P, \NS\right\rangle \hspace{-0.05cm} \rangle, \qquad 
\left|\text{ZZ}_{\R}\right\rangle = \int dP \Psi_{\R}(P)\left|P,\R \right\rangle \hspace{-0.05cm} \rangle, \\
\end{equation}
where the Ishibashi states are expanded in primary states and their descendants as:
\begin{equation}
\left|P, \R\right\rangle \hspace{-0.05cm} \rangle= |\Theta^{++}_{Q/2+iP}\rangle + |\Theta^{--}_{Q/2+iP}\rangle + (\text{descendants}), \qquad \left|P, \NS\right\rangle \hspace{-0.05cm} \rangle= \left|V_{Q/2 + i P}\right\rangle + (\text{descendants})
\end{equation}
Within the double-scaling limit where the cylinder circumference $T\to0$, the (closed-channel) Ishibashi states above are dominated by their primaries. For a single super-Liouville vertex operator insertion between the two branes, we can then compute the amplitude as the three-point function on the sphere. In our case, the operator is in the Ramond sector and can be written as $\Theta^{\epsilon\bar{\epsilon}}_\alpha \equiv \sigma^{\epsilon \bar{\epsilon}} V_\alpha$ where $V_\alpha \equiv e^{\alpha \phi}$ are the Liouville primaries and $\sigma^{\epsilon \bar{\epsilon}}$ is the spin field.  We then denote the sphere three-point functions of interest as:
\begin{align}
\tilde{C}_1 = \left\langle  V_{\alpha_1}\Theta^{\pm\pm}_{\alpha_3}\Theta^{\mp\mp}_{\alpha_2}\right\rangle, \qquad \qquad 
\tilde{C}_2 = \left\langle  V_{\alpha_1}\Theta^{\pm\pm}_{\alpha_3}\Theta^{\pm\pm}_{\alpha_2}\right\rangle,
\end{align}
and the amplitude of interest becomes:
\begin{equation}
\label{sptoco}
\left\langle \text{ZZ}_{\NS}\right| \Theta^{\pm\pm}_{\alpha_3} \left|\text{ZZ}_{\R}\right\rangle = \int dPdP' \Psi^*_{\NS}(P) \Psi_{\R}(P') \left( \tilde{C}_1 + \tilde{C}_2 \right).
\end{equation}
The $\mathcal{N}=1$ DOZZ formula is given by \cite{Rashkov:1996np,Poghosian:1996dw,Fukuda:2002bv}\footnote{$\Upsilon_{\NS}(x)= \Upsilon\left(\frac{x}{2}\right)\Upsilon\left(\frac{x+Q}{2}\right)$, $\Upsilon_{\R}(x)= \Upsilon\left(\frac{x+b}{2}\right)\Upsilon\left(\frac{x+b^{-1}}{2}\right)$.}
\begin{align}
\tilde{C}_1 &= \frac{\Upsilon_{\NS}'(0)\Upsilon_{\NS}(2\alpha_1)\Upsilon_{\R}(2\alpha_2)\Upsilon_{\R}(2\alpha_3)}{\Upsilon_{\R}(\alpha_{1+2+3}-Q)\Upsilon_{\NS}(\alpha_{1+2-3})\Upsilon_{\R}(\alpha_{2+3-1})\Upsilon_{\NS}(\alpha_{3+1-2})}, \\[2mm]
\tilde{C}_2 &= \frac{\Upsilon_{\NS}'(0)\Upsilon_{\NS}(2\alpha_1)\Upsilon_{\R}(2\alpha_2)\Upsilon_{\R}(2\alpha_3)}{\Upsilon_{\NS}(\alpha_{1+2+3}-Q)\Upsilon_{\R}(\alpha_{1+2-3})\Upsilon_{\NS}(\alpha_{2+3-1})\Upsilon_{\R}(\alpha_{3+1-2})}.
\end{align}
In order to implement the double-scaling limit to JT gravity, we need to parametrize the 2d CFT parameters in the following way \cite{Mertens:2017mtv}: $\alpha_1 = Q/2 + 2i bk_1$, $\alpha_2 = Q/2 + 2ib k_2$ and $\alpha_3 = 2b h$, where $k_1$, $k_2$ and $h$ are kept finite as $b\to 0$. Using the limits:
\begin{align}
\Upsilon_{\NS} (Q+2iP_1) \to \frac{1}{\Gamma(2ik_1)} \qquad \Upsilon_{\R} (Q+2iP_2) \to \frac{1}{\Gamma(1/2-2ik_2)}, \qquad \Upsilon_{\R} (4h b) \to \frac{1}{\Gamma(2 h + \frac{1}{2})},
\end{align}
we find the limiting DOZZ-formulas become:
\begin{align}
\tilde{C}_1 &\to \frac{\Gamma(h - i k_2 \pm i k_1) \Gamma(\frac{1}{2} + h +ik_2 \pm i k_1)}{\Gamma(2ik_1)\Gamma(1/2-2ik_2)\Gamma(2h+ \frac{1}{2})}, \\
\tilde{C}_2 &\to \frac{\Gamma(h + i k_2 \pm i k_1) \Gamma(\frac{1}{2} + h - ik_2 \pm i k_1)}{\Gamma(2ik_1)\Gamma(1/2-2ik_2)\Gamma(2h+ \frac{1}{2})},
\end{align}
which matches with \eqref{llast}, with the correspondence of the boundary spin operator $\Theta^{- \beta} \leftrightarrow \tilde{C}_1$ and $\Theta^{+ \beta} \leftrightarrow \tilde{C}_2$. Taking the sum $\tilde{C}_1 + \tilde{C}_2$ is then identified with the sum in the quantity $\tilde{d}(\beta, - |s_+,s'_-) + \tilde{d}(\beta, + |s_+, s'_-)$. \\ 
Using the limits of the brane wavefunctions
\begin{align}
\Psi_{\NS}(P) &\sim \frac{1}{iP \Gamma(-iPb)\Gamma(-iP/b)} \to \frac{1}{\Gamma(-2i k)}, \\
\Psi_{\R}(P) &\sim \frac{1}{ \Gamma(1/2-iPb)\Gamma(1/2-iP/b)} \to \frac{1}{\Gamma(1/2-2i k)},
\end{align}
the amplitude \eqref{sptoco} is then
\begin{align}
\frac{1}{Z} \int_{0}^{+\infty} dk_1 dk_2 &k_1\sinh(2\pi k_1) \cosh(2\pi k_2)  e^{-\tau k_1^2}e^{-(\beta-\tau)k_2^2} \nonumber \\
&\times \left[\frac{\Gamma(h - i k_2 \pm i k_1) \Gamma(\frac{1}{2} + h +ik_2 \pm i k_1)}{\Gamma\left(2h+\frac{1}{2}\right)} +(k_2 \to-k_2)\right],
\end{align}
with a change in spectral density from the NS density for $k_1$ to the R density for $k_2$. Notice the presence of the $\Gamma\left(2h+\frac{1}{2}\right)$ factor, signaling the presence of Ramond degenerate boundary operators when $ h =-\frac{1}{4}, - \frac{3}{4} \hdots $. This expression matches the JT limit of the Liouville supergravity expression we found in \eqref{final3}. \\
Setting $h=0$ gives an insertion of a pure spin field, and this can be viewed as the most basic operator required to change fermionic boundary conditions.
\\~\\
In \cite{Mertens:2019tcm} we framed this approach of computing JT correlators in a larger geometric bulk picture, and it is insightful to generalize that discussion to include the spin operators discussed here. The general picture described there is as follows. The Liouville cylinder is first doubled into a chiral torus. The double-scaling limit then shrinks that torus into a long narrow tube that degenerates into the boundary of the disk. The \emph{exterior} region of the cylinder is in the process projected onto its angular zero-mode and can be identified as the interior of the 2d disk region. We checked this picture in \cite{Mertens:2019tcm} by mapping defects in the JT disk bulk to Chern-Simons Wilson loops encircling the cylindrical tube to Verlinde loops \cite{Verlinde:1988sn} contracted onto the Liouville cylindrical tube.  \\
This story now gets additional ingredients as follows. \\
There is a branch cut between the spin field insertion and the NS-brane (figure \ref{branchcut} left). The latter can also be found by applying a Verlinde loop operator $\mathcal{W}$ on the usual ZZ$_{\R}$ brane, i.e. $\left|\text{ZZ}_{\NS}\right\rangle = \mathcal{W} \left|\text{ZZ}_{\R}\right\rangle$. Since Verlinde loops can be viewed as topological defects in CFT \cite{Drukker:2010jp}, one can view the entire construction as the application of a combined operator on the CFT system, consisting of R spin field plus topological defect to soak up the jump on the branch cut at the other end  (figure \ref{branchcut} right). One can even move the defect up to the vertex operator. The topological defect operator insertion $D(k)$ itself is given by the expression:
\begin{equation}
\left\langle k\right|\mathcal{W}\left|k\right\rangle = D(k) = \frac{S_{0\NS}^{\, P}}{S_{0\R}^{\,P}} = \frac{\left|\Psi_{\NS}(P) \right|^2}{\left|\Psi_{\R}(P)\right|^2} = \frac{\sinh(  \pi P/b) \sinh(  \pi b P)}{\cosh(\pi P/b) \cosh( \pi  b P)} \to \frac{k \sinh(2\pi k)}{\cosh(2\pi k)},
\end{equation}
to be inserted in the amplitude in the momentum $k$ sector. A summary of the situation is given in figure \ref{branchcut2}.
The periodicity / antiperiodicity of the fermions around the small circle in the ZZ-ZZ picture, maps into the local fermionic boundary conditions $\eta=+1$ / $\eta=-1$ respectively in the Liouville disk supergravity picture.

\begin{figure}[H]
\centering
\includegraphics[width=0.99\textwidth]{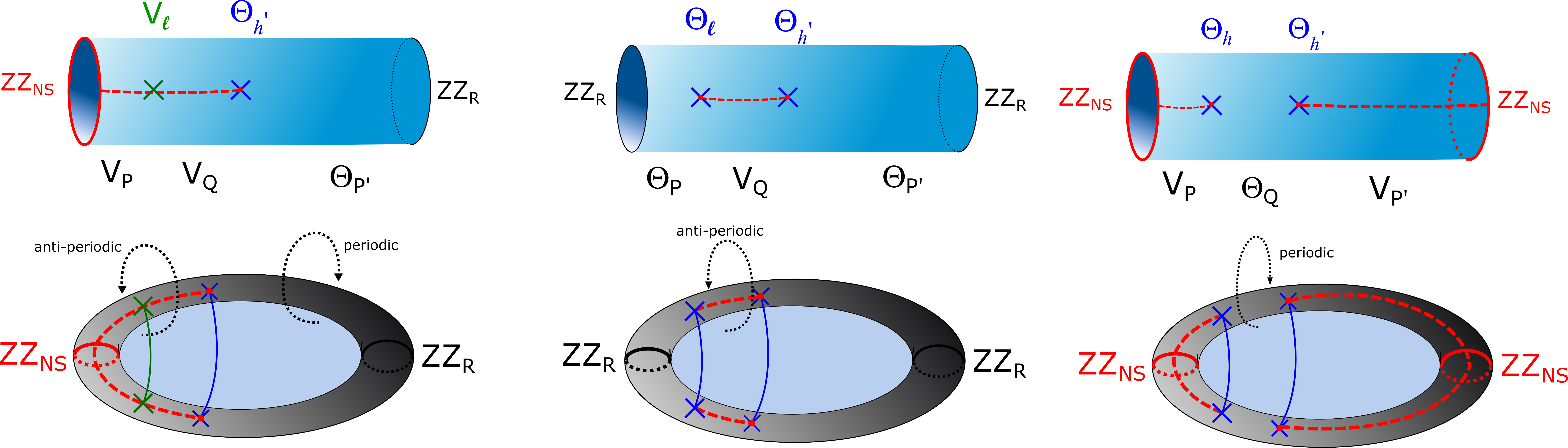}
\caption{NS operators (green) and R operators (blue) inserted in the ZZ-ZZ brane system. R-sector operators create or soak up a branch cut, where the NS operators do not adjust branch cuts. Top: Examples of four-point functions (two Liouville operators) with several spin operators in Liouville language. Bottom: Resulting JT bilocal disk diagram.}
\label{branchcutextra}
\end{figure}

For higher-point functions, we can determine the intermediate set of operators from the fusion rules (figure \ref{branchcutextra}). 
E.g. for the left figure \ref{branchcutextra} we have $V_P \times V_\ell \to V_Q$, whereas for the middle figure we get $\Theta_P \times \Theta_\ell \to V_Q$ and the right figure has $V_P \times \Theta_\ell \to \Theta_Q$. Each time we have a fusion in the NS sector $V_Q$ we have a $k \sinh 2\pi k$ density and a branchcut, whereas when we have an R fusion $\Theta_Q$, we get $\cosh 2 \pi k$ and no branchcut.

This readily generalizes to out-of-time ordered correlators.

\mciteSetMidEndSepPunct{}{\ifmciteBstWouldAddEndPunct.\else\fi}{\relax}
\bibliographystyle{utphys}
{\small \bibliography{references}{}}

\end{document}